\catcode`\@=11
\newif\if@fewtab\@fewtabtrue
{\count255=\time\divide\count255 by 60
\xdef\hourmin{\number\count255}
\multiply\count255 by-60\advance\count255 by\time
\xdef\hourmin{\hourmin:\ifnum\count255<10 0\fi\the\count255}}
\def\ps@draft{\let\@mkboth\@gobbletwo
    \def\@oddhead{}
    \def\@oddfoot
       {\hbox to 7 cm{$\scriptstyle Draft\ version:\ \draftdate$
       \hfil}\hskip -7cm\hfil\rm\thepage \hfil}
    \def\@evenhead{}\let\@evenfoot\@oddfoot}

\def\ceqno{\global\@fewtabfalse
    \ifcase\@eqcnt \def\@tempa{& & &}\or \def\@tempa{& &}
      \or \def\@tempa{&}
      \or\def\@tempa{}\fi\@tempa
{\rm(\theequation)}}

\def\aeqno#1{\global\@fewtabfalse
    \ifcase\@eqcnt \def\@tempa{& & &}\or \def\@tempa{& &}
      \or \def\@tempa{&}
      \or\def\@tempa{}\fi\@tempa
{\rm(\theequation,#1)}}

\def\label#1{\ifnum\draftcontrol=1
 \global\def\draftnote{$\scriptstyle #1$}\fi
 \@bsphack\if@filesw {\let\thepage\relax
   \def\protect{\noexpand\noexpand\noexpand}%
\xdef\@gtempa{\write\@auxout{\string
      \newlabel{#1}{{\@currentlabel}{\thepage}}}}}\@gtempa
   \if@nobreak \ifvmode\nobreak\fi\fi\fi
  \@esphack}

\def\alabel#1#2{\label{#1}\global\@fewtabfalse
    \ifcase\@eqcnt \def\@tempa{& & &}\or \def\@tempa{& &}
      \or \def\@tempa{&}
      \or\def\@tempa{}\fi\@tempa
{\hbox to 3cm{\phantom{\rm(\theequation,#2)}
\draftnote \hfil}\hskip -3cm {\rm(\theequation,#2)}}}

\def\clabel#1{\label{#1}\global\@fewtabfalse
    \ifcase\@eqcnt \def\@tempa{& & &}\or \def\@tempa{& &}
      \or \def\@tempa{&}
      \or\def\@tempa{}\fi\@tempa
{\hbox to 3cm{\phantom{\rm(\theequation)}
\draftnote \hfil}\hskip -3cm{\rm(\theequation)}}}

\def\eqnarray{\def\draftnote{{}}\global\@fewtabtrue
\stepcounter{equation}\let\@currentlabel=\theequation
\global\@eqnswtrue
\global\@eqcnt\z@\tabskip\@centering\let\\=\@eqncr
$$\halign to \displaywidth\bgroup\@eqnsel\hskip\@centering\@eqcnt\z@
  $\displaystyle\tabskip\z@{##}$&\global\@eqcnt\@ne
  \hskip 1\arraycolsep \hfil${##}$\hfil
  &\global\@eqcnt\tw@ \hskip 1\arraycolsep
$\displaystyle\tabskip\z@{##}$
\hfil  \tabskip\@centering&\global\@eqcnt\thr@@\llap{##}\tabskip\z@
\cr}

\def\endeqnarray{\@@eqncr\egroup
      \global\advance\c@equation\m@ne$$\global\@ignoretrue}

\def\@eqnnum{\hbox to 3cm{\phantom{\rm(\theequation)} \draftnote
                         \hfil}\hskip -3cm {\rm(\theequation)}}

\def\@@eqncr{\let\@tempa\relax
    \ifcase\@eqcnt \def\@tempa{& & &}\or \def\@tempa{& &}
      \or \def\@tempa{&}
      \or\def\@tempa{}
\fi\@tempa
\if@eqnsw
\if@fewtab\@eqnnum\fi
\stepcounter{equation}\fi\global
\@eqnswtrue\global\@eqcnt\z@\global\@fewtabtrue\cr}

\def\draftcite#1{\ifnum\draftcontrol=1#1\else{}\fi}

\def\@lbibitem[#1]#2{\item{}\hskip -3cm \hbox to 2cm
{\hfil$\scriptstyle\draftcite{#2}$}\hskip
1cm[\@biblabel{#1}]\if@filesw
     {\def\protect##1{\string ##1\space}\immediate
      \write\@auxout{\string\bibcite{#2}{#1}}}\fi\ignorespaces}

\def\@bibitem#1{\item\hskip -3cm \hbox to 2cm
{\hfil $\scriptstyle\draftcite{#1}$}\hskip 1cm
\if@filesw \immediate\write\@auxout
       {\string\bibcite{#1}{\the\value{\@listctr}}}\fi\ignorespaces}

\catcode `\@=11
\@addtoreset{equation}{section}
\def\theequation{\arabic{section}.\arabic{equation}}
\catcode `\@=12
\def\nsection#1{\section{#1}\setcounter{equation}{0}}

\def\th{\theta}         \def\Th{\Theta}
\def\ga{\gamma}         \def\Ga{\Gamma}
\def\be{\beta}
\def\al{\alpha}
\def\ep{\epsilon}
\def\la{\lambda}        \def\La{\Lambda}
\def\de{\delta}         \def\De{\Delta}
\def\om{\omega}         \def\Om{\Omega}
\def\sig{\sigma}        \def\Sig{\Sigma}

       \def\CB{{\cal B}}       \def\CC{{\cal C}}
\def\CD{{\cal D}}              \def\CF{{\cal F}}
       \def\CH{{\cal H}}       
\def\CJ{{\cal J}}              \def\CL{{\cal L}}
       \def\CN{{\cal N}}       \def\CO{{\cal O}}
\def\CP{{\cal P}}              \def\CR{{\cal R}}
\def\CS{{\cal S}}

\newcommand{\NR}{{{\bf R}}}
\newcommand{\NZ}{{{\bf Z}}}

\newcommand{\Nq}{{{\bf q}}}

\def\qq{ \begin{eqnarray} }
\def\qqq{ \end{eqnarray} }
\def\non{ \nonumber }
\newcommand{\no}{\noindent}
\newcommand{\vs}{\vspace}

\newcommand{\p}{\partial}
\newcommand{\un}{\underline}
\newcommand{\rmp}{\mathrm{p}}
\newcommand{\rmk}{\mathrm{k}}
\newcommand{\rmx}{\mathrm{x}}

\newcommand{\hf}{{_1\over^2}}
\newcommand{\ha}{{1\over 2}}

\def\draftdate{\number\month/\number\day/\number\year\ \ \ \hourmin }

\global\def\draftcontrol{0}
\catcode`\@=12

\documentclass[12pt]{article}

\usepackage{mathbbol}

\renewcommand{\theequation}{\thesection.\arabic{equation}}
\def\theequation{{\thesection.\arabic{equation}}}

\begin{document}

\begin{center}

{\Large{\bf{On the derivation of Fourier's law for coupled anharmonic oscillators}}}

\vs{0.5cm}

{\large{Jean Bricmont}}

UCL, FYMA, chemin du Cyclotron 2,\\ 
B-1348  Louvain-la-Neuve, Belgium\\

\vs{ 0.2cm}

{\large{Antti Kupiainen}}\footnote{Partially supported by the 
Academy of Finland.
}

Department of Mathematics,
Helsinki University,\\
P.O. Box 4, 00014 Helsinki, Finland\\

\end{center}
\vs{ 0.2cm}

\vskip 0.3cm
\begin{center}
\end{center}
\vskip 0.2cm

\begin{abstract}

We study the Hamiltonian system made of weakly coupled anharmonic
oscillators arranged on a three dimensional lattice $\mathbb{Z}_{2N}\times \mathbb{Z}^{2}$, and subjected to a
stochastic forcing mimicking heat baths of temperatures
$T_1$ and $T_2$ on the hyperplanes at $0$ and $N$.
We introduce a truncation of the Hopf equations 
describing the stationary state of the system which
leads to a nonlinear equation for the two-point stationary correlation
functions. We prove that these equations have a unique solution
which, for  $N$ large, is approximately a local equlibrium
state satisfying Fourier law that relates the heat current
to a local temperature gradient. The temperature exhibits
a nonlinear profile.
\end{abstract}

\date{ }

\vskip 0.3cm
\begin{center}
\end{center}
\vskip 0.2cm

\nsection{Introduction}

Fourier's law states that a local temperature gradient
is associated with a flux of heat $J$ which is proportional to the gradient:
\qq
J(x) = -k(x) \nabla T(x)
\label{(I1)}
\qqq
where the {\it heat conductivity} $k(x)$ is a function of
the temperature at $x$ : $k(x) =
\tilde k (T(x))$.

Fourier's law is experimentally observed in a variety of materials from gases to
solids at low and at high temperatures. It also belongs to basic textbook
material. However, a first principle derivation of the law is missing and,
many would say, is not even on the horizon.

The quantities $T$ and $J$ in (\ref{(I1)}) are macroscopic variables,
statistical averages of the variables describing the microscopic dynamics of
matter. A first principle derivation of (\ref{(I1)}) entails a definition of
$T$ and $J$ in terms of the microscopic variables and a proof of the law in
some appropriate limit.

An example of an idealized physical situation would be a crystal occupying the
region $[0,N] \times \mathbb{R}^2$ in $\mathbb{R}^3$. The crystal is heated at
the  two boundaries by uniform temperatures, $T_{1}$ on $\{0\} \times
\mathbb{R}^2$ and $T_{2}$ on $\{N\} \times \mathbb{R}^2$. Then, for large $N$, the
relation (\ref{(I1)}) should hold, with corrections ${o} (1/N)$ and $\nabla T
= \CO (1/N) = J$.

The proper description of the crystal would be quantum mechanical, but, being a
macroscopic law, (\ref{(I1)}) is expected to hold for a classical system as
well and the quantum corrections are expected to be small except at low
temperatures. An example of a classical toy model of such a system is given by 
coupled oscillators organized on a $d$-dimensional lattice $\mathbb{Z}^d$.

Consider a subset $\Lambda=[0,N] \times \mathbb{Z}^{d-1}$ of $\mathbb{Z}^d$ and let
$x \in \Lambda$ index the dynamical variables, coordinates $q_{x}$ and momenta
$p_{x}$. The dynamics in the phase space $\mathbb{R}^{2\Lambda} = \{(q_{x},p_{x})
\mid x \in \Lambda\}$ is defined in terms of the Hamiltonian
$$
H (q,p) = {1\over 2} \sum_{x} p^2_{x} + V(q)
$$
i.e. the Hamiltonian flow is given by the system
\qq
\dot q_{x}& =& p_{x} \label{(2a)}\\
\dot p_{x} &= &- {\p V \over \p q_{x}}.
\label{(2b)}
\qqq
We can think of this model as describing atoms with unit mass with equilibrium
positions at $x \in \Lambda$ and $q_{x}$ being the deviation of the position of
the atom indexed by $x$ from its equilibrium ($q_{x}$ should of course be in
$\mathbb{R}^d$
but we simplify and take $q_{x} \in \mathbb{R}$), while $p_{x}$ is the momentum
(velocity) of the atom.

The potential $V$ should describe the forces between the atoms and possible
restoring or \textit{pinning} forces pulling $q_{x}$ to the equilibrium $q_{x} =
0$. Let, for unit lattice vectors $e_{\al}$
$$
\nabla_{\al} q (x) = q (x+ e_{\al}) - q (x)
$$
denote the discrete derivative. Then, the interactions are described by a potential $U(\nabla q)$ whereas the
pinning is described by $W(q)$, and $V=U+W$. 
In the simplest case, $U$ and $W$ are local:
\qq
W(q) &=& \sum_{x \in \Lambda} w (q_{x}),\non
\\
U(\nabla q) &=& \sum_{x,x+e_{\al} \in \Lambda} u \Bigl(\nabla_{\al} q(x)\Bigr),
\non
\qqq
and, to model small oscillations, $w$ and $u$ are given by a low order
polynomial:
$$
w(q) = aq^2 + bq^3 + cq^4,
$$
and $u$ similarily.

In addition to the Hamiltonian dynamics of the system, we want to model the
heating of the system at the boundary. The simplest way to model this is to add
stochastic forces to (\ref{(2b)}) for $x\in \p \Lambda$, see Section 2 for
details. Then the deterministic flow (\ref{(2a)}, \ref{(2b)}) is replaced by a Markov
process  $(q(t), p(t))$ and the Fourier law will be a statement about
the
stationary state of this Markov process.

The first question one would like to answer is the existence and uniqueness of
the stationary state. When $T_{1} = T_{2} = T$ such a state exists and is  a
Gibbs state, given formally by
\qq
{1\over Z} e^{-\beta H (q,p)} dq \ dp
\label{(I3)}
\qqq
(see Sect 2), with $\beta = 1/T$. This is customarily refered to as the
\textit{equilibrium} state in contrast to the non-equilibrium situation $T_{1} \neq
T_{2}$. In the latter case, there is no simple formula like (\ref{(I3)}) and,
indeed, in our setup, even the existence of a stationary state is an open
problem. In the $d=1$ case of a finite chain of $N+1$ oscillators the
existence is proved  provided the interaction potential $U$ dominates
the pinning one $W$ (see \cite{EPR1, EPR2, EH1, EH2, HN, Ec, RT1, RT2, RB}). In this case,
uniqueness is also proved, i.e. the Markov process converges to this state as
$t\to \infty$.

Supposing that we have a stationary state $\mu$, let us formulate the statement
(\ref{(I1)}). The Hamiltonian flow  (\ref{(2a)}, \ref{(2b)})  preserves the total energy
$H$. In particular, if we write $H$ as a sum of local terms, each one pertaining to
a single oscillator:
$$
H =\sum_{x\in \Lambda} H_{x},
$$
then, under the flow (\ref{(2a)}, \ref{(2b)}),
$$
\dot H_{x} = \sum_{\al} \nabla_{\al} j_{\al} (x) \equiv \nabla \cdot \vec \jmath(x),
$$
where the \textit{microscopic heat current} $\vec \jmath(x)$ will depend on $p_{y}$ and $q_{y}$
for $y$ near $x$ (see (\ref{5.16a})  for a concrete expression). Let also $t(x) =
{1\over 2} p_x^2$ be the kinetic energy of the oscillator indexed by $x$. Then, the
macroscopic temperature and heat current in eq. (\ref{(I1)}) are defined by
\qq
T(x) &=& {\rm E}_{\mu} \ t(x)\non
\\
J(x) &=& {\rm E}_{\mu} \ j(x) \non
\qqq
where $ {\rm E}_{\mu} $ denotes expectation in the stationary
state.

\vskip 2mm

We do not attempt here to give a comprehensive review of the status
of (\ref{(I1)}), but refer the reader to the reviews \cite{BLR},
 \cite{LLP} and \cite{Spohn}. There is also substantial amount of work
on Fourier's law for fully stochastic models (i.e. where there is
noise in the bulk too),  going back to \cite{GKMP}, \cite{KMP}, see eg.
 \cite{BLL}, \cite{Olla1}, \cite{Olla2}, \cite{Olla3}, \cite{EY1},  \cite{EY2}.

The only rigorous results in our model are for the harmonic case where $U$ and
$W$ are quadratic \cite{RLL, SL}. 
In that case, Fourier's law {\it does not hold}: the current $\vec \jmath$ 
is  $\CO(1)$ as
$N\to\infty$ whereas $\nabla T = 0$ except near the boundary. If the model has
pinning, i.e. $W\neq 0$ and $U$ or $W$ are not harmonic,
the law seems to hold
in simulations in all dimensions \cite{AK}. In the unpinned $W=0$ anharmonic case,
conductivity seems anomalous in low dimensions: $k$ in (\ref{(I1)}) depends on $N$
as $N^\al$ in $d=1$ and logarithmically in $d=2$. It is a major challenge to
explain the $\al$ which, numerically, seems to be in the interval $[1/3,2/5]$
(see  \cite{LLP1}, \cite{NR}, \cite{russian} for theories on $\al$).

\vskip 2mm

One way to try to get hold of the stationary state $\mu$ is via its
correlation functions. Denote $u_{x} = (q_{x},p_{x})^T$ and choose $U,W$ even
for simplicity. Then, the Hamiltonian vector field (\ref{(2a)}, \ref{(2b)}) is a sum of a
linear and a cubic term in $u$. Therefore, the correlation functions
$$
G_{n} (x_{1}, x_{2},..., x_{n}) = E_{\mu} u_{x_1} \otimes ... \otimes u_{x_{n}}
$$
satisfy a linear set of equations
\qq
\p_{t} G_{n} = A_{n} G_{n} + V_{n} G_{n+2} + C_{n} G_{n-2},
\label{(I4)}
\qqq
where $A_{n}, V_{n}, C_{n}$ are linear operators coming respectively from the
linear and cubic terms of (\ref{(2a)}, \ref{(2b)}), and the term coming from the noise via Ito's formula.
See Section 3 for a detailed derivation of these equations.

Such equations for the correlation functions are known as the BBGKY hierarchy
in particle systems or the Hopf equations in turbulence. Although linear, the
system (\ref{(I4)}) is intractable due to the appearance of $G_{n+2}$ in the
equation for $G_{n}$.

In this paper we will consider the situation where the \textit{equilibrium} $T_{1} = T_{2}$
Gibbs measure is close to a Gaussian measure. 
This holds if the anharmonicity in $u$ and $w$ (the coefficients $b$ and $c$) is
weak and the harmonic part in $w$ (i.e. the pinning) is large (i.e. the Gibbs
measure is far from critical), as we will assume. In such cases, we expect that
the non-equilibrium measure  is also close to a Gaussian. In such a situation, one
can attempt a \textit{closure} of the Hopf equations, i.e. to express the
higher order correlation functions $G_{n+2}$ in terms of $G_{m}$ with $m \leq n$,
thereby obtaining a finite set of equations for $G_{m}$, $m \leq n$.

We will introduce such a closure and solve the closed equations. The
simplest closure would be to write, for $n=2$, the equation
$$
G_{4} = \sum G_{2} \otimes G_{2} + G^c_{4},
$$
and set $G^c_{4}=0$, thereby obtaining a closed quadratic equation for $G_{2}$.
It turns out that this is too simple: the solution will be qualitatively similar to the one of 
 the harmonic case. Our closure is done to the $G_{4}$-equation by
setting the \textit{connected 6-point function} to zero. This is an
\textit{uncontrolled approximation} that we do not know how to justify rigorously. 
An analogous approximation was studied  in both classical
and quantum systems in \cite{Spohn} 
and  has been used in  \cite{LeS}
in a model similar to ours, but in a translation invariant
setting and in one-dimension; our model, in one-dimension,  was further studied, theoretically and numerically,
 in 
  \cite{ALS}.

Our motivation for studying the closure equations in detail
is on the one hand in the interesting picture of the local
equilibrium state that emerges and on the other hand
in building approaches that go beyond this approximation.
Traditionally one arrives to such a closure in an appropriate
limit, the "kinetic limit", which in our case means taking the
anharmonicity proportional to $N^{-\hf}$ and rescaling
distances by $N$, see \cite{Spohn}. One then arrives
to a Boltzman equation and after a further limit \cite{Spohn}
to the Fourier law. Most of the structure of the stationary
state correlations disappear in these limits. 

In our case we arrive to approximate expressions of stationary
correlation functions whithout any limits (admittedly with no control
of the corrections!) and then can study how the Fourier
law emerges from these expressions. Other interesting
phenomena emerge too. In particular,
one expects the presence of very long range spatial
correlations in the stationary state even though the equilibrium
state has exponential decay of correlations. Although we do not demonstrate
the presence of these long range correlations, because we obtain only upper bounds on the decay rates, not lower ones, we show how to handle the technical
problems caused by these slowly decaying correlations, 
and that could be useful in other contexts.
Finally, we believe some of the methods
developed in this paper could be of use in trying to
prove the existence of the kinetic limit and Fourier's law
therein.

The outline of the paper (to which the reader can return later) is as follows: in the next section, we define our model and we derive the Hopf equation (or BBGKY hierarchy) in section 3. In section 4, we explain the particular closure that we will study, leading to our final equations (see (\ref{4(18)})- (\ref{4(20)}) below). Section 5 is  devoted to several changes of variables: we first apply a Fourier transformation, and then
we introduce variables that could be called slowly and fastly varying, namely the one of which the non translation invariant part of of the correlation functions depends (slow) and the one related to the transtation invariant part (fast). Next, we outline our arguments and state qualitatively our main result (section 6).  To prove the latter, we first derive (in section 7) identities satisfied by our equations, which take the form of current conservation equations, consisting of an energy conservation law and a number conservation law (the presence of the latter being, to some extent, a consequence of our approximations, i.e. of our closure). We also write down the stationary states in the translation invariant case. These do  not reduce, for our closed equations, to the usual Gibbs states, but depend on two parameters, the temperature, as one would expect, but also a  "chemical potential", corresponding to the number conservation law. These conservation laws are related to the presence of zero modes in the linearization of our equations, which are discussed also in section 6. In fact,  the current conservation equations coincide with the projection of the full equations on the zero modes. In section 8, we define precisely the spaces in which our equations are solved and we state our main result in a more technical form. The solution that we shall construct is the sum of a modified stationary state, with coefficients (temperature and chemical potential) slowly varying in space, and a perturbation.

The main technical problem that we face is that the nonlinear terms in our equations involve collision kernels that are delta functions (or principal values) (see (\ref{5(20a)})- (\ref{5(21a)}) below). Since   we want to solve our equations by using a fixed point theorem, we need to show that the nonlinear terms  belong to the space that we introduced, and, because of the presence of the delta functions, this is rather technical. Section 9 is devoted to solving those problems, but most of the proofs of that section are given in Appendix B. Another problem is that the linear operator in  our equations is not invertible, because of the zero modes. In section 10, we show that our linear operator can be inverted in the complement of the zero modes. This uses the fact that this operator is a sum of a multiplication operator and a convolution. To show invertibility, it is useful to know that the convolution operator is compact and this in turn follows from H\"older regularity properties that are proven in section 9. Finally, in section 11, we prove our main result, which consist in using the result of section 10 to solve the equations in the complement of the zero modes with the solution to the current conservation equations, i.e. of the projection of the full equations onto the zero modes. The first equations lead to Fourier's law, namely an expression of the conserved currrents in terms of the parameters of the modified stationary state (temperature and  chemical potential ). And the second equations determine the spatial dependence of those parameters.

\nsection{Lattice dynamics with boundary noise}
\vskip 0.2cm
Let us define the model we consider in more detail. Instead of
working in the strip $[0,N] \times \mathbb{Z}^{d-1}$ it is convenient
to double it to the cylinder
 $V=  \mathbb{Z}_{2N} \times  \mathbb{Z}^{d-1}$ where $ \mathbb{Z}_{2N} $ are the integers modulo $2N$. 
 The noise is put on the "boundary" $ \{ 0 \}\times \mathbb{Z}^{d-1} \cup  \{ N \}\times \mathbb{Z}^{d-1}$.
 We consider the phase space $(q,p) \in
\mathbb{R}^{2V}$ i.e. $q=(q_{x})_{x \in V}$, $q_{x} = q_{x+y}$ for
$y=(2N,0)$, and similarly for $p_x$. The dynamics is given by the stochastic differential
equations
\qq
dq_x &=& p_{x}dt  \non\\
dp_{x} &=& \left(-{\p H \over \p q_{x}} - \ga_{x} p_{x}\right) dt + d\xi_{x}
\label{2(1)}
\qqq
where
\qq
H (q,p) = {1\over 2} \sum_{x\in V} p^2_{x} + {1\over 2} (q,\omega^2q) +
{\la \over 4} \sum_{x\in V} q^4_{x},
\label{2(2)}
\qqq
\qq
\ga_{x} = \ga (\delta_{x_10} + \delta_{x_1N})
\label{2(3)}
\qqq
and the random variables $\xi_{x} (t)$ are  Brownian motions with covariance
\qq
E \xi_{x} \xi_{y} = 4\ga \delta_{xy} (T_{1} \delta_{x_10} + T_{2}
\delta_{x_1N}) t.
\label{2(4)}
\qqq
The Hamiltonian (\ref{2(2)}) describes a system of coupled anharmonic
oscillators with coupling matrix $\omega^2$:
\qq
(q,\omega^2 q) = \sum_{x,y\in V} q_{x} q_{y} \omega^2 (x-y)
\non
\qqq
Our analysis requires that the Fourier transform $\omega^2 (k)$ of $\omega^2$
is smooth and
\qq
\omega^2 (k) = m^2 + \rho (k)
\non
\qqq
with $\rho (k) = \CO (k^2)$ as $k \to 0$, and $m^2 > 2 \| \rho \|_{\infty}$.
Moreover we will need some regularity properties  that will be checked explicitely for
\qq
\omega^2 = (-\Delta + m^2)^2,
\non
\qqq
i.e. 
\qq
\omega (k) = 2 \sum^d_{\alpha=1} (1-\cos k_{\alpha}) + m^2,
\label{2(4b)}
\qqq
see the proof of Proposition 9.3.

\vspace*{4mm}

As explained in the Introduction, the equations (\ref{2(1)}) describe a 
Hamiltonian dynamics subjected to stochastic heat baths on the ``boundaries" of
$V$, with temperature $T_{1}$ on the hyperplane $x_{1}=0$, and
temperature $T_{2}$ on the hyperplane $x_{1}=N$. This defines a Markov process
$(q(t),p(t))$ in the phase space $\NR^{2V}$ and we are interested in the stationary states
for this process.

\vspace*{4mm}

If the temperatures are equal, $T_{1}=T_{2}=T$, an explicit stationary state
is given by the Gibbs state at temperature $T$ of the Hamiltonian $H$. This
probability measure is given as a weak limit
\qq
\nu_{T} = \lim_{M \to \infty} {1\over Z_{M}} \exp \left[- {\la\over T} 
\sum_{x\in V_{M}} q^4_{x}\right] \mu_{T} (dp,dq)
\label{2(5)}
\qqq
where $V_{M}$ is defined by $|x_{i}| < M$, $i=2,\dots,d$, and $\mu_{T}$ is
the Gaussian measure with covariance
\qq
Ep_{x}p_{y} = T\delta_{xy},  \ \ E p_{x}q_{y} = 0
\label{2(6)}
\qqq
\qq
E q_{x} q_{y} = T \omega^{-2} (x-y)
\label{2(7)}
\qqq
 For small $\la$, the Gibbs measure $\nu_{T}$ is nearly Gaussian, with
(\ref{2(6)}) still true and small $\CO (\la)$ corrections in (\ref{2(7)}).
It is very well understood via cluster expansions. Physically, the fact that
the Markov process reaches the stationary distribution $\nu_{T}$ means that
the heat introduced at the boundary spreads inside the system, which reaches
equilibrium at temperature $T$.

\vspace*{4mm}

When $T_{1} \neq T_{2}$, things are very different. Even the existence of a
stationary state is not known rigorously (not even in finite volume, $M <
\infty$). However, physically, one expects a unique stationary state $\nu$ to
exist. In this paper we assume this and inquire about the properties of $\nu$.
In particular one would like to understand how the heat from the boundary now
spreads inside the system: what is its stationary temperature distribution and
what sort of flux of heat exists in it.

\nsection{Hopf equations}
\vskip 0.2cm
Let us introduce a more compact notation for the stochastic differential
equation (\ref{2(1)}). Denote $(q,p)^T = u$, $\La (u) = -\la
(0,q^3)^T$, $(\Ga u)_{x} = (0,\ga_{x}p_{x})^T$ and $\eta = (0,\xi)^T$.
Then (\ref{2(1)}) becomes
\qq
du(t) = \Bigl((A-\Ga)u + \La (u)\Bigr) dt + d\eta (t)
\label{3(1)}
\qqq
where
\qq
A = \left(\begin{array}{ccccc}
0 & 1\\
-\omega^2 & 0
\end{array}\right)
\non
\qqq
Define the correlation functions
\qq
G_{n} (x_{1}, ..., x_{n},t) = {\rm E} \ u_{x_{1}} (t) \otimes ... \otimes u_{x_{n}}
(t) \in \NR^{2nV}
\non
\qqq
By Ito's formula, we get
\qq
\dot G_{n} = (A_{n} - \Ga_{n}) G_{n} + \La_{n} G_{n+2} + C_{n} G_{n-2},
\label{3(2)}
\qqq
where
\qq
A_{n} = A \otimes 1 \otimes ... \otimes 1 + ... 1 \otimes 1 \otimes ...
\otimes A
\non
\qqq
and $ \Ga_{n}$ is defined similarly. Moreover,
\qq
\La_{n} G_{n+2} &=& \sum^n_{i=1} E u_{x_{1}} \otimes ... \otimes \La
(u)_{x_{i}} \otimes ... \otimes u_{x_{n}}, \non \\
C_{n} G_{n-2} &=& \sum_{i<j} \CC_{x_{i}x_{j}} G_{n-2} (x_{1}, ..., \hat x_{i},
..., \hat x_{j} ... x_{n})
\non
\qqq
where the arguments $\hat x_{i}$, $\hat x_{j}$ are missing.
This defines linear operators from $\NR^{2(n+2)V} \to \NR^{2nV}$ and
$\NR^{2(n-2)V} \to \NR^{2nV}$ respectively. $\CC$ equals one-half the time derivative of the
covariance of $\eta$, i.e.
\qq
\CC = \left(
\begin{array}{ccccc}
0 & 0\\
0 & C
\end{array}
\right),
\non
\qqq
with
\qq
C_{xy} = 2\ga \de_{xy} (T_{1} \de_{x_10} + T_{2} \de_{x_1N}).
\label{4(16)}
\qqq

Suppose $\nu$ is a stationary state of the process $u(t)$. Then
equation  (\ref{3(2)}) leads to a linear set of equations for the stationary
correlation functions
\qq
G_{n} (x_{1}, ..., x_{n}) = \int \otimes^n_{i=1} u_{x_{i}}  \ \nu (du)
\label{3(3)}
\qqq
\qq
(A_{n} - \Ga_{n}) G_{n} + \La_{n} G_{n+2} + \CC_{n} G_{n-2} = 0.
\label{3(4)}
\qqq
The equations (\ref{3(4)}) have the drawback that they do not ``close": to solve for $G_{n}$,
we need to know $G_{n+2}$.

For $\la$ small, the equilibrium $T_{1} = T_{2}$ measure is close to Gaussian.
When $T_{1} \neq T_{2}$ we expect this to remain true; however, the measure
will not satisfy $Ep_{x}q_{y}=0$.

We look for a Gaussian approximation to equation (\ref{3(4)}) for small $\la$
by means of a {\it closure}, i.e. expressing the $G_{n}$ in terms of $G_{2}$. Since
$G_{n}\neq 0$ only for $n$ even, the first equation in (\ref{3(4)}) reads:
\qq
(A_{2} - \Ga_{2}) G_{2} + \La_{2} G_{4} + \CC = 0
\label{3(5)}
\qqq
The simplest closure would be to replace $G_{4}$ in (\ref{3(5)}) by the
Gaussian expression
\qq
\sum_{p} G_{2} (x_{i}, x_{j}) \otimes G_{2} (x_{k}, x_{l})
\label{3(6)}
\qqq
where the sum runs over the pairings of $\{1,2,3,4\}$. Equations (\ref{3(5)}) and (\ref{3(6)})
lead to a nonlinear equation for $G_{2}$. It turns out that the solution to
this equation is qualitatively similar to the $\la=0$ case, i.e. $G_{2}$ does
not exhibit a temperature profile nor a finite conductivity. The only effect
of the nonlinearity is a renormalization of $\omega$. We will therefore not
discuss this closure any further.

The next equation is
\qq
(A_{4} - \Ga_{4}) G_{4} + \La_{4} G_{6} + \CC_{4} G_{2} = 0
\non
\qqq
Write
\qq
G_{4} = \sum_{p} G_{2} \otimes G_{2} + G^{c}_{4}
\non
\qqq
and
\qq
G_{6} = \sum_{p} G_{2} \otimes G_{2} \otimes G_{2} + \sum_{p'} G_{2} \otimes
G^c_{4} + G^c_{6}
\non
\qqq
where $G^c_{4}$ and $G^c_{6}$ are the connected correlation functions
describing deviation from Gaussianity and the sums run over the usual partitions
of indices. After some algebra, we may write the first two Hopf
equations in the following form:
\qq
(A_{2} - \Ga_{2} + \Sig_{2}) G_{2} + \La_{2} G^c_{4} + \CC = 0
\label{3(7)}
\qqq
\qq
(A_{4} - \Ga_{4} + \Sig_{4}) G^c_{4} + b (G_{2}) + \La_{4} G^c_{6} = 0,
\label{3(8)}
\qqq
where the operators $\Sig_{2}$ and  $\Sig_{4}$ are
\qq
\Sig_{2} (G_{2}) G_{2} &=& \La_{2} \sum_{p} G_{2} \otimes G_{2}\\
\Sig_{4} (G_{2}) G^c_{4} &=& \sum_{p} \La'_{4} G_{2} \otimes G^c_{4}
\non
\qqq
and $\La'_{4}$ means the following: $G_{2} \otimes G^c_{4}$ belongs to
$(\NR^{2V})^{\otimes 2} \otimes (\NR^{2V})^{\otimes 4}$;  $\La'_{4}$ is a
sum of terms
\qq
\La'_{4} = \sum_{i<j<k} \La^{ijk}_{4}
\non
\qqq
where $\La^{ijk}_{4}$ acts with $\La_{4}$ in the spaces $i,j,k$ and as
identity in the rest. $\La'_{4}$ then has at least one of the indices $i,j,k$ equal to either
1 or 2.

Finally,
\qq
b(G_{2}) = \sum_{p} \La''_{4} G_{2} \otimes G_{2} \otimes G_{2},
\non
\qqq
where $\La''_{4}$ is similar to $\La'_{4}$, but with $\La^{ijk}_{4}$ 
acting on all the three factors $G_{2}$.
Explicitely, denote $i = (\al,x) \in \{1,2\} \times  \NR^{V}$, so that
$u=(u_{i})$, $u_{(1,x)} = q_{x}, u_{(2,x)} = p_{x}$.
Then,
\qq
b(G_{2})_{i_{1}i_{2}i_{3}i_{4}} = 6 \sum_{j_{2}j_{3}j_{4}}
\La_{i_{1}j_{2}j_{3}j_{4}} \prod^4_{a=2} (G_{2})_{i_{a}j_{a}} + (i_{1}
\to i_{b}, b=2,3,4)
\non
\qqq
with
\qq
\La_{j_{1}j_{2}j_{3}j_{4}} = - \la \de_{\al_1 2} \prod^4_{a=2} \de_{\al_{a}1} 
\de_{x_{1}x_{a}}.
\label{3(8a)}
\qqq
Equation (\ref{3(8)}) may be solved for $G^c_{4}$:
\qq
G^c_{4} = - (A_{4} - \Ga_{4} - \Sig_{4})^{-1} \Bigl(b(G_{2}) + \La_{4}
G^c_{6} \Bigr),
\label{3(9)}
\qqq
provided that $A_{4} - \Ga_{4}-\Sig_{4}$ is invertible. Substitution  of
(\ref{3(9)}) in (\ref{3(7)}) yields a nonlinear equation for $G_{2}$ with dependence on
$G^c_{6}$:
\qq
(A_{2} - \Ga_{2}+ \Sig_{2}) G_{2} + \CN (G_{2}, G^c_{6}) +  \CC = 0
\label{3(10)}
\qqq
with
\qq
\CN (G_{2}, G^c_{6}) = - \La_{2} (A_{4} - \Ga_{4} + \Sig_{4})^{-1} \Bigl(b(G_{2}) + \La_{4} G^c_{6}\Bigr)
\label{3(11)}
\qqq
If we set $G^c_{6}=0$ in (\ref{3(10)}) we get a closed equation for $G_{2}$.
However, before defining the closure equation to be studied, let us discuss
in more detail the term $b(G_{2})$.

\nsection{Closure}
\vskip 0.2cm

The leading term (in powers of $\la$ and $\ga$) in (\ref{3(11)}) is
\qq
\CN' = - \La_{2} A^{-1}_{4} b (G),
\label{4(1)}
\qqq
where we write, as we shall do from now on, $G$ for $G_2$, since we shall only deal with $G_2$.
Let us write this more concretely.

To define the inverse of $A_{4}$ we define the stationary state correlation
functions $G_{n}$ as limits $\ep \to 0$ of the stationary state $G^{\ep}_{n}$ where, in
equation (\ref{2(1)}), we add a term $\ep p_x dt$, and in (\ref{2(4)}) a term $4\ep t
\de_{xy}$, i.e. we put a noise and a friction of size $\ep$ everywhere. Then,
 the matrix $A$ becomes
\qq
A = \left(\begin{array}{cccc}
0 & 1\\
-\om^2 & -\ep
\end{array}\right)
\label{4(2)}
\qqq
and, letting
\qq
R(t)= e^{tA},
\label{4(3)}
\qqq
we have,
\qq
-A^{-1}_{4} = \int^\infty_{0} e^{t A_{4}} dt = \int^\infty_{0} R(t)^{\otimes
4} dt.
\non
\qqq
Thus,
\qq
- \Bigl(A^{-1}_{4} b (G)\Bigr)_{i_{1}i_{2}i_{3}i_{4}} = 6 \sum_{{\bf j}}
\int^\infty_{0} dt \Bigl(R(t) \La\Bigr)_{i_{1}j_{2}j_{3}j_{4}}\prod^4_{a=2}
\Bigl(R(t) G\Bigr)_{i_{a}j_{a}} + (3 \ \mbox{permutations}),
\non
\qqq
with ${\bf j}=(j_2, j_3, j_4)$ and, then, with ${\bf i}=(i_2, i_3, i_4)$,
\qq
&&\Bigl(-\La_{2} A^{-1}_{4} b (G)\Bigr)_{ii'} = 6 \Bigl[\sum_{{\bf j}}
\sum_{{\bf i}} \int^\infty_{0} dt \Bigl(R(t) \La\Bigr)_{i{\bf j}} \La_{i' {\bf
i}} \prod^4_{a=2} \Bigl(R(t) G\Bigr)_{i_{a} j_{a}} 
\non\\
&&+ 3 \sum_{{\bf j}}
\sum_{i_{1}i_{2}i_{3}} \int^\infty_{0} dt \La_{i' {\bf i}} \Bigl(R(t)
\La\Bigr)_{i_{1}{\bf j}}
\prod^3_{a=2} \Bigl(R(t) G\Bigr)_{i_{a}j_{a}} \Bigl(R(t)
G\Bigr)_{ij_{4}}\Bigr] + (i \leftrightarrow i')
\non
\qqq
Recalling the expression (\ref{3(8a)}), some calculation gives
\qq
&&\CN'_{\al\be} (x,y) = 6 \la^2\de_{\al 2} \int^\infty_{0} dt \sum_{z}
\Bigl(R(t) G\Bigr)_{11} (x,z)^2
\non\\
&& \cdot \Bigl[3R_{12} (t,x-z) \Bigl(R(t) G\Bigr)_{\beta 1} (y,z) +
\Bigl(R(t)G\Bigr)_{11} (x,z)  R_{\beta 2} (t,z-y)\Bigr]+ tr
\label{4(4)}
\qqq
where tr means that both ${\al,\be}$
and $x,y$ are interchanged and where
the translation invariance of $R$ was used. Since $\p_{t} R=AR$ we have,
from (\ref{4(2)}),
\qq
\p_{t} R_{1\al} = R_{2\al}
\label{4(5)}
\qqq
and, using  $\p_{t} R=RA$, we get:
\qq
R_{\be 2} = - \p_{t} (RG_{0})_{\be 1},
\label{4(6)}
\qqq
where
\qq
G_{0} = \left(\begin{array}{cccc}
\om^{-2} & 0\\
0 & 1
\end{array}\right).
\label{4(7)}
\qqq
Inserting (\ref{4(6)}) in (\ref{4(4)}), and integrating by parts, (\ref{4(4)})
becomes:
\qq
\CN'_{\al\be} (x,y) = \CN_{\al\be} (x,y)+ 6 \la^2\de_{\al 2}\de_{\be 1} 
 \sum_z Q (x,z)^3\om^{-2} (z-y) 
+ tr
\label{4(8)}
\qqq
where we denote $G_{11} = Q$ and
\qq
&&\CN_{\al\be} (x,y) = 18 \la^2 \de_{\al 2} \lim_{\ep \to 0} 
\int^\infty_{0} dt  \sum_{z} \Bigl[\Bigl(R(t) G\Bigr)_{11} (x,z)\Bigr]^2
\non\\
&& \cdot \Bigl[R_{1 2} (t,x-z) \Bigl(R(t)G\Bigr)_{\be 1} (y,z) +
\Bigl(R(t)G\Bigr)_{21} (x,z) \Bigl(R(t) G_{0}\Bigr)_{\be 1} (z-y)
\Bigr]+ tr
\label{4(9)}
\qqq
The closure equation that we will study is the replacement of the exact equation
(\ref{3(10)}) by 
\qq
(A_{2} - \Ga_{2}) G + \CN (G) + \CC = 0
\label{4(10)}
\qqq
i.e. we drop the terms $\Sig_{2}, \Sig_{4}, \Ga_{4}, G^c_{6}$, as well as the
second term in (\ref{4(8)}). Before discussing the motivation for this
approximation, let us write (\ref{4(10)}) explicitely. Let
\qq
G = \left(\begin{array}{cccc}
Q & H\\
H^T & P
\end{array}\right)
\label{4(11)}
\qqq
so, e.g. $H(x,y) = Eq_x p_y$ and $H^T(x,y) = H(y,x)$ . Then, (\ref{4(10)}) can be written as:
\qq
H + H^{T} = 0
\label{4(12)}
\qqq
\qq
P-\om^2 Q - H\Ga + \CN_{12} = 0
\label{4(13)}
\qqq
\qq
-\om^2 H  - H^T \om^2 - P\Ga - \Ga P + \CN_{22} + C =0
 \label{4(14)}
\qqq
where
\qq
\Ga_{xy} = \ga \de_{xy} (\de_{x_10} + \de_{x_1N})
\label{4(15)}
\qqq
and $C$ was defined in (\ref{4(16)}).

Let
\qq
J = {1\over 2} (H-H^T)
\label{4(17)}
\qqq
so by (\ref{4(12)}), $H=J$, $H^T = -J$, and we write:
\qq
G = \left(\begin{array}{cccc}
Q & J\\
-J & P
\end{array}\right).
\label{4(11a)}
\qqq
Then (\ref{4(13)}) and (\ref{4(14)})
can be written as:
\qq
&& 2P = \om^2 Q + Q \om^2 + J \Ga - \Ga J - \CN_{12} - \CN^T_{12}
\label{4(18)}
\qqq
\qq
&& \om^2 Q - Q \om^2 + \Ga J + J \Ga + \CN^T_{12} - \CN_{12} = 0
\label{4(19)}
\qqq
\qq
&&\om^2 J - J \om^2 + P\Ga + \Ga P - \CN_{22} - C = 0
\label{4(20)}
\qqq
(\ref{4(18)}) - (\ref{4(20)}) is a system of
nonlinear equations for the correlation functions $Q,J,P$. 

\vskip 4mm

An important property of $\CN$ in 
equation (\ref{4(9)}) is that, for all $T$,
\qq
\CN (TG_{0}) = 0
\label{4(21)}
\qqq
for 
\qq
G_{0} = \left(\begin{array}{ccccc}
\om^{-2} & 0 \\
0 & 1
\end{array}\right)
\label{4(22)}
\qqq
Indeed, $\Bigl(R(t) G_{0}\Bigr)_{21} = R(t)_{21} \om^{-2} = -R(t)_{12} $ (see
formula (\ref{5(0)}) below). Thus, at $\ga=0$, our set of equations (\ref{4(18)})-(\ref{4(20)})
has a 1-parameter family of solutions 
$$Q=T\om^{-2},\ P=T, \ J=0
$$ 
and, for the
equilibrium case, with $\ga\neq 0$, $T_{1}= T_{2}$, only one of these persists,
namely the one with $T=T_{1}$. 

Note that the true equilibrium Gibbs state has $P=T$, $J=0$, and
 \qq
\widehat Q (k) = T \Bigl(\om (k)^2 + \sig (k,\la,T)\Bigr)^{-1}
\non
\qqq
with $\sig = \CO (\la)$. The terms $\Sig_{2}, \Sig_{4}$ and the second term in
(\ref{4(8)}) would contribute to changing the $\sig=0$ of our closure solution
to a $\sig$ which would agree with the true $\sig$ to $\CO (\la^2)$. Dropping
these terms, as well as $\Gamma_4$, is done for convenience and should not change our analysis
qualitalively. {\it The uncontrolled approximation consists in dropping} $G^c_{6}$.
Presumably, for small $\la$, this would not make a qualitative difference. It
would be interesting to try to prove this in the kinetic limit $\la = \CO
(1/\sqrt{N})$. In the rest of this paper, we will discuss only equations
(\ref{4(18)})-(\ref{4(20)}).

\nsection{Changes of coordinates}
\vskip 0.2cm
For large $N$ we expect the solution to the
equations (\ref{4(18)})-(\ref{4(20)}) 
to be translation invariant in the directions perpendicular to
the 1-direction with  a slowly varying dependence of the first coordinate $x_1$.
It is therefore convenient to represent $G$ in coordinates that
are suited to such behaviour.

We first write eqs. (\ref{4(18)})-(\ref{4(20)}) in terms of
the Fourier transform of $G$. Recall that $x\in
\NZ^d$ with $x_{1} \in \NZ_{2N}$. Introduce momentum variables:
\qq
q = (\mathrm{q}, \Nq),
\non
\qqq
and write
\qq
G(x,y) = \int e^{i (qx+q'y)} \widehat G
(q,q') dq dq'
\non
\qqq
with the shorthand notation 
\qq
\int dq = \sum_{\mathrm{q}} {1\over 2N} \int_{I} {d{\bf q} \over (2\pi)^{d-1}}
\label{5.0}
\qqq
where $\mathrm{q} \in {\pi \over N} \NZ_{2N}$ and $I = [0,2\pi]^{d-1}$. Then $R$
becomes a Fourier multiplier
\qq
\widehat R (t,q) = \left(\begin{array}{cccc}
\p_{t}+\ep & 1\\
\p_{t} (\p_{t} + \ep) & \p_{t}
\end{array}\right) {\sin \tilde \om (q)t \over \tilde \om (q)}
e^{-\ep t/2},
\non
\qqq
where $\tilde \om (q)= (\om^2 (q) -\frac{\epsilon^2}{4})^{1/2}$.
Then, letting $\ep\to 0$ in the matrix, we get
\qq
\widehat R (t,q) &=& \left(\begin{array}{cccc}
\cos \om (q) t & {1\over \om (q)} \sin \om (q) t\\
-\om (q) \sin \om  (q)t & \cos \om (q)t
\end{array}\right) e^{-\ep t/2}\non
\\
&=& {1\over 2} \sum_{s=\pm 1} e^{(is\om (q)-\ep/2)t}
\left(\begin{array}{cccc}
1 & -is\om (q)^{-1} \\
is\om (q) & 1
\end{array}\right),
\label{5(0)}
\qqq
so, 
\qq
\Bigl(R(t) G\Bigr)^{\wedge}_{\cdot 1} (q, q') = {1\over 2}
\sum_{s=\pm 1} e^{(is\om (q)-\ep/2)t} W_{s} (q, q') \left(\begin{array}
{cccc}
1\\
is\om(q)
\end{array}\right)
\non
\qqq
where
\qq
W_{s} (q, q') = \widehat Q (q, q')
+ is\om (q)^{-1} \widehat J (q, q').
\label{Br7}
\qqq
Thus, (\ref{4(9)}) becomes
\qq
\widehat \CN_{\al\be} (q, q') = {9\over 8}(2\pi)^{3d}  \la^2(N_{\al\be} (q, q') + N_{\be\al} (q', q))
\label{5a}
\qqq
with
\qq
&& N (q, q') =i \sum_{\bf s} \int d \mu
  \left(\sum^4_{i=1} s_{i}\om (q_{i}) + i\ep\right)^{-1} 
\prod^2_{i=1} W_{s_{i}}
(q_{i}, q'_{i}) \cdot \left(\begin{array}{ccccc}
0 & 0 \\
1 & is_{4}\om (q')
\end{array}\right)
\non
\\
&& \left[-is_{3}\om (q_{3})^{-1} \de (q_{3} + q'_{3})
W_{s_{4}} (q_{4}, q'_{4}) + is_{3} \om (q_{3}) W_{s_{3}}
(q_{3}, q'_{3}) \om (q_{4})^{-2} \de
(q_{4}+q'_{4})\right]
\label{5(1)}
\qqq
where
\qq
d\mu = \de \left(q-\sum^3_{1}q_{i}\right) \de \left(
\sum^4_{1} q'_{i}\right) \de (q' - q_{4}) \prod^4_{1}
dq_{i} dq'_{i},
\label{5.00}
\qqq
and we replaced $2\ep$ by $\ep$. We got in  (\ref{5a}) a factor $(2\pi)^d$ for each lattice sum 
in the defintion of $\widehat \CN_{\al\be} (q, q') $, and a  factor  $\ha$ for each of the four sums over $s_i$.
Note that, because of the sum over ${\bf s}$ in (\ref{5(1)}), only terms that are even in
 ${\bf s}$ contribute, which means that the factor  $\left(\sum^4_{i=1} s_{i}\om (q_{i}) + i\ep\right)^{-1} $ gives rise to a delta function if the integrand is even in ${\bf s}$, and a principal value if it is odd.
\vspace*{4mm}

We will look for solutions to (\ref{4(18)})-(\ref{4(20)}) which are
translation invariant in the directions orthogonal to the 1-direction. Thus,
we look for solutions of the form
\qq
 \widehat G (q, q') = (2\pi)^{d-1} \de ({\bf q} + {\bf q}') g
(\mathrm{q}, \mathrm{q}', {\bf q})
\non
\qqq
and
\qq
G(x,y) = \int e^{i (\mathrm{q}\mathrm{x} + \mathrm{q}'\mathrm{y}) + i {\bf q}
({\bf x}-{\bf y})} g (\mathrm{q}, \mathrm{q}', {\bf q})
d\mathrm{q}d\mathrm{q}'d{\bf q},
\non
\qqq
where we write $x=(\mathrm{x} , {\bf x})$, and similarly for $y$.
\vspace*{4mm}

It will be convenient to change coordinates in the 1-direction. 
Let $g$ be a $2\pi$ periodic function on $ {\pi\over N} \mathbb{Z}_{2N}
\times {\pi\over N} \mathbb{Z}_{2N}$ . 
Define 
\qq
\tilde g (\mathrm{p}, \mathrm{k})  =  g (\mathrm{p} + \mathrm{k},
\mathrm{p}-\mathrm{k})
\non
\qqq
on the set
\qq
\{\mathrm{p},
\mathrm{k} \in {\pi \over 2N} \mathbb{Z}_{4N}\ |\ \mathrm{p} + \mathrm{k} \in {\pi\over N} \mathbb{Z}_{2N}\}
\non
\qqq
Then, 
\qq
&&\sum_{\mathrm{q},\mathrm{q}'} \left({1\over 2N}\right)^2
e^{i\mathrm{q}x+i\mathrm{q}'y
} \hat g (\mathrm{q},
\mathrm{q}')
=\sum_{\mathrm{p},\mathrm{k}}  \left({1\over 2\sqrt{2}N}\right)^2
e^{i\mathrm{p} (x+y) + i
\mathrm{k} (x-y)} \tilde g (\mathrm{p},\mathrm{k}),
\label{5(4)}
\qqq
 Indeed, each pair
$(\mathrm{q}, \mathrm{q}')$ gets counted exactly twice in the
$(\mathrm{p},\mathrm{k})$-sum (because the pair $(\mathrm{p},\mathrm{k})$ gives the same conribution as the pair $(\mathrm{p}+2N,\mathrm{k}+2N)$, with addition modulo $4N$), which accounts for the factor  $\sqrt{2}$. Note
that, in the $N\to\infty$ limit, both sides tend to $\int_{[-\pi,\pi]^2} \cdot
(2\pi)^{-2}$ since the lattice spacing in the RHS of (\ref{5(4)}) is
$\sqrt{2}\cdot {\pi\over 2N}$.

\vspace*{4mm}

Let $p=(\mathrm{p}, {\bf 0})$. We have then
\qq
G(x,y) = \int e^{ip(x+y)+ik(x-y)} \widetilde G (p,k) dp dk
\label{5(5)}
\qqq
where $\widetilde G (p,k) = g (\mathrm{p} + \mathrm{k}, \mathrm{p} -
\mathrm{k}, {\bf k})$ and the integrals over the first components are 
 Riemann sums, with the same convention as  in (\ref{5(4)}). The function $\widetilde G$ is $2\pi$
periodic in all the variables and, moreover,
\qq
\widetilde G (p + \tilde \pi, k-\tilde \pi) = \widetilde G (p,k)
\label{5(6)}
\qqq
where $\tilde \pi = (\pi, {\bf 0})$. 

We will, from now on, work in the $p,k$
variables, drop, in general, the tilde in $\tilde G(p,k)$, and we shall not distinguish between $p$ and $\mathrm{p}$, unless we need to stress that $\mathrm{p}$ is a number.
Since the components of $p$ (unlike those of $k$) other than the first one are always $0$, this abuse of notation is harmless.

 Equations (\ref{5(1)})  then becomes
\qq
&& N (p, k) = \sum_{\bf s} \int d \nu (\sum s_{i} \om (p_{i} + k_{i}) + i \ep )^{-1}
\prod^2_{i=1}  W_{ s_{i}} (p_{i},k_{i})
\cdot \left(\begin{array}{ccccc}
0 & 0 \\
1 & is_{4}\om (p_{4}+k_{4})
\end{array}\right)
\non
\\
&&s_{3} \om
(p_{3}+k_{3}) 
 \Bigl[\om
(p_{3}+k_{3})^{-2} \de (2p_{3})  W_{s_{4}} (p_{4}, k_{4})
- \om (p_{4}+k_{4})^{-2} \de (2p_{4}) W_{s_{3}}
(p_{3},k_{3})\Bigr]
\label{59a}
\qqq
where
\qq
d\nu &=& \de
(2p - \sum (p_{i}+k_{i})) \de (\sum (p_{i}-k_{i}))\de (p-k-p_{4}-k_{4}) d { \underline  p} d {\underline k},
\label{5(10)}
\qqq
and $\underline k=(k_i)_{i=1}^4$, $\underline p=(p_i)_{i=1}^4$
and where
we used the identity (see (\ref{5.00})) :  
\qq
q-\sum_{i=1}^3 q_i= p+k -(\sum_{i=1}^4 (p_i+k_i) -q_4)=
p+k - (\sum_{i=1}^4 (p_i+k_i) -q')
\non
\\
=p+k - \sum_{i=1}^4 (p_i+k_i) +p-k=2p-
\sum_{i=1}^4 (p_i+k_i) 
\non
\qqq

We can then write
\qq
\CN_{12} (p,-k) \equiv  {9\over 8}(2\pi)^{3d}  \la^2
 n_{1} (p,k) \ \  , \ \   \CN_{22} (p,k) \equiv  {9\over 8}(2\pi)^{3d}  \la^2 (n_{2} (p,k) +
n_{2} (p,-k))
\label{5(20a)}
\qqq
with
\qq
n(W)(p,k)& = &\left(\begin{array}{ccc}
n_{1}\\
n_{2}
\end{array}\right) = \sum_{{\bf s}} \int \prod^2_{i=1} W_{s_{i}} (p_{i},
k_{i}-p_{i}) s_{3} \om (k_{3}) \left(\begin{array}{ccc}
1\\
is_{4}\om (k_{4})
\end{array}\right)
\non
\\
&&\cdot   \left[\om (k_{3})^{-2} \de (2p_{3}) W_{s_{4}} (p_{4}, k_{4}-p_{4}) - 
  (3 \leftrightarrow 4)\right] \nu_{{\bf s} pk} (d \underline{p} \  d
\underline{k})
\label{5(21)}
\qqq
where
\qq
\nu_{{\bf s}pk} (d \underline{p} \  d
\underline{k}) = \Bigl(\sum_{i} s_{i}\om (k_{i}) + i\ep\Bigr)^{-1}\de
\Bigl(2(p-\sum_{i} p_{i})\Bigr) \de (2p-\sum_{i} k_{i})
 \de (p-k-k_{4}) d{\underline p} d{\underline k}
\label{5(21a)}
\qqq
In (\ref{5(21)}) we have shifted the $k_{i}$-integrals by $-p_{i}$
compared to (\ref{59a}).

\vspace*{4mm}

Let us intruduce the convenient notation
\qq
\om(p, k) = (\hf (\om (p+k)^{2} + \om (p-k)^{2}))^\hf   
\label{6(9a)}
\qqq
and
\qq
\de\om^2(p, k) =  \om (p+k)^{2} - \om (p-k)^{2}
\label{6(9aa)}
\qqq
Then the equations (\ref{4(18)})-(\ref{4(20)}) read as follows in the $p,k$
variables:
\qq
&&  \om(p, k)^2Q + \hf((J\Ga -\Ga J)^{\sim} - \CN_{12}
(p,k) - \CN_{12} (p,-k))=P
\label{5(11)}
\\
&& \de\om^2(p, k) Q + (J\Ga + \Ga J)^{\sim} + \CN_{12}
(p,-k) - \CN_{12} (p,k) = 0
\label{5(12)}
\\
&& \de\om^2(p, k)  J + (P\Ga + \Ga P)^{\sim} - \CN_{22}
(p,k) - \widetilde C (p,k) = 0
\label{5(13)}
\qqq
We look for a solution that satisfies in $x$-space $G(x,y) = G (-x,-y),
Q(x,y)=Q(y,x)$ and $J(x,y)=-J(y,x)$ i.e.
\qq
Q(p,k) &=& Q (p,-k) = Q (-p,k)
\label{5(14)}
\\
J (p,k) &=& -J (p,-k) = -J (-p,k)
\label{5(15)}
\qqq
This is consistent since $\CN_{\al\be} (p,k,W)= \CN_{\al\be}
\Bigl(-p,-k,W(-\cdot,-\cdot)\Bigr)$.

We will see below that the $\lim_{\ep\to 0} \CN$ is well defined for bounded
$W(p,k)$. Before that, let us outline our arguments.

\nsection{Heuristics}
Before getting into the details we give an outline of the argument.
Since the nonlinear term $\CN$ depends only on $Q$ and $J$,
$P$ can be solved from (\ref{5(11)}) in terms of them. Then
(\ref{5(12)}) and (\ref{5(13)})  are two equations for the two
unknown functions  $Q$ and $J$ which we write as
\qq
\de \om^2 
\left( \begin{array}{ccc}
0 & 1\\
1 & 0
\end{array}
\right)
 \left(\begin{array}{cc}
J\\
 Q
\end{array}\right)+ \CN(Q,J)+ \CN_\Gamma (Q,J)=\left(\begin{array}{cc}
0\\
 C
\end{array}\right)
\label{8(1)}
\qqq
where
\qq
N (Q,J)=
\left(\begin{array}{cccccc}
\CN_{12} (p,-k) - \CN_{12} (p,k)\\
-\CN_{22} (p,k)
\end{array}\right) 
\label{N}
\qqq
and
\qq
N_{\Ga} (Q,J)=
\left(\begin{array}{cccccc}
J \Ga + \Ga J\\
P\Ga + \Ga P
\end{array}\right) 
\label{Ngamma}
\qqq
where $P$ is expressed in terms of $J$ and $Q$ in (\ref{5(11)}).
$\CN_\Gamma $
collects the friction terms that have both linear and nonlinear
contributions but will be treated as a perturbation.

For large $N$ the solution to this equation will be locally
in the $x_1$ coordinate a perturbation of the equilibrium
state (\ref{4(22)}) given in the new variables by
\qq
Q_T(p,k)=T\om(p+k)^{-2}\de(2\rmp) 
\label{g0}
\qqq
where
\qq
\de(\rmp)=2N\de_{\rmp,0}.
\label{deltafunction}
\qqq
Hence the importance to study
the linearization of  $\CN$ around that state. It turns out
(see Section 9.2) that it is given by an operator which is
a multiplier in the variable $\rm p$:
\qq
D\CN(Q_T,0)\left(\begin{array}{cc}
J\\
Q
\end{array}\right)(p,k)=\CL_{p} \left(\begin{array}{cc}J(p,\cdot)\\
 Q(p,\cdot)
\end{array}\right)(k).
\label{8(2)}
\qqq
$\CL_{p} $ is a matrix of operators
\qq
\CL_{p}=
\left(\begin{array}{cccccc}
\CL_{11} (p)&  \CL_{12} (p) \\
\CL_{21} (p)&  \CL_{22} (p)
\end{array}\right).
\label{8(2.2)}
\qqq
The translation invariant equilibrium (\ref{g0}) has support at
$\rmp=0$ (and due to the periodicity (\ref{5(14)}) at $\rmp=\pi$) .
The nonequilibrium solution will also have most of its mass 
in the neighborhood of these points. Thus it is important
to understand $\CL_{0} $. It will turn out that
\qq
 \CL_{ij} (0) = \CL_{ij} (\pi) = 0, \ \ \ i\neq j\
\label{8(3)}
\qqq
whereas $ \CL_{11} (0) $ is invertible. Invertibility of $\CL_0$
would then follow from invertibility of $ \CL_{22} (0) $. This, however,
is not the case:  $ \CL_{22} (0) $ 
has {\it two zero modes}. 

 One of them is easy to understand. Eq.  (\ref{g0}) is a one-parameter
 family of solutions to the $\gamma=0$ equations. Hence the derivative
 with respect to the parameter $T$ is a zero eigenvalue eigenvector
 of the linearization $ \CL_{22} (0) $ , i.e. 
 \qq
 \CL_{22} (0) \om^{-2}=0.
 \label{0mode1}
\qqq

There is, however, a second zero mode for $ \CL_{22} (0) $ :
\qq
 \CL_{22} (0) \om^{-3}=0.
 \label{0mode2}
\qqq
While the first zero mode has to persist for the full Hopf equations
due to the one parameter family of Gibbs states that solve them
for $\gamma=0$, the second one is an artifact of the closure
approximation. The nonlinear terms in the closure equations
can be interpreted as describing phonon scattering and in our case
only processes where two phonons scatter occur. These processes
conserve phonon energy leading the the first zero mode and also
phonon number, leading to the second one. The connected
six-point correlation function which was neglected in
the closure approximation would produce terms that
violate phonon conservation and remove the second zero mode.
However, for weak anharmonicity its eigenvalue
would be close to zero and should be treated with some
perturbation of the present analysis.

The second zero mode leads one to expect that our equations have in the
$\gamma=0$ limit a two parameter family of stationary solutions
which indeed is the case. These are given by (see Section 7.3)
\qq
{Q}_{T,A}(p,k)=T (\om(p+k)^2 - A \om (p+k))^{-1} \de(2p) .
\label{536a}
\qqq
with $J=0$ and and $P$ given by (\ref{4(18)}). For  (\ref{536a})
to be a well defined covariance we need positivity of the denominator
which holds if $A<m^2$.
The zero-mode  (\ref{0mode2}) is proportional to 
the derivative of ${Q}_{T,A}$ with respect to $A$, at $A=0$.

These considerations lead to the following ansatz for the solution
\qq
Q(x,y)=Q_{0}(x-y)+r(x,y),
\label{Q1}
\qqq
where
\qq
Q_{0}(x-y)=Q_{T(\rmx),A(\rmx)}(x-y)+Q_1(x,y)
\label{Q11}
\qqq
The first term here is of {\it local equilibrium} form with slowly
varying temperature and ``chemical potential" profiles $T(\rmx)$
and $A(\rmx)$ ($A$, 
or more precisely $T\cdot A$, can be considered as being 
related to a chemical potential,  because it arises from 
the conservation of a number current, see section 7.2), and the second is a small perturbation (see 
(\ref{6(9)}) below and the remark following it). 
 $T(\rmx)$
and $A(\rmx)$ are determined from the
{\it current conservation laws} which are projections of
the equations (\ref{8(1)}) as follows. The operator
$ \CL_{22} (0) $ has two left zero modes. Projection
onto them the equation (\ref{8(1)}) yields two
nonlinear elliptic equations for the functions
 $T(\rmx)$
and $A(\rmx)$ coupled to the rest of the
variables i.e. $J$ and $r$. For the latter
taken in a subspace complementary to the
zero modes the linear operator
$
\de\om^2+\CL_{p}
$
is invertible and  $J$ and $r$
 can be determined by fixed point arguments
in a suitable Banach space as functionals of  $T(\rmx)$
and $A(\rmx)$. The projected equations then allow
us to determine the latter.

Our main result, stated more precisely in Section 8, says that $Q(x,y)$ is as above, while
the currents corresponding to the two conservation laws discussed above and defined in section 7, are linearly related, to leading order in $|T_1-T_2|$, to the gradients of $T(\rmx)$
and $A(\rmx)$. $T(\rmx)$ is, to leading order in $|T_1-T_2|$, linear in $\rmx$, and therefore the currents are, to leading order, $\CO ({|T_1-T_2|\over N})$.

\nsection{Current conservation}

In the Introduction we recalled that the Hamiltonian structure
of the dynamics leads to a local conservation law. We will show that
our closure too has such a conservation law.

\subsection{Heat-current}

We start with a simple identity

\vspace*{2mm}

\no{\bf Lemma 7.1} \ $\int \CN_{22} (p,k) dk = 0$ \textit{for all (bounded)} $W$.

\vspace*{2mm}

\no {\bf Proof} From (\ref{5(10)}) we see that $\int dk d\nu_{p,k}$ is
symmetric under the simultaneous interchanges $s_{3} \leftrightarrow s_{4}$, $(p_{3},k_{3})
\leftrightarrow (p_{4},k_{4})$. Hence, the integral over $k$ of the second term in
 (\ref{5(20a)}) vanishes, because the integrand
in (\ref{5(21)}) is antisymmetric 
and the rest of the integrand is symmetric
 under those simultaneous interchanges.  \hfill$
\makebox[0mm]{\raisebox{0.5mm}[0mm][0mm]{\hspace*{5.6mm}$\sqcap$}}$
$
\sqcup$

\vspace*{2mm}

Note that, by (\ref{5(5)}),
\qq
\CN_{22} (x,x) = \int e^{2ipx} \CN_{22} (p,k) dpdk
\label{5(16)}
\qqq
i.e., by Lemma 7.1, \ \ $\CN_{22} (x,x)=0, \ \ \forall x$.
This follows also by inspection from (\ref{4(9)}) since
$\Bigl(R(t)G_{0}\Bigr)_{21} = -R (t)_{12}$ (by (\ref{5(0)})) and $R (t)_{12}(z-y)=R (t)_{12}(x-z)$
if $x=y$.

 Consider now equation (\ref{4(20)}) restricted to the diagonal $x=y$. Let
$J'=\om J + J\om$. Then, $\om^{2}J - J\om^{2} = \om J' - J' \om$.
Defining, for $e_{\mu}$ the unit vector in the $\mu$-direction,
\qq
j_{\mu} (x) =  J' (x-e_{\mu},x),
\label{5.16a}
\qqq
we have, for $\omega$ given in (\ref{2(4b)}):
\qq
(\om^2 J - J \om^2) (x,x) &=& (-\De J' + J' \De) (x,x)
\non
\\
&=& \sum_{\mu} \Bigl(-J' (x-e_{\mu},x) - J' (x+e_{\mu},x)+J' (x,x-e_{\mu}) + 
 J' (x,x+e_{\mu})\Bigr)
\non
\\
&=& 2\sum_{\mu} (J' (x,x+e_{\mu}) -J' (x-e_{\mu},x))
= 2 \nabla \cdot \vec \jmath \  (x),
\non
\qqq
where $(\nabla_{\mu} f) (x) = f (x+e_{\mu}) - f(x)$. Note that, for other 
functions $\omega$, there will be also a current, but its form will depend on $\omega$.
In view of Lemma 7.1,
(\ref{4(20)}), for $x=y$, reads
\qq
\nabla \cdot \vec \jmath \   (x) + \ga P(x,x) (\de_{\mathrm{x} 0}  + 
\de_{\mathrm{x}N}) =
\ga (T_{1}\de_{\mathrm{x}0}+ T_{2} \de_{\mathrm{x} N}),
\non
\qqq
or, since $P(x,x)$ depends only on $\mathrm{x}$, 
\qq
\nabla \cdot \vec \jmath \  (x) = \ga ((T_{1}-P(0,0))
\de_{\mathrm{x}0} +
(T_{2} - P(N,N)) \de_{\mathrm{x} N}),
\label{5(17)}
\qqq
which is a {\it current conservation equation}; the {\it heat current} $\vec
\jmath \  $ has sources on the boundary. Since $\vec  \jmath \  $ depends only on
$\mathrm{x}$, we define $j (\mathrm{x}) = j_{1} (x)$ and, so,
\qq
\nabla \cdot \vec \jmath \   (x) = j (\mathrm{x}+1) - j (\mathrm{x}).
\label{5(18)}
\qqq

It is a useful exercise to rewrite this in the $(p,k)$-variables. We have, see (\ref{2(4b)}),
\qq
\om (p+k) - \om (p-k) &=& 2 \Bigl(\cos (\mathrm{p} - \mathrm{k}) - \cos
(\mathrm{p}+\mathrm{k})\Bigr)
= 4 \sin \mathrm{p} \sin \mathrm{k}
\non
\qqq
and so
\qq
\de\om^2 (p,k)  = 4\sin \mathrm{p} \sin
\mathrm{k} (\om (\rmp+\rmk) + \om (\rmp-\rmk)).
\label{deome}
\qqq
Thus
\qq
\int\de\om^2 (\rmp,\rmk) J(\rmp,k)dk =
2 (e^{2i\mathrm{p}}-1 ) j
(2\mathrm{p})
\label{5(19)}
\qqq
where
\qq
j (\mathrm{p}) = -i \int dk e^{-i\mathrm{p}/2} \sin \mathrm{k} 
(\om (\rmp/2+\rmk) + \om (\rmp/2-\rmk))J(\mathrm{p}/2,k)
\label{5(20)}
\qqq
for $\mathrm{p}\in {\pi \over N}\mathbb{Z}_{2N}$. 
In (\ref{5(19)}) we used the fact that 
the $k$ integral is
$\pi$-periodic in $\rmp$, due to (\ref{5(6)}), to write it as a function
of  $2\rmp$.
It may be checked directly that this is the Fourier transform of $j_1(x)$ given by  (\ref{5.16a}).

\subsection{Number  current }

As stated in Section 6, the
closure equations possess another, approximate, conservation
law which we will derive now. Let $\rho (p,k)$ be given by
\qq
\rho (p,k) = \om (p,k)^{-1},
\label{5(21b)}
\qqq
with $ \om (p,k)$ given by (\ref{6(9a)}),
and project $\CN_{22}$ now onto $\rho$ instead of the function
$1$ as in Lemma 7.1. Let
\qq
\th (\rmp) = \int \rho ({_p\over^2},k) \CN_{22} \Bigl({_p\over^2},k\Bigr) dk = {9\over 4}(2\pi)^{3d}  \la^2 \int
\rho \Bigl({_p\over^2},k\Bigr) n_{2} \Bigl({_p\over^2},k\Bigr)dk
\label{thetadef}
\qqq
Then $\th$ is $2\pi$ periodic. Unlike what happened in Lemma 7.1
$\th$ is not zero, but it will turn out to be very regular, see Proposition 9.7 and Appendix B below,
due to the fact that (\ref{5(21b)}) at $p=0$ is a left zero eigenvector
of the linearization of $\CN_{22}$.

We will now integrate equation
(\ref{5(13)}) multiplied by a linear combination of $\rho$ and 1. For this, write it
in the $(p,k)$ representation. The covariance is
\qq
\widetilde C (p,k) = 2\gamma(T_{1} + T_{2} e^{-2iN p})
\label{5(29)}
\qqq
and the friction term
\qq
(\Ga P + P \Ga)^{\wedge} (q_{1},q_{2}) = \ga\int dq \left[(1+e^{i(q-q_{1})N}) 
 \widehat P (q,q_{2})+ (1+e^{i(q-q_{2})N}) \widehat P (q_{1},q) \right].
\non
\qqq
So, after shifting $q$ in the first integral by $q_2\over 2$ and in the second by $q_1\over 2$, we get:
\qq
(\Ga P + P \Ga)^{\sim} (p,k) &=&\ga \int dq \left[\widetilde P \Bigl({_q\over^2}, {_q\over^2} + k-p\Bigr) + \widetilde P \Bigl({_q\over^2}, - {_q\over^2} 
+ p + k\Bigr) \right]
(1+e^{i(q-2p)N}).
\label{5(30)}
\qqq
Let
\qq
\eta (p,k) = \rho (p,k) - \int \rho (p,k)dk,
\label{5(31)}
\qqq
with $\rho$ given by (\ref{5(21b)}).
Integrating equation (\ref{5(13)}) multiplied by $\eta$, we get
\qq
j' (\rmx+1) - j' (\rmx) - \th (\rmx) + \ga (\rmx) = 0
\label{5(32)}
\qqq
where
\qq
j' (\mathrm{p}) = -i \int dk e^{-i\mathrm{p}/2} \eta (\rmp/2, k) \sin \mathrm{k}
\Bigl(\om (\rmp/2+k) + \om (\rmp/2-k)\Bigr) J (\rmp/2,k)
\label{5(33)}
\qqq
and
\qq
\ga (\rmp) =\ga \int dkd\rmp \psi (\rmp,k,q) \widetilde P \Bigl({_\rmp\over^2}, k\Bigr) (1+e^{i( \mathrm{q}-\rmp)N})
\label{5(34)}
\qqq
where
\qq
\psi (p,k,q) = \eta (p,k+{_p\over^2}-{_q\over^2}) + \eta \Bigl(p,k-{_p\over^2}+{_q\over^2}\Bigr)
\label{5(35)}
\qqq
is a smooth function. Note that the covariance dropped out from (\ref{5(32)})
because $\tilde C(p,k)$ is independent of $k$, and we subtract from $\rho$ its average in the definition of $\eta$. Also,
the latter does not contribute to $\theta$ due to Lemma 7.1.

\vspace*{4mm}

Equation  (\ref{5(32)}) is again a conservation law: $\ga$ is a boundary term and, as we will see,
$\th$ vanishes in the limit of translation invariance. We call the current $j'$ the 
{\it particle number} current.

\subsection{Generalized Gibbs states}

We finish this section by checking that the states   (\ref{536a}), with $J=0$,  are indeed
solutions to the $\gamma=0$ equations.
 Indeed, now,
\qq
W_{s} (p,k-p) = {Q}_{T,A} (k) \de (2p),
\non
\qqq
with ${Q}_{T,A}=T(\om(k)^2-A\om (k))^{-1}$.
So, from (\ref{5(21)}), we get:
\qq
&& n  (p,k) = \de (2p) \sum_{{\bf s}} \int \prod^2_{1} {Q}_{T,A} (k_{i}) s_{3} \om (k_{3})
\left(\begin{array}{ccc}
1
\\
is_{4}\om (k_{4})
\end{array}\right)  \left[\om
(k_{3})^{-2} {Q}_{T,A} (k_{4}) - \om (k_{4})^{-2} {Q}_{T,A} (k_{3})\right]
\non
\\
&&\cdot \Bigl(\sum s_{i} \om (k_{i}) + i \ep\Bigr)^{-1} \de \Bigl(2p-\sum k_{i}\Bigr) \de (p-k-k_{4}) d\underline{k}
\label{5(37)}
\qqq
$n_{1}$ is obviously even in $k$, thus (\ref{5(12)}) holds. As for
(\ref{5(13)}),
 write the [-] in (\ref{5(37)}) as
\qq
- AT^{-1} {Q}_{T,A} (k_{3}) {Q}_{T,A} (k_{4}) \left[\om (k_{3})^{-1}  -\om (k_{4})^{-1} \right],
\non
\qqq
and use 
 $(x+i\ep)^{-1} = \CP \left({1 \over x}\right) -2\pi i \de (x)$
to get:
\qq
n_{2} (p,k) &=& - 2\pi \de (2p) A T^{-1} \tilde n_{2} (p,k), 
\label{5(37a)}
\\
\tilde n_{2}(p,k) &=& \sum_{\bf s} \int \prod^{4}_{1} {Q}_{T,A} (k_{i}) s_{3}
s_{4} \left[\om (k_{4}) - \om (k_{3})\right] \non
\\
&&
\cdot\de \Bigl(\sum^4_{1} s_{i} \om (k_{i}) \Bigr)  \de (2p-\sum k_{i})
\de (p-k-k_{4}) d\underline{k} 
\label{5(38)}
\qqq
(By $\bf s\to -\bf s$ symmetry only the delta function contributes).
 The integral in (\ref{5(37)}) is supported on
 \qq
\sum s_{i} \om (k_{i}) = 0.
\label{5(24)}
\qqq
We will choose $m^2$ in (\ref{2(4)}) large enough so that (\ref{5(24)}) forces 
\qq
\sum s_{i} = 0
\label{5(25)}
\qqq
By symmetry, we may replace $s_{3}$ by ${1\over 3} \sum^3_{i=1} s_{i}$ and, by
 (\ref{5(25)}) also by $-{1\over 3} s_{4}$. Again, by symmetry, $s_{3} \om
(k_{3})$ may be replaced by  ${1\over 3} \sum^3_{i=1} s_{i} \om (k_{i})$
and by  (\ref{5(24)}) by $-{1\over 3} s_{4} \om (k_{4})$. Doing the first replacement for $s_3\om(k_4)$
in the first term in the $[\cdot]$ in (\ref{5(38)}), and the second for $s_3\om(k_3)$ in the second term, we see that (\ref{5(38)}) vanishes. Note that
$\tilde n_{2} $ vanishes for all $p$ and not only for $2p=0$ mod $2\pi$. We shall need this
later (see the derivation of (\ref{7.3a}) below). Hence (\ref{536a}) solves (\ref{5(11)})-(\ref{5(13)}) if $\ga=0$.

\vspace*{4mm}

Note however that these are {\it not} equilibrium $T_{1}=T_{2}$
solutions, {\it unless} $A=0$. This is because for $A\neq 0$, the noise and
friction terms in (\ref{4(20)}) will not any more balance each other.

\nsection{The space of local equilibrium solutions}
\vskip 0.2cm

We will now describe the space where (\ref{4(18)})-(\ref{4(20)}) are solved.
To motivate our choice, consider the current conservation equation (\ref{5(17)}).
Summing over $\mathrm{x}$, we get, since $j (\mathrm{x})$ is periodic:
\qq
T_{1} - P (0,0) = - \Bigl(T_{2} - P (N,N)\Bigr) \equiv j_{0}
\label{6(1)}
\qqq
and so
\qq
j (\mathrm{x} + 1) - j (\mathrm{x}) = j_{0} (\de_{x0} - \de_{xN})
\label{6(2)}
\qqq
which is solved by
\qq
j(\mathrm{x}) =
\left\{
\begin{array}{lllll}
j(0) + j_{0}  &\ \ \ \ & \mathrm{x} \in [1,N] \\
&&
\\
j(0) & & \mathrm{x} \in [-N+1, 0]
\end{array}
\right.
\label{6(2a)}
\qqq
Since $j(-\mathrm{x}) = -j (\mathrm{x})$ we get $j(0)= - {1\over 2} j_{0}$. 
The Fourier transform of this is
\qq
j(\mathrm{p}) = \left\{
\begin{array}{lllll} \displaystyle 
j_{0} {1-e^{i\mathrm{p}N} \over e^{i\mathrm{p}}-1} & \ \ \ \ &\displaystyle  \mathrm{p} \in {\pi \over N}
\mathbb{Z}_{2N} \ \ , \ \mathrm{p} \neq 0  \\
&&\\
0 & & \rmp =0
\end{array}
\right.
\label{6(3)}
\qqq
(note that $e^{i\mathrm{p}N}$ takes only the values $\pm 1$). The current $j_{0}$ will
turn out to be $\CO (1/N)$. 

We now describe the space to which functions such as $J(p,k)$ belong.
This space has to encode the $1/\rmp$ singularity at origin in
the equation (\ref{6(3)}) as well as the factor  $e^{i\mathrm{p}N}$ 
coming from the fact that in $x$ space there are two special
points in the first coordinate, the origin and $N$.

Let $\CH$ be the space of continuous functions $f(p,k)$ on 
\qq
\Omega= \{(p,k)\ | \ \mathrm{p},
\mathrm{k} \in {\pi \over 2N}\mathbb{Z}_{4N}
, \ \mathrm{p}+\mathrm{k} \in {\pi \over N}\mathbb{Z}_{2N}
, \
{\bf k} \in
[-\pi,\pi]^{d-1}\}
\label{6.3a}
\qqq
that are 
  2$\pi$-periodic in all the variables, and invariant under
$(\mathrm{p}, \mathrm{k}) \to (\mathrm{p} + \pi, \mathrm{k} + \pi)$. 
We denote by $\|f\|_{\infty}$ the sup  norm in $\CH$.

Let $$\Omega_+=\{(p,k)\in\Omega \ | \ 
\mathrm{p} \in {\pi \over N} \mathbb{Z}_{2N}\}$$ and
$\Omega_-=\Omega\setminus\Omega_+$. Then
$$
\CH=\CH_+\oplus \CH_-
$$
with
\qq
f(p,k)=f_+(p,k)\sigma_+(2\mathrm{p})+f_-(p,k)\sigma_-(2\mathrm{p})
\label{6(6a)}
\qqq
with
\qq
\sigma_\pm (\rmp)=1\pm e^{i\mathrm{p}N}
\qqq
and $f_\pm={\hf}f|_{\Omega_\pm}$.

Define the discrete $C^{\al}$-norm for $f\in\CH_\pm$ as
\qq
\|f\|_{\al} = \sup_{p,k,\la,\mu} \Bigl(|f(p,k)| + |\la|^{-\al}
|f(p+\la e_{1}, k) - f (p, k)|+ |\mu|^{-\al}
|f(p, k+\mu e_{1}) - f (p, k)|\Bigr)
\label{discrete}
\qqq
where $\la,\mu \in {\pi \over N} \mathbb{Z}_{2N} \setminus 0 $ (note that
$\rmp+\la\in\Om_\pm$ if $\rmp\in\Om_\pm$ ). Set then, for $f\in\CH$,
\qq
\| f \|_\al \equiv  \| f_{+} \|_{\al} + \| f_{-} \|_{\al}.
\non
\qqq
Let
\qq
d(\rmp)^{-1}\equiv\left\{
\begin{array}{lllll} \displaystyle 
 (e^{i\mathrm{\rmp}}-1)^{-1} & \ \ \ \ &\mathrm{p} \neq 0  \\
&&\\
0 & & \rmp =0,
\end{array}
\right.
\label{dminus}
\qqq
and let, with some abuse of notation,
\qq
d(\rmp)=e^{i\mathrm{\rmp}}-1.
\qqq
We will then consider a space $\CS$ of functions $J(\rmp, k)$ of the form
\qq
J =  N^{-1}\de(2\rmp)J_0+(N (d(2\rmp))^{-1} J_{1} + N^{\al/2-1}J_{2} +( N
d(2\rmp))^{-3/2}J_{3}
\label{6(4b)}
\qqq
where $J_i(\rmp,k)$ are in $\CH$.
Define
\qq
\|J\|_\CS=\max\{\|J_i\|_\al , \|J_3\|_\infty ,\ \ i=0,1,2\}.
\label{snorm}
\qqq
More properly, $\CS$ is the space $\CH^{\oplus 4}$ of 4-tuples $(J_0,\dots,J_3)\equiv \bf J$ which is a Banach space with this norm.
We identify $J$ and $\bf J$ when no confusion arises.

\vskip 0.4cm

\no {\bf Remarks}.  1. Due to the friction and the noise terms (see
(\ref{5(29)}) and (\ref{5(30)})) we end up with the factors $e^{2i\rmp N}$.  This 
leads to  a factor
$e^{i\rmp N}$ in functions such as $j(\rmp )$ and $j'(\rmp )$ . A H\"older continuous function $f$, of exponent $\al$, 
decays in $x$-space
at least as $|\mathrm{x}|^{-\al}$ whereas $e^{i\rmp N}f$ produces decay away from
$\mathrm{x}=N$ i.e. $|\mathrm{x}-N|^{-\al}$. Thus $\| f \|_\al\leq \CO(1)$ gives rise to
functions of $\mathrm{x}$ that are localized near the boundaries, $\mathrm{x}=0$
or $N$.

\vskip 2mm

\no 2. The Fourier transform of (\ref{6(4b)}) is as follows.
The first term is constant in the 1-direction and $\CO (1/N) $.
The second one is also of $\CO (1/N) $ because it involves a
(discrete) Hilbert transform of a H\"older continuous function. However,
e.g. for the current it can produce opposite constant values
for positive and negative $\rmx$.
The third term
is also of this order whereas the second is $\CO 
(N^{-1+\al/2} (\mid \mathrm{x} \mid^{-\al} + \mid N - \mathrm{x}
\mid^{-\al}) )$ i.e. $\CO (N^{-1-\al/2})$ for $\mathrm{x}$
far away from the boundaries. The $J_{2}$ and $J_{3}$ terms will be subleading
corrections to the current and the profile. Note that no smoothness
in the transverse variable $\bf k$ is assumed so that correlations
will have a very slow decay in $x$ space. This is due to the
poor regularity properties of the "collision kernel" that enters
the nonlinear terms, see Proposition 9.3.

\vskip 2mm

\no 3. The $J$ correlation function is odd in $\rmp$
and hence $J_0=0$. However, with $Q$, we will
encounter functions with a nonzero first term in (\ref{6(4b)}).

\vspace*{2mm}

Finally, we will extend these definitions to the functions like $j(\mathrm{p})$ defined on ${\pi \over N} \mathbb{Z}_{2N}$,
the only difference being the occurence of $\sigma_\pm(2\mathrm{p})$
in (\ref{6(6a)}) and $d(2\rmp)$ in (\ref{6(4b)}), instead of  
 $d(\rmp)$ here. We denote the space now by $S$. Hence $j\in S$ is of the form
\qq
j(\rmp) =  N^{-1}\de(\rmp)j_0+(N d(\rmp))^{-1} j_{1}(\rmp)
 + N^{\al/2-1}j_{2}(\rmp) +( N
d(\rmp))^{-3/2}j_{3}(\rmp)
\label{jdef}
\qqq
with $j_0$ a constant. Of course, here, since $j$ is defined on ${\pi \over N} \mathbb{Z}_{2N}$, the space is finite dimensional and all norms are equivalent, but  (\ref{jdef}) is a convenient way to record the dependence on $N$ of various terms.

\vskip 0.4cm

\vspace*{4mm}

We will look for solutions to (\ref{4(18)})-(\ref{4(20)}) with
$J \in S$ and $Q$ as follows.

$Q$ will have a leading ``local equilibrium" term which we now describe.
Recall that our equations have  the 2-parameter family (\ref{536a}) of solutions when the friction vanishes.
 The leading term in $Q$ will be of this form, where the constants $T$ and
$A$ will be replaced by $\rmp $-dependent functions. More precisely let $T(\rmp ), A(\rmp )$ be functions on ${\pi \over N} \mathbb{Z}_{2N}$, of the form 
\qq
T(\rmp ) = T_{0} \de (\rmp ) + (d(-(\rmp))^{-1} t (\rmp )
\label{6(7)}
\\
A(\rmp ) = A_{0} \de (\rmp ) + (d(-(\rmp))^{-1} a(\rmp )
\label{6(8)}
\qqq
where 
$\de(\rmp )
$
is defined  in (\ref{deltafunction}) and   $t, a \in S$.  
Define
\qq
Q_{0} (p,k) = \sum^\infty_{n=0} (T \ast A^{\ast n}) (2\rmp ) \om
(p,k)^{-2-n}
\label{6(9)}
\qqq
where $\ast$ is the convolution and $\om
(p,k)$ is defined in (\ref{6(9a)}).

\vskip 0.4cm

\no {\bf Remark} \ Let $Q_{T,A}$ be the generalized Gibbs state
in (\ref{536a}), at $p=0$,
\qq
Q_{T,A}(x-y)=T\int dk e^{ik(x-y)} \Bigl(\om(k)^2 - A \om (k)\Bigr)^{-1}.
\label{gg}
\qqq
Comparing (\ref{6(9)}) with 
\qq
T (\om^2 - A\om)^{-1} = \sum^\infty_{n=0} T A^n \om^{-2-n}
\label{6.10}
\qqq
it is not hard to show that the Fourier transform of (\ref{6(9)}) is
\qq
Q_{0}(x,y)=Q_{T(x),A(x)}(x-y)+Q_{1}(x,y)
\label{gg1}
\qqq
where (the Fourier transform of) $Q_{1}$ is in $\CS$, and, as it will turn out,
\qq
Q_{1}(x,y)=\CO(\tau/N).
\label{gg2}
\qqq
The first term is of a local (generalized) equilibrium form.

\vspace*{4mm}

We make now the following assumptions. We take 
\qq
\ga = N^{-1+\al/4}
\label{gamma}
\qqq
and $m^2$ large enough that 
$$\sum s_{i} \om (k_{i}) = 0 \Rightarrow \sum s_{i} =
0$$
(recall  (\ref{5(24)}) and (\ref{5(25)})). We shall use below the set of $p's$ that are close to singularities: 
\qq
E_{0}=[-p_0,p_0]\cup [\pi-p_0,\pi+p_0]
\label{E_0}
\qqq
with $p_0=B\la^2$, 
where $B$ is a number  that will be chosen large below (see (\ref{8(14)})).
Then, we have the

\vspace*{4mm}

\no{\bf Theorem} \ \textit{There exist} $\la_{0}$ \textit{such that
 for}  $0<\la < \la_{0}$, $\mid T_{1} - T_{2} | < \tau$ \textit{, with } $\tau = \tau (\la)$,
 \textit{and for} $N>N(\la)$
 \textit{the equations}
(\ref{4(18)})-(\ref{4(20)}) \textit{have a unique solution with} $J \in S$ \textit{and}
\qq
Q = Q_{0} + r
\non
\qqq
 \textit{with} $r \in \CS$. $Q_{0}$ \textit{is given by}  (\ref{6(7)})-(\ref{6(9)})
\textit{with} $T_{0} = \hf(T_{1} +T_2)+ \CO (\la^2 \tau)$, 
\qq
t(\rmp ) = t_0(\rmp ) + \tilde t (\rmp ),
\label{t}
\qqq
\textit{where, in $x$ space,}
\qq
\hf(T_{1} +T_2)+ t_0 (x)= T_{1} + {|x|\over N} (T_{2} - T_{1}),
\label{t_0}
\qqq
and
$\tilde t  \in S$, \textit{with} 
\qq
\| \tilde t \|_S= \CO (\la^2\tau).
\non
\qqq
\textit{ Moreover,} $A_0 =\CO (\tau^2)$, $a  \in S$, \textit{and }
\qq
\| a \|_S= \CO (\tau^2).
\non
\qqq
\textit{Finally, the currents} $j$ \textit{and} $j'$ \textit{are given
by}
\qq
(j(\rmp ),j'(\rmp ))^T =\kappa(\rmp )(t(\rmp ),s(\rmp ))^T+ \CO (\tau^2).
\non
\qqq
\textit{where $s(\rmp )=d(-\rmp)T\ast A(\rmp)$, the conductivity matrix $\kappa$ 
is H\"older continuous in
$\rmp$ of exponent $\al$, for some $\al>0$, and  is invertible and $\CO(\la^{-2})$
for $\rmp\in E_0$.
 }.
 
 \vskip 0.4cm

\textbf{Remarks }
1. The shape of the profile, to leading order, is actually given by
\qq
\beta(x) \equiv T(x)^{-1} = T^{-1}_1 + {|x|\over N} (T^{-1}_2 - T^{-1}_1)
\label{beta}
\qqq
This can be seen, from the proofs, as follows: The full relation between $J$
and $t(p)$, which, in $x$ space, is equal to $-\nabla T(x)$, is given by
(\ref{12(11)}) with ${\cal D}_{p}$ replaced by the full $DN(q_{0}) = {\cal L} +
{\cal L}^{\prime}$, see (\ref{7(7)}). The leading term in ${\cal L}^{\prime}$
comes from the $n=0$ term in (\ref{6(3)}) inserted in (\ref{7(6)},
\ref{7(7)}). Going back to $x$-space, we get, within those approximations:
$T(x)^2 j(x) = - {c\over \lambda^2} \nabla T(x)$, which, since $j(x)$ is
constant, implies $\nabla(T^{-1})(x)$ constant, i.e. (\ref{beta}). Of course, for $\tau$ small,
we can expand (\ref{beta}) in $\tau$ and obtain  (\ref{t_0}), to leading order in $\tau$.

 \vskip 0.4cm
 2. The choice of constants.  There are four parameters in this model: $m^2$, the parameter in $\om(k)$, see
(\ref{2(4b)}), $\la$ the strength of the nonlinearity, $\tau$ that bounds 
the temperature difference and $N$, the size of the system in the direction where there is a temperature difference. We choose $m^2$
enough so that (\ref{5(24)}) implies (\ref{5(25)}). Next we choose $\tau$
small compared to $\la^2$, say $\CO(\la^3)$ and we choose $N$ large so that we
can use bounds like $N^{-\al} \leq \tau$, for any $\alpha >0$. 

In the proofs, $C$ or $c$ will denote constants that can change from place to
place. We also use in various proofs auxiliary functions denoted $f,g$, whose
meaning is given in the proofs where they are used.

Since $\la^2$ enters as a multiplication factor in the nonlinear terms, see
(\ref{5(20a)}), it will be convenient to discuss in the next section the $\la$
independent nonlinearities $n(W)$, and introduce explicitely $\la^2$ in
section 10 (because only part of the linear operator there is multiplied by
$\la^2$) and in section 11, where $\la^2$ is used to make some Lipschitz
constants less than 1.

The value of $\al$ is determined by the degree of H\"older continuity that one obtains in Proposition 9.3 below. We do not try to optimize that value (any $\al>0$ suffices); in fact, it will convenient sometimes to assume  that $\al$ is not too close to one, so that, e.g. the power on $N$ in (\ref{12(7)}) is negative, and we shall implicitely assume that.


\nsection{Nonlinear terms}

In this Section we study the nonlinear terms of our equations given in
(\ref{5(20a)}) and (\ref{5(21)}). 
Our goal is to show that 
their linearization defines a bounded operator on $\CS$ and that the remaining nonlinearities
 define suitable Lipschitz functions on $\CS$,

 It is convenient to introduce a space for
the functions $T$, $A$ , analogous to $S$ but
stronger singularity at $\rmp=0$. We let
$E$ denote the pairs $(T_0, t)$ with $T_0\in \mathbb{R}$ and
$t\in \CS$  with $t_0=0$ in (\ref{jdef}).
They  parametrize functions
\qq
T(p)=T_0 \ \de(p)+d(-p)^{-1} t(p),
\label{E}
\qqq
(from now on, we shall identify $p$ and $\rmp$).
We use the norm
\qq
\|T\|_E = |T_0|+ \| { t} \|_S ,
\label{Enorm}
\qqq
which has the convenient property, used often below,  that $\|dT\|_S  \leq \|T\|_E $.
In this section, we shall work with $(T,A) \in \CB_{\ep}$, where
\qq
\CB_{\ep} = \{(T,A) \in E \times E | \ \|T\|_{E} \leq C, \|A\|_{E} \leq \ep\}
\label{Br2}
\qqq
where $C$ is arbitrary and $\ep$ is chosen small enough so that various series
below converge.
\vskip 2mm

We expand (\ref{5(21)}) around $Q_{0}$ defined by (\ref{6(9)}). Let $W=Q_{0}+ w$
\qq
n (W)  = n(Q_{0}) + Dn (Q_{0}) w + \tilde n (w)
\label{7(1)}
\qqq
We discuss the three terms on the right hand side in turn.

\subsection{$n(Q_{0})$}

Consier first $n(Q_{0})$. Let us start with a  lemma, whose proof is given in Appendix B, and which controls convolutions between elements of $E$, $S$ and among themselves:

\vspace*{2mm}

\no{\bf Lemma 9.1.} (a) {\it Let $ T\in E$ and $ j\in S$. Then
$T\ast j\in S$ and}
\qq
\| T\ast j\|_S \leq C \|T\|_E\| j\|_S
\label{l1}
\qqq
\no  (b) {\it Let $ T,A\in E$. Then
$T\ast A\in E$ and}
\qq
\| T\ast A\|_E \leq C \|T\|_E\| A\|_E
\label{l2}
\qqq
\no  (c) {\it Let $ j,k\in S$. Then
$j'\equiv j\ast k\in S$  with}
\qq
|j'_0| &\leq& CN^{-1} \|j\|_S\| k\|_S,
\label{Br50}\\
\| j'_1\|_\al &\leq &CN^{-1} \|j\|_S\| k\|_S,
\label{l3a}\\
  \| j'_2\|_\al &\leq& CN^{-1+\al/2} \|j\|_S\| k\|_S,
\label{l3b}\\
 | j'_3 (p)|&\leq& CN^{-1}\log (N|p|) \|j\|_S\| k\|_S,
\label{br52}
\qqq
and
\qq
 \| j'\|_S \leq CN^{-1+\al/2} \|j\|_S\| k\|_S
\label{br51}
\qqq
\vskip 2mm

Using this  Lemma, we get:

\vskip 0.4cm

\no {\bf Proposition 9.2} {\it The function $n(Q_{0})$ satisfies
the following bounds, for $(T,A) \in \CB_{\ep}$.
\qq
n_1(Q_{0})=  \sum^\infty_{n=1} T^{\ast 3} \ast A^{\ast n} (2p) 
g_{n} (p,k)+m
\label{n1}
\qqq
with $g_n$ smooth functions bounded together with their derivatives (to any given order)
by $C^n$ and $m\in\CS$. Moreover
\qq
\| m\|_\CS\leq C(\|t\|_S+\|A\|_E)
\label{nm}\\
\|n_1(Q_{0})(p,k)-n_1(Q_{0})(-p,k)\|_\CS\leq C(\|t\|_S+\|A\|_E)
\label{n11}\\
\| n_2(Q_{0})\|_\CS\leq C(\|t\|_S+\|A\|_E)
\label{n2}
\qqq 
Furthermore  the functions on the LHS of (\ref{nm}),  (\ref{n11}) and
 (\ref{n2}) are uniformly Lipschitz in 
$T$ and $A$, for $(T,A) \in \CB_{\ep}$.
}

\vskip 0.4cm

\no {\bf Proof.}
From (\ref{5(21)}) and (\ref{6(9)}), we get 
\qq
&& n(Q_{0}) (p,k) = \sum_{{\bf n}} \sum_{{\bf s}} \int \prod^2_{i=1} 
T {\ast } A^{\ast n_i}
(2p_{i}) \om(p_{i}, k_{i}-p_{i})^{-2-n_i} \ s_{3} \om (k_{3})   \left(\begin{array}{ccc}
1\\
is_{4}\om (k_{4})
\end{array}\right)
\non
\\
 &&\cdot \left[\om (k_{3})^{-2} \de (2p_{3}) T {\ast } A^{\ast n_4}(2p_{4})
 \om (p_{4},k_{4}-p_{4})^{-2-n_4} - (3 \leftrightarrow 4)\right] \nu_{{\bf s}pk}
(d \underline{p} \ d \underline{k})
\label{7(2)}
\qqq
with $\nu_{{\bf s}pk}$ given by (\ref{5(21a)}), ${\bf n}=(n_i)_{i=1}^4$ and $\om(p,k)$ given in (\ref{6(9a)}).
We have, see (\ref{6(9a)}),
\qq
\om(p,k-p) = \om (k) + (e^{2ip}-1)  \CO (1) 
\label{7(3)},
\qqq
where the $\CO (p) $  term, is written in an unusual form, which records the fact that it vanishes also at $p=\pi$, and which will be convenient later.
Inserting this into (\ref{7(2)}), the leading term is
\qq
&&  \sum_{{\bf n}} \sum_{{\bf s}} \int \prod_{i=1}^2 T {\ast } A^{\ast n_i} (2p_{i}) \om
(k_{i})^{-2-n_{i}} s_{3}  \om (k_{3}) \left(\begin{array}{ccc}
1\\
is_{4}\om (k_{4})
\end{array}\right) \non
\\
&& \cdot \left[\om (k_{3})^{-2} \de (2p_{3}) T {\ast } A^{\ast n_4}
(2p_{4}) \om (k_{4})^{-2-n_{4}} -
[3 \leftrightarrow 4]\right] \nu_{{\bf s}pk} (d \underline{p} \ d \underline{k})
\non
\qqq
We may do the $p_{i}$-integrals to get
\qq
&=& \sum^\infty_{n=0} T^{\ast 3} \ast A^{\ast n} (2p) \sum_{{\bf s}}
\sum_{\Sigma n_{i}=n} \int \prod^2_{i=1} \om (k_{i})^{-2-n_{i}}
s_{3} \om (k_{3}) \left(\begin{array}{ccc}
1\\
is_{4}\om (k_{4})
\end{array}\right)
\non
\\
&& \cdot  \left[\om (k_{3})^{-2}  \om  (k_{4})^{-2-n_{3}} -
(3 \leftrightarrow 4) \right] \tilde \nu_{{\bf s}p k} (d
\underline{k})
\non
\\
& \equiv& \sum^\infty_{n=1} T^{\ast 3} \ast A^{\ast n} (2p) 
\left(\begin{array}{ccccc}
g_{n} (p,k)\\
f_{n} (p,k)
\end{array}\right)
\label{gn}
\qqq
where $\tilde \nu_{{\bf s}p k}$ is like $ \nu_{{\bf s}p k}$ above, but without $\de (2(p-\sum p_i))$, and where, in the last equality, we used the fact that the $n=0$ term has $n_3=0$ and
therefore the $[\cdot]$ factor vanishes.
Using (\ref{6.10}), we see that $f_{n} (p,k)$ are the Taylor coefficients of the
expansion in $A$ of $\tilde n_{2} (p,k)$, given by (\ref{5(37a)}, \ref{5(38)}). Since $\tilde n_{2}$
vanishes identically, we get
\qq
f_{n} (p,k) = 0.
\label{7.3a}
\qqq

For $g_{n}$ we need to study $n_{1}$ in (\ref{5(37)}) in more detail.
Proceeding as with $n_{2}$, we see that $g_n$ are the Taylor 
coefficients of the expansion in powers of the constant $A$ of
\qq
&&
\tilde n_{1} (p,k) = \sum_{{\bf s}} \int \prod^4_{1} Q_{0} (k_{i}) \CP
\left(\sum^4_{1} s_{i} \om (k_{i})\right)^{-1}
\non
\\
&&
\cdot \  s_{3} \Bigl[1-\om (k_{3}) \om (k_{4})^{-1} \Bigr] \de \left(2p-\sum
k_{i}\right)
\de (p-k - k_{4} ) d \underline{k}.
\non
\qqq
Here, again because of the ${\bf s} \to {-\bf s}$ symmetry, only the principal value contributes.
Consider first those terms with $\sum s_{i} = 0$. We may replace $s_{3}$ by
$-{1\over 3} s_{4}$ and $s_{3} \om (k_{3})$ by 
\qq
{1\over 3} \sum^3_{1} s_{i}
\om (k_{i}) = - {1\over 3} s_{4} \om (k_{4}) + {1\over 3} \sum^4_{1} s_{i} \om
(k_{i}). 
\non
\qqq
Thus these terms give 
\qq
{1\over 3} \sum_{\Sig s_{i}=0} \int \prod^4_{1} Q_{T,A} (k_{i}) \de \left(2p-\sum
k_{i}\right)
\de (p-k - k_{4} ) d \underline{k}
\non
\qqq
which is smooth in $p$. For the terms with $\sum s_{i} \neq 0$, $\CP(\cdot)$ has
no singularity, (see (\ref{5(24)}), \ref{5(25)})), and they are smooth. 
Thus, the functions  $g_n$ are smooth and bounded together with their derivatives (to any given order)
by $C^n$.

The contribution to (\ref{7(2)}) of the $(e^{2ip}-1) \CO (1) $ term in (\ref{7(3)}) is
of form
\qq
m&=&\sum_{{\bf n}} \sum_{{\bf s}} \int \prod^2_{i=1} T {\ast } A^{\ast n_i}  (2p_{i})
(e^{2ip_{3}}-1) T {\ast } A^{\ast n_3}  (2p_{3}) \de (2p_{4}) \non
\\
&
\cdot &G_{{\bf s}} (p,k, \underline{p},
\underline{k}) 
 \de \left(2p-2\sum p_{i}\right) 
\nu_{{\bf s}pk} (d \underline{p} \ d \underline{k})
\label{7(4)}
\qqq
where $G_{{\bf s}}$ is smooth. By an easy extension of Lemma 9.1 to convolutions of functions in $E$ and $\CS$, each summand
is in $\CS$ and (\ref{nm}) follows easily. The series over ${\bf n}$ converge for $\|A\|_E\leq \ep$ small enough.

\vspace*{4mm}

To get (\ref{n11})  we use the smoothness and periodicity to get
$$
g_n(p,k)-g_n(-p,k)=(e^{2ip}-1)\CO (1),
$$
and hence each summand is now in $\CS$. The bound (\ref{n2}) is
obtained in the same way as the bound on $m$ (by (\ref{7.3a}), the only contribution to $n_2 (Q_0)$ is similar to $m$ ). 
\hfill$
\makebox[0mm]{\raisebox{0.5mm}[0mm][0mm]{\hspace*{5.6mm}$\sqcap$}}$
$
\sqcup$

\subsection{Linearization}

Let us turn to the second term in (\ref{7(1)}) i.e. the linearization of $n$
at $Q_{0}$. We wish to separate a leading term when $T_{2}-T_{1}$ is small.
Therefore, we write, using (\ref{6(7)}) and (\ref{6(9)}),
\qq
Q_{0} = T_{+} \de (2p) \om^{-2} (p,k) + \tilde Q_{0}
\label{7(6)}
\qqq
with $\omega$ given by (\ref{6(9a)}), and $T_+=\ha(T_1+T_2)$ being the average  temperature .
Insert this into (\ref{5(21)}). The [ - ] term vanishes for the first term
in (\ref{7(6)}), so 
\qq
Dn (Q_{0}) w = \CL w + \CL' w,
\label{7(7)}
\qqq
with
\qq
&& (\CL w) (p,k) = 2 T^2_{0} \sum_{{\bf s}} \int \Bigl(\om (k_{1}) \om
(k_{2})\Bigr)^{-2} s_{3} \om (k_{3}) \left(\begin{array}{ccc}
1\\
i s_{4}\om (k_{4})
\end{array}\right) 
\non
\\
&&
\Bigl[ \om (k_{3})^{-2}
w_{s_{4}} (p,k_{4}-p)-\om (k_{4})^{-2} w_{s_{3}} (p,k_{3} - p) \Bigr]
\non
\\
&&
\left(\sum s_{i} \om (k_{i}) + i \ep\right)^{-1} \de \left(2p - \sum k_{i}\right)
\de (p-k-k_{4}) d \underline{k}
\label{7(8)}
\qqq
(where we used $\omega(0, k)= \omega (k)$, and 
 $w (p+\pi, k_{i} -p -\pi) = w (p,k_{i}-p)$, ${i} = 3,4$). Since
$k_{4}-p=-k$ and $k_{3}-p = k-k_{1}-k_{2}$ (coming from $k_1+k_2+k_3=2p-k_4=2p - (p-k)=p+k$), we see that $\CL$ is a
multiplication operator in the $p$-variable
\qq
(\CL w) (p,k) = \Bigl(\tilde \CL_{p} w (p,\cdot)\Bigr) (k)
\label{7(9)}
\qqq
and $\tilde  \CL_{p}$ in turn is a sum of an operator which acts as a  multiplication operator
in the subspace of even or of odd functions, and an integral
operator:
\qq
\tilde \CL_{p} = \tilde M_{p} +\tilde  K_{p},
\label{7(10)}
\qqq
where, after shiflting $k_3$ by $p$,
\qq
(\tilde K_{p} w) (k) =- 2 T^{2}_{0} \om (p-k)^{-2} \sum_{{\bf s}} \int w_{s_{3}}
(p, k_{3}) 
{s_{3} \om (k_{3} + p) \over \om (k_{1})^2 \om (k_{2})^2}
\left(\begin{array}{ccc}
1\\
i s_{4}\om (p-k)
\end{array}\right) d\mu_{{\bf s}pk} ,
\label{7(11)}
\qqq
with
\qq
d\mu_{{\bf s}pk}  = \left(\sum^2_{i=1} s_{i} \om (k_{i}) + 
s_{3} \om  (k_{3}+p) + s_{4} \om (p-k) + i \ep\right)^{-1} \de \left(k -
\sum^3_{1} k_{i}\right) \prod_{i=1}^3 d {k_i}
\label{7(12)}
\qqq
and (remember that $k_4-p=-k$ so that the operator acts as a multiplication operator only on functions of definite parity):
\qq
\tilde M_{p} (k) = - \om (p-k)^2 \Bigl(\tilde K_{p} \om^{-2} (p+ \cdot)\Bigr) (k),
\label{7(13)}
\qqq
when it acts on even functions, and 
\qq
\tilde M_{p} (k) = \om (p-k)^2 \Bigl(\tilde K_{p} \om^{-2} (p+ \cdot)\Bigr) (k),
\label{7(13a)}
\qqq
 when it acts on odd functions.
Here, we integrated over $k_4=p-k$, and we used:  
$$2p-\sum_{i=1}^4k_i=2p-\sum^3_{i=1} k_{i}-p+k ,
$$
which, after shifting $k_3$ by $p$ equals $k-\sum^3_{i=1} k_{i}$. We shall discuss $\CL'$ in the next susection, but let now analyze further $\CL$.

\vspace*{4mm}

Separating the real and imaginary parts of $w_s$ (see (\ref{Br7})), one can also view 
the operator $\CL_{p}$ as a $2\times 2$ matrix of operators the $(J,Q)$ variables, 
 where here $Q$ denotes an arbitrary even function, and $J$
 an arbitrary odd one.  We write it as:
\qq
\tilde \CL_{p} \left(\begin{array}{cc}
J \\
Q
\end{array}\right)=
\left(\begin{array}{cccccc}
\tilde \CL_{11} (p) J +\tilde  \CL_{12} (p) Q \\
\tilde \CL_{21} (p) J +\tilde  \CL_{22} (p) Q
\end{array}\right);
\label{8.2}
\qqq
this defines the operators $\tilde \CL_{ij} (p)$, $i,j=1,2$. From (\ref{7(10)}), we
see that each $\tilde \CL_{ij} (p)$ is a sum of a multiplication operator $\tilde  M_{ij}
(p)$ and a integral  operator $\tilde K_{ij} (p)$.

\vspace*{4mm}

The operators here are acting on functions defined on $\Om$, see (\ref{6.3a}). However, it is convenient to consider them as acting on functions defined on $[-\pi,\pi]^d$, which is always possible, by extending, say in a piecewise linear way, function on $\Om$ to functions on $[-\pi,\pi]^d$. With that identification, we may consider these operators to be acting, for all $N$, on the same space. Moreover, a discrete H\"older continuous function (see (\ref{discrete})) becomes, with such an extension, an ordinary  H\"older continuous function. The main property of these operators is:  

\vspace*{4mm}
\no {\bf Proposition 9.3.} {\it $\tilde M_{p} (k)$ is $C^\al$ in ${p},{k}$, for some $\al>0$,
and the operators $\tilde K_{p}$ are compact operators
mapping $C^\al \left(\Om_p (k)\right)$, where $\Om_p (k)= \{ k | (p,k) \in \Om \}$,
into itself. Moreover, in the norm of bounded
operators on $C^\al $, they are uniformly bounded in $N$
and $ C^\al$ in ${p}$, uniformly in $N$.
$\tilde M_{p} (k)$  converges as
$N\to\infty$ to a $C^{\al}$ function, while $\tilde K_{p}$ converges to a compact
operator mapping $C^\al \Bigl([-\pi,\pi]^d\Bigr)$ into itself.}

\vskip 4mm

Obviously, this  proposition implies that the operators $\CL_{ij} (p) $ define also bounded operators from $\CS$ into itself.

\vskip 4mm
\par\noindent
\textbf{Proof.}
Looking at (\ref{7(11)}-\ref{7(12)}, \ref{8.2}), we see that $\tilde K_{p}$ is a two by two
matrix of integral operators whose kernel is a sum of terms of the form
\qq
A_{p} (k,k') &=&   \int \De
\Bigl(\sum^2_{i=1} s_{i}\om (k_{i}) +  s_{3}\om (k'+p) + s_{4} \om (p-k)\Bigr) 
\non\\
&&\cdot\de (k-k_{1}-k_{2}-k') \rho_{{\scriptsize \textbf{s}}} (k_{1},k_{2},k',k,p)
dk_1 dk_{2}
\label{1}
\qqq
where $\De (x) = \de (x)$ or $\CP \left({1\over x}\right)$, $\rho_{{\scriptsize
 \textbf{s}}}$ is $C^\infty$ in all its arguments and we write $k'$ for $k_{3}$.

\vspace*{4mm}

Let us consider first $\De (x) = \de (x)$, integrate over $k_{2}$, and choose
$s_{1} = +1$, $s_{2}=-1$ (all other terms can be treated similarly). 
We obtain the integral
\qq
    \int \de \Bigl(\om(k_{1}) -\om (k-k_{1}-k') + s_{3} \om (k'+p) + s_{4} \om 
 (p-k)\Bigr)
  \rho_{{\scriptsize \textbf{s}}} (k_{1}, k-k_{1}-k', k',k,p) dk_{1}.
\label{2}
\qqq
Now, by (\ref{5.0}), $\int dk_{1}$ is actually a discrete sum over the first
component of $k_{1}$ and an integral over the last two components. Fix a
value of $k^1_{1}$, replace $k'-k$ by $k'$,  write $k^2_{1} = -{\pi\over 2}
+ q_{2}$, $k^3_{1} = -{\pi\over 2} + q_{3}$ and use lower indices, $k'_1, k'_2, k'_3$
for the components of $k'$. We get, using the explicit formula (\ref{2(4b)}) for $\omega$,
\qq
(\ref{2}) = \int \de \Bigl(\sin q_{2}-\sin (q_{2} - k'_{2}) + \sin q_{3} -
\sin (q_{3} - k'_{3}) + f(k'_2, k'_3, \la)\Bigr)  \rho_{{\scriptsize \textbf{s}}}
(q_{2},q_{3},\la) dq_{2}dq_{3}
\label{3}
\qqq
where we write $\la$ for the set of variables $(k^1_{1}, k'_1,k,p)$, the integral is an ordinary, not discrete, one and
\qq
f(k'_2, k'_3, \la)=
\sin q_{1}-\sin (q_{1} - k'_{1}) +s_{3} \om (k'+p) + s_{4} \om 
 (p-k)
\label{3a}
\qqq

We
shall now study the singularities of (\ref{3}) in $k'_2, 
k'_3$ and the smoothness of
(\ref{3}) in $\la$ away from the singularities. For notational simplicity, we
set $\rho_{{\scriptsize \textbf{s}}}=1$; since $\rho_{{\scriptsize
\textbf{s}}}$ is smooth, this does not affect our arguments.
Let us denote by $\Om$ the argument of the delta function
and shift the integration variables: let
$q_{i} =r_{i-1}+y_{i-1}$ where $y_{i-1}=\hf {k'_{i}}$. Then
\qq
\Om =\sum_{i=1}^2( \sin (r_{i}+y_i) - \sin (r_{i}-y_i)) + f(2y,\la)
=2\sum_{i=1}^2\cos r_i\sin y_i + f(2y,\la).
\label{3b}
\qqq
Next, change variables to $2-2\cos r_i=x^2_i$. Our integral
becomes
\qq
I(y,\la) = \int \de(x_1^2\sin y_1+ x_2^2\sin y_2 -g(y,\la))\prod_1^2
h(x_i)dx_i
\label{6}
\qqq
with $h(x)=(1-\hf x^2)^{-\hf}$, $x_i^2\in [0,2]$ and 
\qq
g(y,\la))=f(2y,\la)+2(\sin y_1+ \sin y_2) .
\label{a6}
\qqq
For $y_i\neq 0$ and $g\neq 0$ $I$ is bounded by
\qq
|I(y,\la)| \leq C|\sin y_1\sin y_2|^{-\hf}(1+|\log |g(y,\la)||)
\label{a7}
\qqq
(the $\log$ term is absent when $y_1y_2>0$). From (\ref{3a})
and (\ref{a6}) 
we have 
$$
g(y,\la)=2s_3(\cos(2y_1)+\cos(2y_2)+2(\sin y_1+ \sin y_2)+g'(\la)
$$
with $g'$ smooth. Thus the bound (\ref{a7}) is integrable in
$y$, uniformly in $\la$. Moreover, the
H\"older derivative of order $\al $ in $\la$ also remains integrable for $\al$ small enough. Hence  for all bounded 
functions $f$, $\int A_{p} (k,k') f(k')d\textbf{k}'$ is $C^{\al}$ in
$\la=(k^1_{1}, k'_1,k,p)$. This in turn implies that $\int A_{p} (k,k') f(k')dk'dk_1^1$
(where now the integral includes a Riemann sum over $k'_1$, $k_1^1$,
 is $C^{\al}$ in $k,p$, since each term in the Riemann sum is $C^{\al}$ in $k,p$.
 This means that each matrix element of
$K_{p}$ maps bounded functions into H\"older continuous
ones. 
Moreover, all the bounds are uniform in $N$, since the bound (\ref{a7})
 is independent of $N$,
and taking the Riemann sums preserves that property.

To obtain compactness, let 
$\al'$ the H\"older exponent
obtained above. Using the Arzel\`a-Ascoli's theorem, one easily shows that
 the unit ball in $C^{\al'}$ is compactly embedded in
$C^\al$ for $\al' > \al$. Since  $A_{p}$ 
is bounded from $C^\al$ into itself and 
maps $C^\al$ into 
$C^{\al'}$ it   a compact operator from  $C^\al$ into itself .

Obviously $ \int A_{p} (k,k') dk'$ is also $C^\al$ in $p,k$, so that each matrix element of $M_{p} (k) $
is $C^\al$ in $p,k$.

Next, we get:
\qq
| A_{p} f(k)- A_{p'} f(k)- (A_{p} f(k')- A_{p'} f(k'))|
&\leq& C \|f \|_\infty \min(|k-k'|^{\al'}, |p-p'|^{\al'})\non\\
& \leq & C \|f \|_\infty |k-k'|^{\al'/2} |p-p'|^{\al'/2},
\qqq
so that, choosing $\al=\al'/2$, we get:
 \qq
\|A_{p} - A_{p'}\|_{\al} \leq C|p-p'|^{\al}
\label{20}
\qqq
where $\|\cdot\|_{\al}$ is the operator norm on bounded operators of $C^{\al}$
into itself.

Finally, it is easy to see that the Riemann sum of a $C^{\al}$ function converges to the corresponding integral, with, for the sum in (\ref{5.0}), an error $\CO(N^{-\al})$. This then implies  the claims on convergence as $N\to \infty$ made in the Proposition.\hfill$
\makebox[0mm]{\raisebox{0.5mm}[0mm][0mm]{\hspace*{5.6mm}$\sqcap$}}$
$
\sqcup$ 

\vspace*{2mm}

In Appendix A, we will extend this result.

\vspace*{2mm}

\subsection{Nonlinearities}
 Let us now turn to 
the $n_i$'s defined in (\ref{5(21)}). They are linear combinations of functionals of
 the following form
\qq
u(f_1,f_2,f_3)(p,k) &=&
 \int G \prod^3_{i=1} f_{{i}} (p_{i},
k_{i}-p_{i})
\de(2p-\sum_1^3 2p_i)d\underline{p}
\mu ( d
\underline{k})
\label{udef}
\\
v(f_1,f_2,f_3)(p,k)& =&
 \int G\prod^2_{i=1} f_{{i}} (p_{i},
k_{i}-p_{i}) f_{{3}} (p_{3},
p-k-p_{3})
\de(2p-\sum_1^3 2p_i)d\underline{p}
\mu ( d
\underline{k})
\label{vdef}
\qqq
where $G$ is some smooth function of all the variables (defined
in the continuum ($N=\infty$) and restricted to the discrete) and
\qq
\mu_{p,k} ( d
\underline{k})
=\Bigl(\sum^3_{1} s_{i}\om (k_{i}) + s_{4}\om (p-k)+ i\ep\Bigr)^{-1}\de(\sum^3_{1} k_{i}
-p-k)dk_1dk_2dk_3.
\label{mu}
\qqq
Indeed, (\ref{udef}) corresponds to the second term in the bracket in (\ref{5(21)}), after integrating over $k_4$, $p_4$, while (\ref{vdef}) corresponds to the first term,  after integrating over $k_4$ (i.e. replacing $k_4$ by $p-k$) and $p_3$, and  relabelling $p_4$ as $p_3$.
Since $Q=Q_0+r$,  $r\in \CS$ and  $Q_0$ is given by
the series (\ref{6(9)}) we only need to
consider $f_i(p,k)=F(2p)$ for $F\in E$ or $f_i\in \CS$.
 We have then the

\vspace*{4mm}

\no{\bf Proposition 9.4.} {\it  Let $m$ be of the form (\ref{udef})
or (\ref{vdef}). Then}

\vskip 2mm

\no (a) {\it Let $ f_{{i}} (p,k)=F_i(2p)$ with 
$F_i\in E$. Then}
\qq
\| m(f_1,f_2,f_3)\|_E \leq C \prod_{i=1}^3\|F_i\|_E
\label{e1}
\qqq
\no (b) {\it Let $ f_i\in \CS$, $ f_{{j}} (p,k)=F_j(2p)$ with 
$F_j\in E$ for $j\neq i$. Then}
\qq
\| m(f_1,f_2,f_3)\|_\CS \leq C \|f_i\|_\CS\prod_{j\neq i}\|F_j\|_E
\label{e2}
\qqq
\no (c) {\it Let $ f_i, f_j\in \CS$, $ f_k\in E$. Then}
\qq
\| m(f_1,f_2,f_3)\|_\CS \leq C\ N^{-\ha+\al}\|f_k\|_E\prod_{l\neq k}\|f_l\|_\CS
\label{e3}
\qqq

\vskip 2mm

For the proof, see Appendix B.
\vskip 2mm

The operators $\CL'$ in (\ref{7(7)}) are not multiplication operators in $p$.
Inserting the expansion (\ref{6(9)}) in (\ref{5(21)}), we see that $\CL' w$
is a sum of terms of the form discussed in Proposition 9.4.(b)
with $F_i$  of form $T {\ast } A^{\ast n} $ with $n>0$ or 
$ d(-2p)^{-1}t(2p)$.  Proposition 9.4.(b) then 
gives the

\vspace*{4mm}

\no {\bf Proposition 9.5.}  {\it The operator  $\CL' :\CS\to \CS$ is bounded, for $(T,A) \in \CB_{\ep}$, in operator norm by
\qq
\| \CL' \| \leq C(\| t\|_{S}+
\| A\|_{E}),
\label{lprime}
\qqq
 and it is uniformly Lipschitz in $T$ and $A$ for $(T,A) \in \CB_{\ep}$.
}

\vspace*{4mm}


\no Proposition 9.4.(c) gives immediately for the term $\tilde n (w)$ in (\ref{7(1)}) :

\vspace*{4mm}

\no {\bf Proposition 9.6.} \ {\it For $ \| w 
\|_S\leq \CO(1)$, and  $(T,A) \in \CB_{\ep}$} $\| \tilde n (w) \|_{S} \leq C N^{-\ha+\al}
 \| w 
\|^2_{S}$.
{\it and it is Lipschitz in $w$, $T$ and $A$ with constant $CN^{-\ha+\al}$.}

\vskip 4mm

We still need to discuss the function $\th$ given by (\ref{thetadef}).
Its main property, proven in Appendix B, is:

\vskip 4mm

\no {\bf Proposition 9.7.} {\it $d^{-1}\th$ is in
$S$, for $(T,A) \in \CB_{\ep}$,
with, for $ \| w 
\|_S\leq \CO(1)$,
\qq
\| d^{-1} \th \|_{S} &\leq& C \la^2 (\|t\|_S+\|A\|_E+\|r\|_S\|J\|_S+\|J\|_S^2).
\label{theta12}
\qqq
and it is uniformly Lipschitz in $T$ and $A$ for $(T,A) \in \CB_{\ep}$, and in $J$ and $r$, with constants $C\la^2$.Moreover,} 
\qq
|\th (0)|\leq CN^{-1}.
\label{Br33}
\qqq

\nsection{Solution of the linear problem}
\vskip 0.2cm

 In order to solve equations (\ref{5(12)}), (\ref{5(13)}) we need to study the
invertibility of the linear operator 
\qq
\de \om^2 (p,k) \left( \begin{array}{ccc}
0 & 1\\
1 & 0
\end{array}
\right)
 + \CL_{p}
\label{8.1}
\qqq
where $\de \om^2 (p,k) = \om^2 (p+k) - \om^2 (p-k)$ and $\CL_{p}$ denotes the linearization
around the first term in (\ref{7(6)}) of the nonlinear terms in (\ref{5(12)}, \ref{5(13)}).
 $\CL_{p}$  is given explicitely by adding or subtracting to (\ref{7(8)}) a term
 with $k \rightarrow -k$, see (\ref{5(12)}, \ref{5(20a)}) , and multiplying it by ${9\over 8}(2\pi)^{3d}  \la^2$, see (\ref{5(20a)}). 
The fact that $\CL'$ in (\ref{7(7)}) is a small
perturbation of $\CL$ follows from Proposition 9.5, for $\tau$ small enough, since, as we shall see in the next section, we shall solve our equations in a space where the RHS of (\ref{lprime}) is of order $\tau$. So, to
invert $\de \om^2 (p,k) \left( \begin{array}{ccc}
0 & 1\\
1 & 0
\end{array}
\right) + DN
(Q_{0})$, it is enough to concentrate our attention to (\ref{8.1}), where $\CL_p$ is written as a two by two matrix as in (\ref{8.2}).

\vspace*{4mm}

Because of the zero modes, in order for (\ref{8.1}) to be invertible, it
needs to be restricted to the orthogonal complement of the zero modes (which occur at $p=0, \pi$). 
Let $\CH_{p}$ be the Hilbert space $L^2 \Bigl({\pi \over N} \mathbb{Z}_{2N} \times
[-\pi,\pi]^{d-1} ,\om(p,k)^2dk\Bigr)$ and let
$P^{\perp}$ be the projection to the orthogonal complement of
$\Bigl\{\om(p,k)^{-2},\om(p,k)^{-3}\Bigr\}$ in  $\CH_{p}$, and  $P=1-P^\perp$.
Note that the scalar product of $f(k)$ with $\om(p,k)^{-2}$ equals $\int f(k) dk$, which implies
\qq
P^\perp 1=0.
\label{Br3}
\qqq
 We shall study the
operator $\CD_p$, defined as:
\qq
\CD_p =\Pi
 \left(\de \om^2 
\sigma_1 + \CL_{p}
\right) \Pi,
\label{CD1}
\qqq
where we use the shorthand notations $\sigma_1= \left(\begin{array}{ccc}
0 &1\\
1 & 0
\end{array}\right)$, 
\qq
\Pi = \left(\begin{array}{ccc}
1 & 0\\
0 & P^\perp
\end{array}\right),
\non
\qqq
for $p\in E_0$, with $E_0$ defined in (\ref{E_0}).
\qq
\CD_p =
 \left(\de \om^2 
\sigma_1 + \CL_{p}
\right) ,
\label{CD2}
\qqq
for  $p \notin E_0$.

\vspace*{4mm}

We shall make the following assumptions, that we shall verify later for the
operator $\CL_{p}$ :
\qq
1.  \hspace{50mm} \CL_{ij} (0) = \CL_{ij} (\pi) = 0, \ \ \ i \neq j\hspace{50mm}
\label{8.3}
\qqq
2. $\exists \ \ c > 0$ such that
\qq
\CL_{11}(0) &<&  -c\la^2 \ \ 
\label{8(4)}\\
P^{\perp } \CL_{22} (0) P^{\perp }& > & c\la^2
\label{8(5)}
\qqq
where the inequalities holds for operators in $L^2 \left({\pi\over N}
\mathbb{Z}_{2N} \times [-\pi,\pi]^{d-1}, \omega^2(k) dk \right)$ restricted to 
 functions that are odd in $k$ (for $\CL_{11}$) or even in $k$ (for
$\CL_{22}$). The same inequalities hold
for $\CL_{11}(\pi) $, $\CL_{22}(\pi) $.
\\
3. $\exists \ C < \infty$, independent of $N$, such that, $\forall p,p',\forall i,j=1,2$,
\qq
\| \CL_{ij} (p) - \CL_{ij} (p') \|  \leq C |p-p'|^\al
\label{8(6)}
\qqq
where $\| \cdot \|$ is the operator norm of bounded operators mapping $\CC^\al 
\left({\pi\over N}
\mathbb{Z}_{2N} \times [-\pi,\pi]^{d-1}\right)$ into itself.
\\
4.  $\exists \ c>0$ such that, $\forall p, \forall k, \forall f \in
\mathbb{R}^2$,
\qq
\left | \Bigl(\de \om^2  \sigma_1 + M (p,k)\Bigr) f  Ê\right| \geq c (\la^2
+ |\sin p| \ |\sin k| \ ) |f|
\label{8(7)}
\qqq
where $|f| = |f_{1}| + |f_{2}|$
\\
5. The kernels 
$ K_{ij}(k,\cdot) \in L^{1+\eta} ({\pi \over N} \mathbb{Z}_{2N} \times
[-\pi,\pi]^{d-1})$, for some $\eta>0$, and a norm $\CO (\la^2)$.

\vspace*{4mm}

\no {\bf Proposition 10.1} \ Under assumptions (1-5) above, for any $B$ in (\ref{E_0}),
$\exists \la_{0}$ such that, for $\la \leq \la_{0}$ and for all $p \in {\pi
\over 2N} \mathbb{Z}_{2N}$, $\CD_{p}$ is
invertible, and $\exists C < \infty$ such that $\forall \la \leq \la_{0}$,
\qq
\|  \CD_p^{-1} \| \leq {C \over \la^2},
\label{8(8)}
\qqq
where the norm is the one of operators in $\CC^\al 
\left({\pi\over N}
\mathbb{Z}_{2N} \times [-\pi,\pi]^{d-1}\right) \oplus \CC^\al 
\left({\pi\over N}
\mathbb{Z}_{2N} \times [-\pi,\pi]^{d-1}\right)$.
\vspace*{4mm}

\no {\bf Proof}. \ We consider $|p| \leq {\pi \over 2}$. For $|p| > {\pi \over
2}$ the proof below can be repeated with $0$ replaced by $\pi$.

Let us first consider $|p| \leq B \la^2$, where the constant $B$,
independent of $\la$, will be specified later.
Write $p=a \la^2$, with $|a| \leq B$.
We have, for $|p| \leq B \la^2$,
\qq
\de \om^2 (k,p) = a \la^2 \varphi (k) + \CO (\la^4)
\label{8(9)}
\qqq
where we expand $\de \om^2 (k,p)$ (which vanishes at $p=0$) in $p$, the
first term in (\ref{8(9)}) is the one linear in $p$, with $\varphi$ an odd function
of $k$ alone, and the second is $\CO(p^2)$ for $|p| \leq B \la^2$.
Moreover, due to (\ref{8(6)}),
\qq
\CL_{ij} (p) = \CL_{ij} (0) + \CO (\la^{2+\al})
\label{8(10)}
\qqq
where $\CO (\la^{2+\al})$ is a bound on the operator norm ($\CL_{ij} (p)$ has a factor $\la^2$). So, we have
\qq
\de \om^{2} \sigma_1+ \CL_{p} = \la^2
\Bigl(a\varphi (k) \sigma_1 +
\tilde \CL_{0}\Bigr) + \CO (\la^{2+\al})
\label{8(11)}
\qqq
with $\tilde \CL_{0} = \CL_{0}/\la^2$, which is $\la$-independent.
Thus if we show that $\exists \ C(B)$ such that $\forall a$, $| a | \leq B$,
\qq
\|  \left[\Pi \left(a \varphi (k)\sigma_1 + \tilde \CL_{0} \right)\Pi\right]^{-1} \| \leq C (B) <
\infty,
\label{8(12)}
\qqq
we obtain, from (\ref{8(11)}) and a resolvent expansion the bound (\ref{8(8)})
for $|p| \leq B \la^2$, provided that $\la$ is small enough, given $B$.

 To prove (\ref{8(12)}), we observe that the spectrum of the multiplication
part of $a\varphi (k)$ $\sigma_1  + \tilde \CL_{0}$, i.e. $a\varphi (k)  \sigma_1 + {M (0,k) \over \la^2}$ lies outside a ball of fixed radius
around zero. This
follows from (\ref{8(7)}), using the approximations (\ref{8(9)}) and
(\ref{8(10)}) for $\de \om^2$ and $M (p,k)$. Hence, since adding to this a
compact operator does not change the essential spectrum  (note that the
projection operator $P^{\perp }$ adds a rank 2 operator), (\ref{8(12)})
will hold, provided that we show that $a \sigma_1+ \tilde \CL
(0)$ has no zero eigenvalue. But solving for $f_{2}$ the first of the two
equations
\qq
\Pi
\left(\begin{array}{ccccccc}
\tilde \CL_{11} (0) & \ \ & a \varphi (k) \\
a\varphi (k)
 & & \tilde \CL_{22} (0)
\end{array}\right) \ \ 
\Pi
\left(
\begin{array}{cc}
f_{1} (k) \\
f_{2} (k)
\end{array}
\right) = 0
\non
\qqq
(we use the fact that, by  (\ref{8.3}), $\tilde \CL_{ij} (0) = 0, i \neq j $),
 substituting in the second equation, and taking a scalar product in $L^2
\Bigl({\pi\over N}
\mathbb{Z}_{2N}
\times
[-\pi,\pi]^{d-1}, \omega^2(k)dk\Bigr)$, with $f_1$, we get:
\qq
\Bigl(P^{\perp}f_{1}, \tilde \CL_{11} (0) P^{\perp} f_{1}\Bigr) - a^2 \Bigl(P^{\perp} 
f_{1}, \varphi \tilde \CL_{22}^{-1} (0) \varphi P^{\perp} f_{1}\Bigr) = 0
\non
\qqq
which is impossible (for all $a$'s) because of  (\ref{8(4)}) and (\ref{8(5)}).
This finishes the proof of (\ref{8(12)}) and therefore of (\ref{8(8)}) for
$|p| \leq B\la^2$.

 \vspace*{4mm}

 Consider now $|p| > B \la^2$. Since adding a compact operator does not change the
 essential spectrum, which is bounded away from zero by (\ref{8(7)}), it is enough to show that  the equation
\qq
 (\de\om^2 \sigma_1 + \CL_{p})f=\mu f
\label{8(13)}
\qqq
with  $f= \left(
\begin{array}{cc}
f_{1} (k) \\
f_{2} (k)
\end{array}
\right)$, does not have non zero solutions for $|\mu|\leq c\la^2$, and $c>0$.
 Let $E=\{k\|\sin k| \leq b\}$ for some constant $b$ to be specified later.
 Then, from (\ref{8(7)}), (\ref{8(13)}) and the fact that $\CL_{p} = M_{p} +
 K_{p}$, we get, for $k \not \in E$: 
\qq
c'Bb\la^2 |f(k)| &\leq & |(\de \om^2 \sigma_1 +M_{p})f (k) |
\leq \| K_{p} \| \ \| f\|_{\infty} + |\mu| \  \| f \|_{\infty}
\non\\
&\leq &(C \la^2 + |\mu|)  \| f \|_{\infty},
\non
\qqq
using the fact (see the
proof of Proposition 9.3) that $K_{p}$ maps bounded functions into $C^\al$ ones which, in particular, are bounded, and that $\| K_{p}\| \leq C \la^2$.
 Given $\beta > 0$, $c>0$,  $c'>0$ $C<\infty$ and $b>0$, we can choose $B=B(\beta,c, c', C, b)$, independent of
$\la$, so that this implies, if $|\mu| \leq c\la^2$, 
\qq
|f(k)| \leq \beta \| f \|_{\infty}
\label{8(14)}
\qqq
for $k \not \in E$.

 Consider now $k \in E$. Then, we get from (\ref{8(7)}), (\ref{8(13)}) :
\qq
&&c\la^2 |f(k)| \leq |(\de \om^2\sigma_1  +M_{p})f (k)|
\non\\
&&\leq \left|\int_{k'\in E} K (k,k') f (k')\right| + \left|\int_{k' \not \in E} K
(k,k') f(k') dk'\right| 
+ |\mu| \|f \|_{\infty}
\label{8(15)}
\qqq 
We bound, using  (\ref{8(14)}),
\qq
\left|\int_{k'\not \in E} K (k,k') f (k') dk'\right| \leq C \la^2 \beta
\|f\|_{\infty}
\label{8(16)}
\qqq
and since $k'\in E$ means, for $b$ small, that $k'$ must be close to zero or
close to $\pi$, we get, using assumption 5 and H\"older's inequality 
\qq
&&\left|\int_{k' \in E} K (k,k') f (k') dk'\right| \leq 
 \| K (k,\cdot) \|_{1+\eta} (cb)^{\eta/1+\eta} \|f\|_{\infty}
\leq C \la^2 b^{\eta/1+\eta} \|f\|_{\infty}.
\label{8(17)}
\qqq
 
 Inserting  (\ref{8(16)}) and  (\ref{8(17)}) in  (\ref{8(15)}), we get $|f(k)|
\leq \left(C' \beta + {|\mu| \over c\la^2}\right) \| f \|_{\infty}$, for $k\in
E$, by choosing $b$ small enough. 
 This, combined with (\ref{8(14)}) implies $f=0$, for $\beta$ small enough, if,
say, $|\mu| < {c\la^2 \over 2}$. Thus, there is no non-zero solution of
(\ref{8(13)}) in that ball and (\ref{8(8)}) holds.

  \hfill$
\makebox[0mm]{\raisebox{0.5mm}[0mm][0mm]{\hspace*{5.6mm}$\sqcap$}}$
$
\sqcup$ 

 \vspace*{4mm}

Let us now check that the operator $\CL_{p}$ has the properties (1-5) above.

To do that, we must first write explicitely the operators $\CL_{ij}(0)$,
$\CL_{ij}(\pi)$. Let us start with $p=0$. Using (\ref{7(8)}, \ref{8.2}), we get,
since only terms that are even in $\textbf{s}$ give  a non-zero contribution
(see (\ref{5(1)}) and comments afterwards):
\qq
\CL_{12} (0) Q (k)& =& 2 T^{2}_{0} \sum_{{\scriptsize  \textbf{s}}} \int
\Bigl(\om (k_{1})
\om (k_{2})\Bigr)^{-2} s_{3} \om (k_{3}) \left[\om (k)^{-2} Q (k_{3}) - \om (k_{3})^{-2} Q (-k)\right]
\non
\\
&&\cdot\CP
\biggl(\Bigl(\sum s_{i} \om (k_{i})\Bigr)^{-1}\biggr)\de \Bigl(\sum k_{i}\Bigr) \de (k+k_{4}) d \underline{k} - (k \to -k),
\label{8(18)}
\qqq
where the $(k \to -k)$ term comes from  (\ref{5(12)});
(\ref{8(18)}) vanishes because $Q$ and $\om$ are even in $k$, see (\ref{5(14)}).
\qq
\CL_{11} J(k)& =& 2 T^2_{0} \sum_{{\scriptsize  \textbf{s}}} \int \Bigl(\om
(k_{1})
\om (k_{2})\Bigr)^{-2} s_{3} \om (k_{3})  \left[\om (k)^{-2}  {s_{3} J(k_{3}) \over \om (k_{3})} - \om (k_{3})^{-2}
{s_{4} J(-k) \over \om (k)}\right]
\non
\\
&& \cdot\de \Bigl(\sum s_{i} \om (k_{i})\Bigr) \de \Bigl(\sum k_{i}\Bigr) \de
(k+k_{4}) d \underline{k} - (k \to -k)
\label{8(19)}
\qqq
which equals twice the first term, since $J$ is odd in $k$ (see (\ref{5(15)}))
and $\om$ even.

$\CL_{21} (0) J$ vanishes by symmetry, like (\ref{8(18)}) (there is a + ($k\to
-k$) term in (\ref{5(20a)}) and $J$ is odd). This proves property 1, for $p=0$.

To prove point 2, write:
\qq
 (\CL_{22} Q) (k)&=& 4 T^2_{0} \sum_{{\scriptsize  \textbf{s}}} \int \Bigl(\om
(k_{1})
\om (k_{2})\Bigr)^{-2} s_{3} \om (k_{3}) s_{4} \om (k_{4})
 \left[\om (k_{4})^{-2} Q(k_{3}) - \om (k_{3})^{-2} Q
(k)\right]
\non
\\
&& \cdot\de \Bigl(\sum s_{i} \om (k_{i})\Bigr) \de \Bigl(\sum k_{i}\Bigr) \de
(k+k_{4}) d \underline{k}.
\label{8(20)}
\qqq

Taking the scalar product of (\ref{8(20)}) with $Q(k)$ in $L^2 \Bigl({\pi\over
N} \mathbb{Z}_{2N} \times [-\pi,\pi]^{d-1}, \om^2 (k) dk\Bigr)$, and replacing
in the second term in [---], 
 $s_{3} \om (k_{3})$ by $\displaystyle \left({1\over 3} \sum^3_{i=1} s_{i} \om
(k_{i})\right)$, using  symmetry, and then by    $\displaystyle \left(- {1\over 3}
s_{4} \om (k_{4}
\right)$, using the delta function, we get:
\qq
  (Q,\CL_{22} Q)& = &4 T^2_{0} \sum_{{\scriptsize  \textbf{s}}} \int
\prod^4_{i=1} \om (k_{i})^{-2}
 \left[s_{3} \om^3 (k_{3}) s_{4} \om^3  (k_{4}) Q (k_{3}) \bar Q (k) +
{1\over 3} \om^6 (k_{4}) \Bigl|Q(k) \Bigr|^{2}\right]
\non
\\
&& \cdot \de \Bigl(\sum s_{i} \om (k_{i})\Bigr) \de \Bigl(\sum k_{i}\Bigr) \de
(k+k_{4}) d \underline{k}.
\label{8(21)}
\qqq

Using $k=-k_{4}$, the symmetry between the $k_i$'s and
 the evenness of $\om, Q$, we can write
(\ref{8(21)}) as:
\qq
 {T^2_{0} \over 3} \sum_{{\scriptsize  \textbf{s}}} \int \prod^4_{i=1} \om
(k_{i})^{-2} \left| \sum s_{i} \om (k_{i})^3 Q (k_{i})\right|^2
\de \Bigl(\sum s_{i} \om (k_{i})\Bigr) \de \Bigl(\sum k_{i}\Bigr) \de
(k+k_{4}) d \underline{k}.
\label{8(22)}
\qqq
A similar computation, starting with (\ref{8(19)}), leads to
\qq
(J,\CL_{11} J) =- {T^2_{0} \over 3}  \sum_{{\scriptsize  \textbf{s}}}
\int \prod^4_{i=1} \om (k_{i})^{-2} \left|\sum \om (k_{i})^2 J
(k_{i})\right|^2
 \de \Bigl(\sum s_{i} \om (k_{i})\Bigr) \de \Bigl(\sum k_{i}\Bigr) \de
(k+k_{4}) d \underline{k}.
\label{8(23)}
\qqq

To conclude the proof of (\ref{8(4)}), (\ref{8(5)}), we need the following

 \vspace*{4mm}
 
\noindent {\bf Lemma 10.2}.{\it
Let $f$  be a H\"older continuous function from $\mathbb{T}^d$
  into $\mathbb{R}$, with $d\geq 3$, satisfying
\qq
f(k_{1}) + f(k_{2}) = f(k_{3}) + f(k_{4})   
\non
\qqq
on the set 
\qq
\{ (k_{i})^4_{i=1} | \ k_{i} \in \mathbb{T}^d,  \ i=
1,..., 4, \
k_{1} + k_{2} = k_{3} + k_{4}, \ \om (k_{1}) + \om (k_{2}) = \om (k_{3}) + \om (k_{4})\},
\non
\qqq
then,}
\qq
f(k) = a \om (k) + b.
\qqq

 \vspace*{4mm}
 
 Now, consider first (\ref{8(22)}) and  
(\ref{8(23)}) in the $N\to \infty$ limit,  i.e. with the sum in (\ref{5.0}) replaced by an integral.

To apply the Lemma, use the fact that, by (\ref{5(15)}) and relabelling
indices, we may assume that $s_{1}=s_{2}=1$, $s_{3}=s_{4}=-1$ and  change
$k_{3} \to -k_{3}$, $k_{4} \to - k_{4}$. Then, the Lemma applied to $f=\om (k)^3 Q(k)$ or $f=\om (k)^2 J(k)$ implies  that
(\ref{8(22)}) and  
(\ref{8(23)})
cannot equal zero unless $J=0$, since $J(k)$ is odd, or unless
$Q(k) = a \om (k)^{-2} + b \om (k)^{-3}$ which, for  $Q=P^\perp Q$, i.e. for $Q$
orthogonal to $\om (k)^{-2}$ and to $\om (k)^{-3}$ implies $a=b=0$.
Since each $\CL_{ii} (0)$ is the sum of a multiplication operator which
is bounded away from zero (see (\ref{J12}), (\ref{J13}) below for $p=0$) and a compact operator, either there is a zero
eigenvalue or inequalities (\ref{8(4)})-(\ref{8(5)}) hold.  Since we 
 showed
that there cannot be a zero eigenvalue (\ref{8(4)})-(\ref{8(5)}) hold for $p=0$, and $N=\infty$.

To conclude the proof for $N$ large, but finite, we simply use the convergence results stated in
Proposition 9.3, which imply the convergence of (\ref{8(22)}) and  
(\ref{8(23)}) to their $N=\infty$ limit.

The statements for $p=\pi$ follow from (\ref{7(8)}) and the fact that
$w(\pi,k-\pi) = w (p,k)$.

The bound (\ref{8(6)}) follows from Proposition 9.3,
and point 5 follows easily from the bounds (\ref{a7}) on the kernels
$K_{ij} (p) (k,\cdot)$, given by (\ref{1}) .

 Finally (\ref{8(7)}) holds because, for given $p,k$, $\Bigl(\de \om^2
  \sigma_1 + M (p,k)\Bigr)$ is a $2\times 2$ matrix, and the lower bound
(\ref{8(7)}) holds if the eigenvalues of that matrix satisfy
\qq
|\mu_{i} (p,k)|  \geq c (\la^2 + |\sin p | |\sin k|), \  \ i=1,2.
\non
\qqq

To prove this lower bound, it is enough to prove it for the square root of the
absolute value of the determinant of the matrix $\de\om^2 \sigma_1+ M (p,k)$, which equals
\qq
M_{11} (k,p) M_{22} (k,p) - \Bigl(\de\om^2 + M_{12} (k,p)\Bigr)
\Bigl(\de\om^2 + M_{21} (k,p)\Bigr)
\label{J11}
\qqq
We can check, from the explicit formulas (\ref{8(19)}), (\ref{8(20)}), in
which the multiplication operator corresponds to the last term in the [---],
that, for all $p$,
\qq
M_{11} (k,p) &<& -c\la^2
\label{J12}
\\
M_{22} (k,p) &>& c\la^2
\label{J13}
\qqq
for $c>0$. The signs and the factor $\la^2$ are obvious, and to get a non zero
contribution, we need only to check that $\sum {s_i} \om (k_i)$ vanishes for some
$k_{1}, k_{2},k_{3},k_{4}$, with the constraints $k_{4} = p-k$, $\displaystyle
\sum^3_{i=1} k_{i} = p+k$.
Choosing $s_{1} =+1$, $s_{2} = -2$, $s_{3} = +1$, $s_{4} = -1$ (which can
always be obtained by relabelling indices), and inserting the constraints,
this means
$$
\om (k_{1}) - \om (k_{2}) + \om (p+k-k_{1}-k_{2}) - \om (p-k)
$$
which vanishes for $k_{1} = k_{2} = k$.

Now, since $M_{ij} (k,p)$ for $i \neq j$ vanishes at $p=0$ or $\pi$ and at
$k=0$ or $\pi$, we get, using Proposition 9.3, 
\qq
|M_{ij} (k,p)| \leq C\la^2 \Bigl(|\sin k| \ |\sin p| \Bigr)^{\al/2}
\label{J14}
\qqq
for $i\neq j$. 

Inserting (\ref{J12}), (\ref{J13}), (\ref{J14}) into (\ref{J11}), using
$|\de\om^2| = 4 |\sin k| \ |\sin p|$, we get that $M_{ij}$, $i\neq j$ is small compared to 
$|\de\om^2|$ if $|\sin k| \ |\sin p|\geq c'\la^2$, for $c'$ small (in which case, both terms in (\ref{J11}) are negative), and that $  \Bigl(\de\om^2 + M_{ij} (k,p)\Bigr)$ is small
compared to $\la^2$, i.e. compared to $M_{ii}$, otherwise.

We are left with the
 \vspace*{4mm}

\noindent
{\bf Proof of Lemma 10.2.}
The proof follows closely the one of Proposition 12.1 in \cite{Spohn}, which itself is
inspired by \cite{CeKr}.

Let us first assume that $f$ is twice continuously differentiable.
The hypothesis of the Lemma imply that $f(k) + f(k')$ is constant $\forall k,
k' \in \mathbb{T}^d$, with $k+k'$ constant and $\om(k) + \om (k')$ constant. 
Therefore, there exists  $g : \mathbb{R} \times 
\mathbb{T}^d \to \mathbb{R}$, of
class $C^2$, such that
$$f(k) + f(k') = g \Bigl(\om (k) + \om (k'),
k+k'\Bigr).$$
Writing $k = (k^\al)^d_{\al=1}$, $k^\al \in \mathbb{T}$, $\om=\om (k'), \om' =
\om (k')$, we get 
\qq
&&
\p_{\al} f(k) = \p_{\om} g (\om + \om', k+k') \p_{\al} \om +
\p_{\al} g (\om+\om', k+k'),
\non
\\
&&
 \p_{\al} f(k') = \p_{\om} g (\om + \om', k+k'),
\p_{\al} \om' + \p_{\al} g (\om+\om', k+k'),
\non
\qqq
where $\p_{\al} = {\p\over \p
k^\al}$, $\p_{\om} = {\p \over \p \om }$.
Subtracting these two equations, we get 
\qq
\Bigl(\p_{\al} f(k) - \p_{\al} f (k')\Bigr) =
\p_{\om} g (\om + \om', k+k') \Bigl(\p_{\al} \om (k) - \p_{\al} \om (k')\Bigr)
\label{8(24)}
\qqq

Multiplying first (\ref{8(24)}) by $\Bigl(\p_{\be} \om(k) - \p_{\be} \om
(k')\Bigr)$, then rewriting the resulting equation by exchanging $\al$ and $\be$
we get, for all $ \al, \be$:
\qq
\Bigl(\p_{\al} f(k) - \p_{\al} f(k')\Bigr) \Bigl(\p_{\be} \om (k) - \p_{\be} 
 \om (k')\Bigr)
 = \Bigl(\p_{\be} f(k) - \p_{\be} f(k')\Bigr) \Bigl(\p_{\al} \om (k) -
\p_{\al} \om (k')\Bigr).
\non
\qqq

If we differentiate this identity with respect to $k_{\ga}$, we get:
\qq
&&\p_{\al} \p_{\ga} f(k) \Bigl(\p_{\be} \om (k) - \p_{\be} \om (k')\Bigr)
+ \Bigl(\p_{\al} f(k) - \p_{\al} f(k') \Bigr) \p_{\be} \p_{\ga} \om (k)
\non
\\
&&= \p_{\be} \p_{\ga} f(k) \Bigl(\p_{\al} \om (k) - \p_{\al} \om (k')\Bigr)
+ \Bigl(\p_{\be} f(k) - \p_{\be} f (k') \Bigr) \p_{\al} \p_{\ga} \om (k).
\non
\qqq
Differentiating now this with respect to $k'_{\de}$, we get:
\qq
\p_{\al} \p_{\ga} f(k) \p_{\be} \p_{\de} \om (k') + \p_{\al} \p_{\de} f(k')
\p_{\be} \p_{\ga} \om (k)= \p_{\be} \p_{\ga} f(k) \p_{\al} \p_{\de} \om (k') + \p_{\be} \p_{\de}
f(k') \p_{\al} \p_{\ga} \om (k).\hspace{4mm}
\label{8(25)}
\qqq
Now, for $\om (k)$ as in (\ref{2(4b)}), we have $\p_{\al} \p_{\be} \om (k) = 
\de_{\al\be} \cos k_{\al}$.
Using this and choosing in (\ref{8(25)}) $\al \neq \ga \neq \be = \de$, we
get $\p_{\al} \p_{\ga} f(k)=0$ for $\al \neq \ga$. This holds first on a dense
set $k_{\ga} \neq \pm {\pi\over 2}$ ($\cos k_{\ga} \neq 0$) and then, by
continuity, everywhere on $\mathbb{T}^d$.
Then, choosing $\al=\ga \neq \be=\de$, we get 
\qq
\p^2_{\al} f(k) \p^2_{\be} \om (k') = \p^2_{\be} f(k') \p^2_{\al} \om (k).
\non
\qqq
which implies that $\p^2_{\al} f (k) = a \p^2_{\al} \om (k)$ for a constant
$a$. Integrating, we get 
\qq
f(k) = a \om (k) + b + ck, 
\non
\qqq
for $a,b,c \in  \mathbb{R}$, and we get $c=0$ from the fact that $f$ is a
continuous function on $\mathbb{T}^d$.

This finishes the proof of $f$ of class $C^2$. For $f$ merely H\"older
continuous, we interpret all the (linear) identities above, and all the
derivatives in the sense of distributions, and we obtain the same conclusion. \hfill$
\makebox[0mm]{\raisebox{0.5mm}[0mm][0mm]{\hspace*{5.6mm}$\sqcap$}}$
$
\sqcup$

\vspace*{5mm}

\noindent
{\bf Remark}. The Lemma holds by assuming only that $f$ is a distribution, but we
do not need this.
It is crucial here that the dimension $d\geq 3$  in order to
be able to choose $\al \neq \ga \neq \be$. 
For counterexamples in $d=1$, see
\cite{LeS}.
\nsection{Proof of the Theorem}
\vskip 0.2cm

In this section we solve  equation (\ref{8(1)}). Due to the presence of zero modes
in the operator $\CL_{22}(0)$ we need to consider separately (\ref{8(1)}) in
the complement of these zero modes, at least for  $p\in E_{0}$,
 where 
 $$E_{0}=[-p_0,p_0]\cup [\pi-p_0,\pi+p_0],$$ with $p_0=B\la^2$, was defined in (\ref{E_0}),
 and the projection of (\ref{8(1)}) onto the zero modes. The
solution to the complementary equations leads to the \textit{Fourier Law},
i.e. an expression of the currents $(j,j')$ in terms of the temperature $T$
and chemical potential $A$. The solution to the projected equation determines
finally $T$ and $A$.

We look for a solution of the form 
$$W=Q_{0}+w, \ \ \
$$
where $Q_0=Q_0 (T, A)$,
for some functions $T,A$, is given by (\ref{6(9)}), and $w$,
also written as a pair $(J,r)$, is as follows. Let
\qq
w_{s} = w \ \chi (p\in E_{0})
\non
\label{}
\qqq
Let 
$P$ be the projection in $L^2 \Bigl(\om(p,k)^2dk\Bigr)$
to the span of
$\Bigl\{\om(p,k)^{-2},\om(p,k)^{-3}\Bigr\}$ and    $P^{\perp}=1-P$.
We demand 
 $$P^{\perp} r_s(p,\cdot)=r_s(p,\cdot), \ \ p\in E_0$$ 

Given a function $f$ that is H\"older
continuous in $p$ on $ E_{0}$, let  $\tilde f$ denote a linear extension of  $f$ to
${\pi\over N} \mathbb{Z}_{2N}$. We have $\|\tilde f\|_\al\leq C\|f\|_\al$.
We proceed similarly with elements of $\CS$ or of
$E$. Now write 
$$w=\tilde w_{s} + w_{\ell} $$ where  
$\tilde w_{s}$ is the extension of $w_{s}$ defined above. The function $w_{\ell}$
satisfies $w_{\ell} (p)=0$ for $p\in E_{0}$.

\vspace*{4mm}

Since from (\ref{5(29)}, \ref{Br3}) we have $PC=C$,
 eq. (\ref{8(1)}) can be written, for $p\in E_0$, as a pair of equations:
\qq
\Pi \left(\left(\begin{array}{ccc}
0 & \de\om^2\\
\de\om^2 & 0
\end{array}\right) \left(\begin{array}{c}
J\\
Q
\end{array}\right) + \CN (Q,J) + \CN_{\Ga} (Q,J)
\right) = 0
\label{12(1)}
\qqq
where
\qq
\Pi = \left(\begin{array}{ccc}
1 & 0\\
0 & P^\perp
\end{array}\right)
\non
\qqq
and
\qq
P(\de\om^2J + \CN_{2} + \CN_{\Gamma 2}) = PC.
\label{12(2)}
\qqq
For $p \notin E_0$, we will solve directly (\ref{8(1)}):
\qq
\left(\begin{array}{cccc}
0&\de\om^2\\
\de\om^2 &0
\end{array}\right)
\left(\begin{array}{c}
J\\
Q
\end{array}\right) + N (Q,J) + N_{\Ga} (Q,J) = \left(\begin{array}{c}
0\\
C
\end{array}\right) 
\label{JB2}
\qqq

We look for solutions where $(T,A)$, defined for $p\in E_0$,
 is such that $(\tilde T,\tilde A) \in \CB_{1}$, where
$\CB_{1} \subset E \times E$ is the ball
\qq
&& \Bigl\|T- T _+\de (p)\Bigr\|_{E} \leq B_1\tau
\label{12(15)}
\\
&& \|A\|_{E} \leq B_1\tau^2
\label{12(16)}
\qqq
with
\qq
 T_+ = \hf (T_{1} + T_{2})
\label{12(17)}
\qqq
and $B_1$ will be fixed later to be $\CO(1)$. For $\tau$ small enough, we have $\CB_1\subset  \CB_{\ep}$, defined in (\ref{Br2}), so that the estimates of section 9 can be used.
 As for $w$, 
we look for $w_s$ with $\tilde w_s$ in the ball $\CB_{2}\subset S \oplus S$, given by
\qq
\CB_{2} = \{(J,r)\ |\ 
\|(
\tilde J, 
\tilde r) - (
\tilde  J_{0}, 
\tilde  r_{0})\|_{\CS} \leq B_2\tau\}
\label{12(18)}
\qqq
where  $(J_{0},   r_{0})$ is given in (\ref{12(13)}) and bounded in  (\ref{12(14)}).
Finally, we shall choose
$w_{\ell} \in \CB_{3}$, where 
\qq
\CB_{3} = \{(J,r)
\  | \  \|J\|_{\CS} + \|r\|_{\CS} \leq B_3 \la^{-2}\tau,\ \
J_{p} = r_{p} = 0 \ \textrm{for} \ p\in E_{0}\}.
\label{JB1}
\qqq

Our Theorem is a consequence of the Proposition below and the Remark following it.

 \vspace*{4mm}
 
\noindent
{\bf Proposition 11.1}.   {\it Let $ \la$, $\tau$, $N$ be as in the Theorem}. 
\vskip 2mm
\no(a) {\it 
 Given $(T,A)$ defined for $p\in E_0$, such that $(\tilde T,\tilde A) \in \CB_{1}$, and given $w_\ell \in \CB_3$, there exists a unique  $w_s$, such that
$\tilde w_{s}\in \CB_{2}$ and such that $Q_{0} (\tilde T, \tilde A)+ \tilde w_{s} + w_{\ell}$ solves}
(\ref{12(1)})  {\it for $p\in E_{0}$, 
Moreover, $\tilde w_{s}$ is Lipschitz in $(T,A, w_\ell)$ with
Lipschitz constant $\CO(\la^2)$.}

\vskip 2mm

\no (b)  {\it  Given $ w_{\ell} \in \CB_3$,
 there exists a unique $(T,A) $ defined for $p\in E_0$, such that $(\tilde T,\tilde A) \in \CB_{1}$,
 and such
that $Q_{0} (\tilde T, \tilde A)+ \tilde w_{s} +
w_{\ell}$ solves} (\ref{12(2)}), {\it  for $p\in E_{0}$,
 where $w_{s} ( T,  A,
w_{\ell})$ is the solution obtained in }(a).  {\it The pair $(T,A)$ is Lipschitz in
$w_{\ell}$ with Lipschitz constant $\CO(\tau)$.}

\vskip 2mm

\no (c)  {\it There exists a unique $w_{\ell} \in \CB_{3}$ such that
$Q_{0} (\tilde T, \tilde A) + \tilde w_{s} ( T, A, w_{\ell}) +
w_{\ell}$ solves} (\ref{JB2}), {\it  for $p\notin E_0$, where $(T,A) = (T,A) (w_{\ell}) $ is the solution
obtained in }(b).

\vspace*{4mm}
\no
{\bf Remark}.   Moreover, the precise bounds stated in the Theorem will be given in the course of the proof: see (\ref{B60}), 
(\ref{Br5}), for the statements about $t$, (\ref{12(16)}) for the ones on $A$, and, for the currents, see (\ref{12(18)})  the Remark at the end of  subsection 11.1, specially (\ref{12(22)}).

\vspace*{4mm}

\subsection{Fourier's Law}
Let us prove first part (a) of Proposition 11.1.
For simplicity of notation, we shall write
 $Q_{0}$ for $Q_{0} (\tilde T,\tilde A)$, and drop the tilde on $T, A$.
Recall that $Q=Q_{0}+r$. The leading inhomogeneous term in (\ref{12(1)}) is
 $-\de\om^2 Q_{0}$. Using (\ref{6(9)}), we write:
\qq
\de\om^2 Q_{0} = t(2p) \rho_{1} (p,k) + d(-2p)
(T \ast A) (2p) \rho_{2} (p,k) +
\rho_{3} (p,k)
\label{Br13}
\qqq
where $t(p)=d(-p) T(p)$,
\qq
\rho_{1} (p,k) &=& d(-2p)^{-1} \de \om^2 (p,k) \om (p,k)^{-2} 
\label{rho1}\\
\rho_{2} (p,k) &=& d(-2p)^{-1} \de \om^2 (p,k) \om (p,k)^{-3}
\label{rho2}
\qqq
are smooth functions and
\qq
\rho_{3} (p,k) = \de \om^2 (p,k) \sum^\infty_{n=2} T \ast A^{\ast n} \om
(p,k)^{-2-n}
\non
\qqq
is in $\CS$ with
\qq
\| \rho_{3} \|_{\CS} \leq C \|A\|^2_{E}.
\label{Br11}
\qqq

The nonlinear term $\CN (Q,J)$ in (\ref{12(1)}, \ref{JB2}) was studied in Section 9.
It is given by (see eq. (\ref{5(20a)}, \ref{5(12)}, \ref{5(13)}))
\qq
\CN (p,k) = {9\over 8}(2\pi)^{3d}  \la^2 \left(\begin{array}{cc}
n_{1} (p,k) -n_{1} (p,-k)\\
n_{2} (p,k)+n_2(p,-k)
\end{array}\right)
\label{12(3)}
\qqq
and $n$ is the sum (\ref{7(1)}). From Proposition 9.2, we infer
\qq
\|\CN (Q_{0},0)\|_{\CS} \leq C\la^2 (\|t\|_{\CS} + \|A\|_{E}).
\non
\qqq
From (\ref{7(7)}), (\ref{7(9)}), the definition of $\CL_{p}$ in section 10 and Proposition 9.5, we get 
\qq
\| D\CN (Q_{0},0) - \CL_{p}\| \leq C\la^2 (\|t\|_{\CS} + \|A\|_{E})
\label{12(4)}
\qqq
where on the LHS the norm is the operator norm in $\CS \oplus \CS$. 
Finally, Proposition 9.6 gives 
\qq
\| \CN-\CN (Q_{0},0) - D\CN (Q_{0},0) 
(J, r)^T
\|_{\CS} \leq CN^{-\ha+\al} (\|J\|_{\CS} + \|r\|_{\CS})^2.
\label{12(5)}
\qqq
So, combining those estimates, we get that , for $T$, $A$, $J$, $r$, as in the theorem:
\qq
\| \CN (p,k)-\CL_p(J,r)^T \|_\CS \leq C\la^2 \tau
\label{Br8}
\qqq
and is Lipschitz  as a function of $T$, $A$ with constant $C\la^2$, and as a function of $J$, $r$, 
with a constant $CN^{-\ha+\al}$.
Consider next the function $\CN_{\Ga}$, defined in (\ref{Ngamma}):
\qq
\CN_{\Ga} =(\Ga J + J \Ga, \Ga P + P\Ga)^T.
\label{12(6)}
\qqq
Let us start with
 $\Ga J + J\Ga$, given by (\ref{5(30)}) with $P$ replaced by
$J$. Recalling the definition (\ref{6(4b)}),  using
\qq
\int |d(p)|^{-1} {dp} &\leq& C \log N
\label{Br1},
\qqq
 that follows from
$|d(p)|^{-1}\leq C|p|^{-1}$, for $p\neq 0$ and $|p|\geq cN^{-1}$,
we get
\qq
\int |J(q,q+k-p)| dq \leq CN^{-1+\al/2}  \|J\|_{\CS},
\label{L1boundS}
\qqq
uniformly in $k,p$, since the singularities in (\ref{6(4b)}) affect only the first variable and $\log N \leq N^{\al/2}$.
Now, use the fact that, for functions on the $\pi/2N$
lattice, 
\qq
\|f\|_\al\leq CN^{\al}\|f\|_\infty,
\label{Br200}
\qqq
 and  identify $\Ga J + J\Ga$ as an element of $\CS$ of the form
$\Ga J + J\Ga = (0,0,\star, 0)$, to get:  
\qq
&&\|\Ga J + J\Ga\|_{\CS} \leq C \ga N^{\al} N^{1-\al/2} \|\int |J(q,q+k-p)| dq\|_\infty
 \non
\\
&&\leq C\ga N^{\al} \|J\|_{\CS}\leq  C N^{-1+5\al/4} \|J\|_{\CS}
\label{12(7)}
\qqq
where in the last inequality, we use the definition (\ref{gamma}) of $\ga$. 
$\Ga J + J\Ga$ is of course Lipschitz, with constant $ C N^{-1+5\al/4}$.
For the term $\Ga P + P\Ga$, we need to study $P$, given in (\ref{5(11)}):
\qq
P=\om (p,k)^2 Q + \hf(J\Ga - \Ga J) - {9\over 8}(2\pi)^{3d}  \la^2\Bigl(n_{1}(p,k) +
n_{1}(p,-k)\Bigr).
\label{12(8)}
\qqq
We have
\qq
\om (p,k)^2 Q = T(2p) + (T \ast A) (2p) \om (p,k)^{-1} + p_{1} + \om (p,k)^2
r
\non
\qqq
where $p_{1}$ collects the $n \geq 2$ terms coming from the expansion
(\ref{6(9)}). From Proposition 9.2, the last term in (\ref{12(8)}) at $J=r=0$
is given by (\ref{n1}). Propositions 9.3, 9.5 and 9.6 control the corrections.
The $(J\Ga - \Ga J)$ term is bounded as in  (\ref{12(7)}).
To summarize, let us call $P_S$ the sum of $\hf(J\Ga - \Ga J)  + \om (p,k)^2
r$, of the corrections to the $J=r=0$ term in $-{9\over 8}(2\pi)^{3d}  \la^2 \Bigl(n_{1}(p,k) +
n_{1}(p,-k)\Bigr)$ and of the term corresponding to $m$ in (\ref{n1}). Let $P_E$ be the rest, i.e.
$T(2p) + (T \ast A) (2p) \om (p,k)^{-1} + p_{1}$, and what corresponds to
 the sum in (\ref{n1}). Collecting the bounds established for these various terms, we get (remember that inverse powers of $N$ are small compared to $\la^2$):

\vspace*{4mm}

\no {\bf Proposition 11.2}. {\it  $P$ can be written as
\qq
P=P_{E} + P_{S}
\non
\qqq
with
\qq
P_{E} (p,k) = T(2p) + (T \ast A) (2p) \om (p,k)^{-1} +{\overline P_{E}} (p,k)
\label{12(9)}
\qqq
where
\qq
{\overline P_{E}} (p,k)=
 \sum^\infty_{n=1}
F_n (2p) h_{n} (p,k)
\label{12(9a)}
\qqq
where $F_n \in E$, with $ \|F_n\|_E \leq C^n \|A \|_E^n$, for $n \geq2$,
$\|F_{1}\|_{E}\leq C \la^2 \|A \|_{E}$, and where
 the functions $h_{n}$ are smooth with
\qq
\|h_{n}\|_{\infty} \leq C^n,
\label{12(10)}
\qqq
$P_{S} \in \CS$ and}
\qq
\|P_{S}\|_{\CS} \leq C(\|r\|_{\CS} + \la^2( \|J\|_{\CS} + \|t\|_{\CS} + \|A\|_{E})).
\label{Br9}
\qqq

\vspace*{4mm}

We may now bound $P^{\perp} (P\Ga + \Ga P)$. The $P_{S}$ term is in $\CS$, like $J$
and can be bounded, as in (\ref{12(7)}), by the RHS of (\ref{Br9}) times $N^{-1+5\al/4}$. The $T(2p)$ in (\ref{12(9)}) drops out from
$P^{\perp}$, since it is constant in $k$ and we use (\ref{Br3}). The other terms in (\ref{12(9)}) are bounded using
\qq
\int dq |F(q)| \leq C\|F\|_{E},
\label{Br6}
\qqq
which follows  easily from:
\qq
|F(p)|\leq \Bigl(\de(p)+{1\over{N|d|^2}}+{1\over{N^{1-\al/2}|d|}} + {1\over{N^{3/2}|d|^{5/2}}}
\Bigr)\|F\|_E,
\label{tbound}
\qqq
using (\ref{Br1}) and:
\qq
\int |d(p)|^{-k} {dp} &\leq& C N^{k-1} \ \ \ \ k> 1,
\label{Br0}
\qqq
which is proven like (\ref{Br1}).
Thus $P^{\perp} (\Ga P_{E} + P_{E} \Ga) (p,k)$ is smooth, since, looking at (\ref{5(30)})
we see that  the dependence on $p,k$
in $P^{\perp} (\Ga P_{E} + P_{E} \Ga) (q, q+k-p)$ is only through the second argument
of $P_{E} $, in which $P_{E} $ is smooth, by Proposition 11.2.
Moreover, using  (\ref{Br6}) and  (\ref{12(9)}), (\ref{12(9a)}),
\qq
|\int P^{\perp} (\Ga P_{E} + P_{E} \Ga) (q, q+k-p)|  \leq C\ga \|A\|_{E}.
\label{12aa}
\qqq
Since $P^{\perp} (\Ga P_{E} + P_{E} \Ga) (p,k)$ is smooth, let us identify it with an element of $\CS$ of the form 
$(0,0, \star, 0)$. Then,
by (\ref{gamma}), $P^{\perp} (\Ga P_{E} + P_{E} \Ga) \in \CS$ and
\qq
\|P^{\perp} (\Ga P_{E} + P_{E} \Ga) \|_{\CS} \leq CN^{-1+\al/4} N^{1-\al/2} \|A\|_{E} = CN^{-\al/4} \|A\|_{E}.
\label{12ab}
\qqq
Combining the  above bounds, we get:
\qq
\|P^{\perp} (\Ga P + P \Ga) \|_{\CS} \leq CN^{-\al/4} \|A\|_{E}
+CN^{-1+5\al/4}(\|r\|_{\CS} + ( \|J\|_{\CS} + \|t\|_{\CS})
\label{Br10}
\qqq
and $P^{\perp} (\Ga P_{E} + P_{E} \Ga)$ is Lipschitz in $A$, $E$, $J$, $r$, 
with constants $\CO(N^{-\al/4})$.
\vspace*{4mm}

We may summarize this discussion by rewriting equation
(\ref{12(1)}); consider $T, A$ given, for $p\in E_0$ and $w_\ell$ given for $p\notin E_0$.
Let us denote
\qq
s(p)=d(-p)(T \ast A)(p)\equiv d(-p) S(p).
\label{sdef}
\qqq
Then
\qq
(J_s, r_s)^T
= -\CD_p^{-1} \Pi \left[
( \rho_{1}t(2p) +  \rho_{2}s (2p),
0)^T
 + \CR\right]
\label{12(11)}
\qqq
for $p\in E_0$, using the fact that  $r_s=P^{\perp}r_s$, and therefore, $(
J_s,
 r_s)^T=-\Pi \ (
J_s,
 r_s)^T$. Here,  $\CR \in \CS$ includes  all the  terms in (\ref{12(1)}), apart from the first two in (\ref{Br13}). Combining (\ref{Br11}), (\ref{Br8}), (\ref{12(7)}), (\ref{Br10}), $\|\CR\|_{\CS}$
 is bounded, for $T$, $A$, $J$, $r$ as in the Theorem (using the fact that $\tau \la^{-2}\leq 1$), by:
\qq
\|\CR\|_{\CS} &\leq& C \la^2B_1 \tau
\label{12(12)}
\qqq
and is Lipschitz $J,r$ with constant
$\CO (\tau)$, and in $T,A$ with constant $\CO (\la^2)$.
 $\CD_p^{-1}$ is the operator defined by (\ref{CD1}), and  bounded in Proposition 10.1.

Let, for $p\in E_0$,
\qq
(J_{0}, r_{0})^T
 = -\CD_p^{-1} \Pi
(\rho_{1}t(2p) + \rho_{2}s(2p),0)^T \label{12(13)}
\qqq

Then, by Proposition 10.1 and Lemma 9.1.b,
\qq
\|(
\tilde J_{0}, 
\tilde r_{0})
\|_{\CS} \leq C\la^{-2} (\|t\|_{\CS} + \|s\|_{\CS}) \leq C\la^{-2} (\|t\|_{\CS} 
+ \|A\|_{E}).
\label{12(14)}
\qqq
Now, given $\tilde w_s \in \CB_2$ and $w_\ell \in \CB_3$ we have from
Proposition 10.1. and  (\ref{12(12)})
\qq
\| \CD_p^{-1} \Pi \CR\|_{\CS} \leq CB_1\tau,
\label{12(19)}
\qqq
i.e.  $\CD^{-1}\Pi \CR \in \CB_{2}$ for  $B_2>CB_1$.  It is Lipschitz in $J,r, w_\ell$ with constant
$\CO (\tau)$, and in $T,A$ with constant $\CO (\la^2)$. Thus we get, from the contraction mapping principle,

\vspace*{4mm}

\no {\bf Proposition 11.3}. {\it Given $T, A, w_\ell$ as above, 
 equation }(\ref{12(11)}) {\it has a unique solution $(\tilde J_s,\tilde r_s)\in  \CB_2$,
  which is Lipschitz in $(T,A)$ with Lipschitz constant $\CO
(\la^2)$ and in $ w_\ell$ with Lipschitz constant $\CO
(\tau)$ .}

\vspace*{4mm}

This proves part (a) of Proposition 11.1. Now assume that we have proven part (b) of that Proposition. This will be done in the next subsection. We want to prove here part (c), since it is done in the same spirit as part (a). 

Thus, consider equation (\ref{JB2}), for $p \notin E_0$. Following the argument that led from 
(\ref{12(1)}) to (\ref{12(11)}), we may rewrite it as:
\qq
(J, r)^T
= -\CD_p^{-1} \left[ (\rho_{1}t (2p)+ \rho_{2}s (2p),
0)^T
 + \CR\right]
\label{12a}
\qqq
where $\CD_p$ was defined in  (\ref{CD2}), for $p\notin E_0$, and is invertible by Proposition 10.1.
We can write (\ref{12a}) as:
\qq
(J_\ell, r_\ell)^T
= -(\tilde J_s, \tilde r_s)^T-\CD_p^{-1} \left[ (\rho_{1}t + \rho_{2}s,
0)^T
 + \CR\right]
\label{12b}
\qqq
where $(\tilde J_s,
\tilde  r_s)$ is the Lipschitz function of $w_\ell$, given by Proposition 11.3.
The $\CS$-norm of the RHS is bounded by 
$
C\la^{-2}B_1\tau
$
and so, taking $B_3=CB_1$, (\ref{12b})
is in $\CB_3$, provided we show it vanishes for $p \in E_0$.

By assumption, $T, A$ here is such that (\ref{12(2)})  holds  for $p \in E_0$ (part (b) of the Proposition, to be proven below), and, by part (a), we know that (\ref{12(1)})  holds  for $p \in E_0$. Putting (\ref{12(1)})  and (\ref{12(2)})  together, we see that the full equation, (\ref{JB2}), is satisfied by $Q_{T, A}+\tilde w_s + w_\ell$  for $p\in E_0$.  But (\ref{12a})
is merely a rewriting of (\ref{JB2}), whenever $\CD_p$ is invertible. Since we can assume, by choosing, if necessary, $B$ in Proposition 10.1 larger than here, that $\CD_{\pm p_0}$ and $\CD_{ \pi \pm p_0}$ are invertible, and since, by assumption, 
$$Q_{T, A}+\tilde w_s + w_\ell = Q_{T, A}+\tilde w_s 
$$ 
for $p\in E_0,
$
we know that $Q_{T, A}+\tilde w_s $ solves  (\ref{12a}) for $p=\pm p_0$ or $p=\pi \pm
p_0$. But that means that the RHS of  (\ref{12b}) vanishes for those values of $p$. We  can define both sides to be zero for other values of $p \in E_0$, if we want, without affecting the fact that $J_\ell, r_\ell$ are in $\CS$, and thus we obtain part (c) of Proposition 11.1, by applying  the contraction mapping principle to  the fixed point equation  (\ref{12b}) to $(J_\ell, r_\ell) \in \CB_3$.

\vspace*{4mm}
\no
{\bf Remark}.  The function $J(T,A)$ is a general form of the
\textit{Fourier Law}, expressing the $E q_x p_y$ correlation function as a
function of the local temperature and chemical potential. In particular, see  (\ref{12(13)})  and remember that $\Pi$ is the identity of the first component,
\qq
J_{0} (p,k) = \kappa_{1} (p,k) t (2p) + \kappa_{2} (p,k) s (2p)
\label{12(20)}
\qqq
with
\qq
\kappa_{j} =-\CD_p^{-1} 
( \rho_{j} ,
0)^T.
\label{12(21)}
\qqq
Inserting to (\ref{5(20)}) and (\ref{5(33)}) we get
\qq
\Bigl(j_{0} (p), j'_{0} (p)\Bigr)^{T} = \kappa_0(p) \Bigl(t(p),s(p)\Bigr)^{T}
\label{12(22)}
\qqq
where $\kappa_0(p)$ is  $C^{\al}$ in $p$. Since $\CD_p$ is of order $\la^2$, 
$\kappa_0(p)$ is of order $\la^{-2}$.
The full $\kappa (p)$ introduced in the Theorem has
corrections $\CO(1)$ coming from $\CD_p^{-1}$ applied to the parts of $\CR$ in (\ref{12(11)})
that are linear in $t, s$. Since $\CR$ is or order $\la^2$, the result is  $\CO(1)$, i.e. small compared to 
$\kappa_0$, at least if the latter  does not vanish.
 In the next subsection, we shall need the fact that $\kappa(p)
$ is invertible for $p\in E_0$.
 To show that, let us first compute $\kappa_0(0)$, which we shall denote for simplicity $\kappa^0$.

From (\ref{CD1}) and (\ref{8(3)}), we get
\qq
\CD_{0} = \left(\begin{array}{ccc}
\CL_{11} (0) & 0\\
0 & \CL_{22} (0)
\end{array}\right).
\non
\qqq
Inserting (\ref{deome}) to (\ref{rho1}) and (\ref{rho2}) we have
\qq
\rho_{j} (0,k) = 4 i  \sin k \om (k)^{-j}
\non
\qqq
for $j=1,2$, and so, defining
\qq
\psi_{j} (k) = 4 \sin k \om (k)^{-i},\ \ \ j=1,2,
\non
\qqq
we have
\qq
\kappa_{j} (0,k) =- i \Bigl(\CL_{11} (0)^{-1} \psi_{j}\Bigr) (k).
\non
\qqq
Inserting to (\ref{5(20)}) we have, for $j=1,2$,
\qq
\kappa^0_{1j} =- 2\int dk \ \sin k\ \om (k) (\CL_{11} (0)^{-1} \psi_{j}) (k).
\label{12(23)}
\qqq
From (\ref{5(33)}) we get
\qq
\kappa^0_{2j}  =- 2 \int dk \  \sin k\ \eta  (0,k) \om (k) (\CL_{11} (0)^{-1} \psi_{j}) (k).
\label{12(24)}
\qqq
where $\eta (0,k)$ equals $\om (k)^{-1} - \int
dk\om(k)^{-1}$.
Now, note  that
\qq
\kappa^0_{1j}  =- \ha \Bigl(\psi_{1}, \CL^{-1}_{11} (0) \psi_{j}\Bigr)
\label{12(25)}
\qqq
where $(\cdot,\cdot)$ is the scalar product in $\CH_{0}$, and that
\qq
\kappa^0_{2j}  = -\ha \Bigl(\psi_{2}, \CL^{-1}_{11} (0) \psi_{j}\Bigr)+\be_0 \ha \Bigl(\psi_{1}, \CL^{-1}_{11} (0) \psi_{j}\Bigr),
\label{12(26)}
\qqq
where $\be_0 =\int
dk\om(k)^{-1}$.
Computing the determinant of the $2\times 2$ matrix $\kappa^0$, we see that it equals the one with $\be_0=0$, and the latter does not vanish because $\CL_{11} (0)$ is a strictly negative operator (see (\ref{8(4)})). Now, from H\"older continuity, we get that $\kappa_0 (p)$, and therefore $\kappa (p)$,
is invertible for $p$ small, i.e. for $p\in E_0$, for $\la$ small enough.

Finally, observe that, since $\CL_{11} (0)$ is a strictly negative operator, $\kappa (p)$, is a positive matrix, for $p\in E_0$. To understand the connection with the Fourier law (\ref{(I1)}), note that $t(p)=d(-p)T(p)$, and, in $x$-space, $d(-p)$ is $-\nabla$.

\subsection{Solving the conservation laws}
\vskip 0.2cm
We are left with the proof of part (b) of Proposition 11.1. This reduces to solving
 the two conservation laws, equations (\ref{5(17)}) and 
(\ref{5(32)}), which are equivalent to (\ref{12(2)}). Indeed, (\ref{5(17)}) is  (\ref{4(20)})  for $x=y$, which is the same as  integrating (\ref{5(13)}) with $dk$, or taking the scalar product of (\ref{5(13)}) with $\om(p,k)^{-2}$ in $\CH_p$. 
For (\ref{5(32)}), it amounts to taking the scalar product of (\ref{5(13)}) with a linear combination of $\om(p,k)^{-2}$  and $\om(p,k)^{-3}$ .

Let us introduce a more compact notation. We set $\CJ = (j,j')$
and write  (\ref{5(17)}) and  (\ref{5(32)}) as
\qq
d(p) \CJ (p) + \CF (p) - \Theta (p) = 2\ga (T_{1} + e^{iNp} T_{2}, 0)^T
\label{9(1)}
\qqq
for $p\in E_0$ with $ E_0$ given by  (\ref{E_0}), where we can assume that $\kappa (p)$ is invertible.
Equation (\ref{9(1)}) has the friction term, see (\ref{5(34)})
\qq
\CF (p) = \ga \int dkdq P ({q}/2, k) (1+ e^{i(q+p)N}) (2, \psi (p,k,q))^T,
\label{9(2)}
\qqq
where we use $e^{ipN}=e^{-ipN}$,
and the projection of $N_{22}$:
\qq
\Th (p) = (0,\th (p))^T,
\label{9(3)}
\qqq
where $\th$ is given by (\ref{thetadef}). 

$\CJ$ is given by the Fourier law i.e. the solution of (\ref{12a})
with leading term (\ref{12(22)}) which we may write as (see the Remark at the end of subsection 11.1):
\qq
\CJ (p) = \kappa (p) [
(t(p), s(p))^T + z (p)],
\label{9(4)}
\qqq
where  $s$ is given in eq. (\ref{sdef}).
$\CJ, \CF$ and $\Th$ are functions of $T,A \in E$, and $w_\ell\in \CB_3$ (including, indirectly, as functions of $w_s$, which is a function of $T$, $A$, $w_\ell$, by part a of Proposition 11.1).
So, for $w_l$ fixed 
(\ref{9(1)}) is a nonlinear, nonlocal elliptic equation for $T$ and $A$.  We look for a solution to (\ref{9(1)}) in the ball
$\CB_{1} \subset E \times E$, defined in (\ref{12(15)}, \ref{12(16)}).
For such $T$ the map $A \to T \ast A\equiv S$ is invertible (because $T$ in (\ref{12(15)}) is close to a delta function, which is  the identity for the convolution) and
\qq
\| S - T_{+} A \|_{E} \leq C\tau^3.
\non
\qqq
Thus we can use $T,S$ as the unknowns, in the ball 
$\CB_{1} $. The following
Proposition collects the properties of the functions $z$ and $\th$, studied
in Propositions 11.3 and 9.7 (since $|\theta(0)|\leq CN^{-1}$, 
we have that $d^{-1}( \theta-\theta(0) )\in S$) :

\vskip 0.6cm

\no {\bf Proposition 11.4.}\ {\it $z$ and $d^{-1}( \Th-\Theta(0))$ are Lipschitz
functions $\CB_{1} \times \CB_{3} \to S \oplus S$ with
\qq
&&\| z \|_{S} \leq C\la^2 \tau,
\non
\\
&&\| d^{-1}( \Theta-\Theta(0) )\|_{S} \leq C\la^2 \tau ,
\label{9(6)}\\
&&|\Theta(0)|\leq CN^{-1}.
\label{Br53}
\qqq
and Lipschitz constants bounded by $C\la^2$.}

\vskip 0.6cm

Recall next that $f\in \CS$ is decomposed as (see (\ref{6(6a)}))
\qq
f(p,k) = f_{+} (p,k) \sig_{+} (2p) + f_{-} (p,k) \sig_{-} (2p),
\non
\qqq
with $\sig_{\pm} (p) = 1 \pm e^{iNp}$, and $f_\pm=\ha f$. 
Insert 
\qq
1+e^{i(q+p)N} = {1\over 2} \Bigl(\sig_{+} (p) \sig_{+} (q) + \sig_{-} (p)
\sig_{-} (q)\Bigr)
\label{Br12}
\qqq
into eq. (\ref{9(2)}) and use $\sig_{+} (q) \sig_{-} (q) = 0$ for $q \in {\pi
\over N} \mathbb{Z}_{2N}$ to get:
\qq
\CF (p) = \CF_{+} (p) \sig_{+} (p) + \CF_{-} (p) \sig_{-} (p),
\non
\qqq
with
\qq
\CF_{\pm} (p) = \ga \int_{\pm} dq \int dk P_{\pm} ({q/ 2},k)
(2,
 \psi (p,k,q))^T
\label{9(8')}
\qqq
where $\int_{\pm} dq = \ha \int dq \sig_{\pm} (q)$ i.e. the $q$-sum in
$\int_{\pm}$ runs over the odd (for $-$) or even, non zero (for $+$), multiples of
${\pi\over N}$. We used here the fact that $\sig_{\pm} (q)^2=0, 4$ to cancel the factors
of $\ha$ in (\ref{Br12}) and in $P_\pm=\ha P$.

Thus, (\ref{9(1)}) becomes two equations, one $(+)$ valid on the even sublattice, and the other one $(-)$ on the odd sublattice:
\qq
d \CJ_{\pm} + \CF_{\pm} - \Theta_{\pm} = 2\ga T_{\pm}
(1,0)^T
\label{9(9')}
\qqq
with $T_{+}$ the average temperature (\ref{12(17)}) and
\qq
T_{-} = \hf(T_{1} - T_{2})
\label{9(10)}
\qqq
It is useful to separate from (\ref{9(8')}) the part which is  constant in $p$:
\qq
\CF_{\pm} (p) = \CF_{\pm} (0) + f_{\pm} (p),
\label{9(11)}
\qqq
and similarly for $\Theta_{\pm} (p)$:
\qq
\Theta_{\pm} (p) = \Theta_{\pm} (0) +{\Theta'}_{\pm} (p),
\label{9.111}
\qqq
Since $\psi$ in (\ref{5(35)}) is smooth we may write:
\qq
f_{\pm} (p) = d(p)  g_{\pm} (p).
\label{9(12)}
\qqq
and we may  estimate $g_\pm$ as in the derivation of (\ref{12ab}), and obtain (there is no $P^\perp$
here, so we have  a term linear in $T$):

\vskip 0.4cm

\no {\bf Proposition 11.5} \ {\it The functions $g_{\pm} \in S$ are Lipschitz in $(T,S) \in
\CB_{1}$ with $\| g_{\pm} \|_{S}$ and the Lipschitz constants $\CO (N^{-\al/4})$.}

\vskip 0.2cm

Using (\ref{9(4)}), we may write (\ref{9(9')}) as
\qq
d(p) \kappa (p) \left((t_{\pm} (p), 
s_{\pm} (p))^T
 + W_{\pm} (p)\right)= 2\ga T_{\pm} (1, 0)^T
 - \CF_{\pm} (0)+\Theta_{\pm} (0)
\label{9(13)}
\qqq
where $W_{\pm}(p)= z_{\pm} (p)+  \kappa (p)^{-1} (g_{\pm}(p)- d(p) ^{-1}\Th'_{\pm}(p))$, is defined for $p\in E_0$, where $\kappa (p)$ is invertible (see Remark at the end of subsection 11.1). So,  combining Propositions 11.4 and 11.5, $\tilde W_{\pm} \in S$, with
\qq
\|\tilde W_{\pm}\|_{S} \leq C\la^2\tau,
\label{Br14}
\qqq
and the functions $\tilde W_{\pm}$ are Lipschitz in $(T,S)$, with constants $C\la^2$. 

Let us next
analyze $\CF_{\pm} (0)$. Using Proposition 11.2,  $S=T\ast A$ and (\ref{5(21b)}), we write 
\qq
 \CF_{\pm} (0)-\Theta_{\pm} (0)& =& \phi_{\pm} + \psi_{\pm}
\non
\\
\phi_{\pm} &=& \ga \int_{\pm} dq \int dk \Bigl(T(q) + \rho ({q/2
},k) S(q)\Bigr)(2,  \psi(0,k,q))^T
\label{9(14)}
\\
\psi_{\pm}& =& \ga \int_{\pm} dq \int dk \Bigl({\overline P_{E}} ({q/ 2
}, k) + P_{S}({q/ 2
}, k) \Bigr) 
(2, \psi(0,k,q))^T-\Theta_{\pm} (0).
\label{9(15)}
\qqq
For $\psi_{\pm}$, we proceed as in the proof of 
 (\ref{12ab}).  The  contribution coming from $P_{S}$ is small, see  (\ref{12(7)}), while the one coming from $P_{E}$ is to leading order, see (\ref{12(9a)}), $\CO (\ga\la^2\tau^2)$. The contribution of 
 $\Theta_{\pm} (0)$ is $\CO (N^{-1})$, by Proposition 9.7. So,
$\psi_{\pm}$ are bounded by
\qq
|\psi_{\pm}| \leq C\ga\la^2\tau^2
\label{9(16)}
\qqq
and are Lipschitz in ($T,S$) with constant $C\ga \la^2 $. 

It is instructive to solve first  the simplified equation (\ref{9(13)}) with
$W_{\pm}$ dropped and $\CF_{\pm} (0)$ replaced by $\phi_{\pm}$. Then, for
$p\neq 0$,
\qq
\kappa (p) \Bigl(t_{\pm} (p), s_{\pm} (p)\Bigr) = d(p)^{-1} \xi_{\pm}
\label{9(17)}
\qqq
with
\qq
  \xi_{\pm} = 2 \ga T_{\pm}
  (1, 0)^T - \phi_{\pm}
\label{9(18)}
\qqq
and, for $p=0$ (where obviously only the equation with index $+$ holds),
\qq
2\ga T_{+}  (1, 0)^T= \phi_{+}.
\label{9(19)}
\qqq
Equations (\ref{9(18)}) and (\ref{9(19)}) imply
\qq
\xi_{+} = 0
\label{9(19a)}
\qqq
so, by  (\ref{9(17)}),  
\qq
(t_{+}, s_{+}) = 0
\non
\qqq
and, from (\ref{9(17)}) for $-$, we can write
\qq
\Bigl(t_{-} (p), s_{-} (p)\Bigr) = d(p)^{-1} \Bigl(\tau (p), \zeta (p)\Bigr)
\label{9(20)}
\qqq
with 
\qq
\kappa (p) \Bigl(\tau (p), \zeta (p)\Bigr)^T= \xi_-
\label{Br26}
\qqq
constant in $p$, i.e.:
\qq
\Bigl(\tau (p), \zeta (p)\Bigr)^T = \kappa (p)^{-1} \kappa (0) \Bigl(\tau (0), \zeta
(0)\Bigr)^T,
\label{9(21)}
\qqq
where $(\tau (0), \zeta
(0)\Bigr)$ is defined by extending (\ref{Br26}) (which was derived on the odd sublattice) to $p=0$.
Hence, since $s(p)=d(-p)S(p)$, $t(p)=d(-p)T(p)$,
\qq
U(p) \equiv  \Bigl(T(p), S(p)\Bigr)^T  = U_{0} \de (p) + |d(p) |^{-2} \Bigl(\tau
(p), \zeta (p)\Bigr)^T \sig_{-} (p).
\label{9(22)}
\qqq

The unknowns are $U_0=(T_0, S_0)^T$, $\tau (0)$ and $\zeta (0)$ and they will be determined
from (\ref{9(18)}) and (\ref{9(19)}), where the functions $\phi_{\pm}$, given by
(\ref{9(14)}), are functions of $U_0$, $\tau (0)$, $\zeta (0)$ via  (\ref{9(22)}):
\qq
\phi_{+} &=& \ga \int dk \Bigl(T_{0} + \rho (0,k) S_{0} \Bigr) \ \Bigl(2,\psi
(0,k,0)\Bigr)^T
\label{9(23)}
\\
\phi_{-} &=& 2\ga \int_{-} d{ q} \int dk |d({ q})|^{-2} \Bigl(\tau({ q}) + \rho
\Bigl({{ q}/ 2},k\Bigr) \zeta ({ q})\Bigr) 
\
\Bigl(2,\psi (0,k,q)\Bigr)^T,
\label{9(24)}
\qqq
where the prefactor of $2$ comes from the fact that, in (\ref{9(22)}), $\sig_{-} (p)=0,2$.

Equations (\ref{5(21b)}), (\ref{5(31)}) and  (\ref{5(35)}) imply 
$\psi (0,k,0)=2( \rho (0,k)-\int dk \rho (0,k))$, with $\rho (0,k)=\om(k)^{-1}$. So,
\qq
\int \psi (0,k,0) dk& = &0,
\non
\\
 \int \rho (0,k) dk& \equiv& \be_1 > 0,
\non
\\
\hf \int \rho (0,k) \psi (0,k,0) dk& = &\int \om (k)^{-2} dk -
\Bigl(\int \om (k)^{-1}dk\Bigr)^2
\equiv \be_2 > 0.
\non
\qqq

Hence (\ref{9(23)})  gives $\phi_{+}=2\ga (T_0 + \be_1 S_0, \be_2 S_0)^T$ and, from (\ref{9(19)}), we get:
\qq
S_{0} = 0  \ \ \ , \ \ \ T_{0} = T_{+} = \hf (T_{1} + T_{2}).
\label{9(25)}
\qqq

To analyze (\ref{9(24)}), we need the straightforward

\vskip 0.4cm

\no{\bf Lemma 11.6}  {\it Let $a$ be  a function on ${\pi \over N} \mathbb{Z}_{2N}$.
Then
\qq
&&2 \int_{\pm} |d({ q})|^{-2} a({ q}) d{ q} = N \Bigl(I_{\pm} a(0) + \CO
(N^{-\al})\|a\|_\al\Bigr)
\label{B70}
\qqq
 where $ I_{+} ={1\over 12}$ and $I_{-} = 3I_{+}={1\over 4}$.}

\vskip 0.4cm

The constants $I_+$, $I_-$ follow from the fact that (see (\ref{5.0}))
$2\int_{\pm} |d({ q})|^{-2}  d{ q} $ equals, to leading order in $N$, $N$ times the sum over even or odd non-zero integers $n$ (positive and negative) of $\frac{1}{\pi^2n^2}$. The even sum equals a quarter of the sum over all integers, and the latter equals $1\over 3$.
 
 \vskip 0.4cm
 Inserting (\ref{9(21)}) into (\ref{9(24)}), and applying Lemma 11.6 to the result, we obtain 
 \qq
 \phi_{-} = 2\ga NI_{-} \left[\left(\begin{array}{ccc}
1 & \be_1
\\
0 & \be_2
\end{array}\right) + \CO (N^{-\al})\right] \left(\begin{array}{c}
\tau(0)
\\
\zeta (0)
\end{array}\right).
 \label{9(26)}
 \qqq
Recalling that $\xi_{-} = \kappa (0) \Bigl(\tau(0),\zeta(0)\Bigr)^T= 2 \ga T_{-}
  (1, 0)^T - \phi_{-}
$ (see (\ref{Br26}), (\ref{9(18)})), we can write:
\qq
2\ga NI_{-} \left[\left(\begin{array}{ccc}
1 & \be_1
\\
0 & \be_2
\end{array}\right) + \CO (N^{-\al})\right] \left(\begin{array}{c}
\tau(0)
\\
\zeta (0)
\end{array}\right)+ \kappa (0) \left(\begin{array}{c}
\tau(0)
\\
\zeta (0)
\end{array}\right)= 2\ga T_-\left(\begin{array}{c}
1
\\
0
\end{array}\right).
\qqq
Since  $\ga =
N^{-1+\al/4}$, we get:
\qq
\tau(0) &=& (NI_{-})^{-1} (T_{-} + \CO(N^{-\al/4} ))
\non
\\
\zeta(0) &=& \CO (N^{-1-\al/4} ).
\non
\qqq
So, the simplified problem is solved by
\qq
(T^0, S^0) \equiv  (T_{+},0) \de(p) + {T_{-}\over NI_{-}} |d(p)|^{-2}\sigma_-(p)
\left[(1,0) + \CO \Bigl(N^{-\al/4}\Bigr)\right].
\label{9(26a)}
\qqq
In $x$-space, this is a linear profile: use, for $p\neq 0$ the explicit formula (\ref{6(3)}) with $j_0$ replaced by
$ {T_{-}\over NI_{-}}=  {2(T_1-T_2) \over N}  $, and observe that, in (\ref{6(2a)}), $j(x)= {j_0\over 2} $ for $x\in [1,N]$;  remember also that  $t(p)=d(-p)T(p)$, and, in $x$-space, $d(-p)$ is $-\nabla$. So, we get:
\qq
T^{0} (x) = T_{1} + {|x|\over N} (T_{2} - T_{1})
\label{B60}
\qqq
plus a $\CO (N^{-1-\al/4})$ correction to the slope $1/N$. The first term comes from integrating 
(\ref{9(26a)}) over $p$, using (\ref{B70}) for the second term, and $T_1=T_++T_-$.

\vskip 0.2cm

Let us now return to the full eq. (\ref{9(13)}) and incorporate the corrections
$W_{\pm}$ in (\ref{9(13)}) and $\psi_{\pm}$ given by (\ref{9(15)}).
Let
\qq
u_{\pm} (p) = \Bigl(t_{\pm} (p), s_{\pm} (p)\Bigr)^T
\non
\qqq
and 
\qq
U_{\pm} (p) = U_{0} \de (p) + 
d(-p)^{-1} u_{\pm} (p).
\non
\qqq
Write
\qq
u_{\pm} (p) = \kappa(p)^{-1} \Bigl(d(p)^{-1} \xi_{\pm}\Bigr) + v_{\pm}(p).
\label{9(27)}
\qqq
In the absence of $W_{\pm}, \psi_{\pm}$, $v_{\pm}=0$ and $\xi_{\pm}$ are as
above. Consider the equation
\qq
v_{\pm} (p) = - W_{\pm} (p).
\label{9(28)}
\qqq

Since the functions $W_{\pm}$ are Lipschitz in $U$, (\ref{9(28)}) has a unique
solution $v_{\pm} \in S$, with
\qq
\|v_{\pm}\|_{S} \leq C\la^2\tau,
\label{Br15}
\qqq
and $v_{\pm}$ is Lipschitz in $U_{0}, \xi_{\pm}$. With this $v_{\pm}$,
(\ref{9(13)})
becomes, as in (\ref{9(18)}) and (\ref{9(19)}) and (\ref{9(19a)}):
\qq
&&\xi_{+} = 0,\ \  \ \ 2\ga T_{+}(1,0)^T
 = \phi_{+} + \psi_{+},
 \non
 \\
&&  \xi_{-} = 2\ga T_{-} 
(1,0)^T
 -\phi_{-}-\psi_{-} 
\label{9(29)}
\qqq
Proceeding as in the simple case,
\qq
\phi_{+} &=& 
2\ga  \left(
T_{0}, 
 \be_1 S_{0}\right)^T 
+ \CO (\ga \tau\la^2)
\non
\\
\phi_{-} &=& 2\ga NI_{-} \left[\left(\begin{array}{ccc}
1 & \be_1
\\
0 & \be_2
\end{array}\right) + \CO (N^{-\al})\right] \kappa^{-1}(0) 
\xi_{-}
 + \CO (\ga \tau \la^2)
\non
\qqq
where the term $ \CO (\ga \tau \la^2)$ collects the contributions from $v_\pm$ and $\psi_\pm$ that are bounded by 
(\ref{9(16)}), (\ref{Br15}), and 
is Lipschitz in $T_{0},S_{0},\xi_{\pm}$. Thus
$T_{0},S_{0},N\xi_{-}$ have $ \CO (\tau \la^2)$ corrections and
\qq
\|(T,S) - (T^0, S^0) \|_{E} \leq C\tau\la^2
\label{Br5}
\qqq
which, combined with  (\ref{9(26a)}), yields the claim of Proposition 11.1 (b).\hfill$
\makebox[0mm]{\raisebox{0.5mm}[0mm][0mm]{\hspace*{5.6mm}$\sqcap$}}$
$
\sqcup$

\vskip 0.4cm

\appendix{}

\nsection{ H\"older regularity. }
\vskip 0.2cm

In this Appendix, we prove some refinements of  the H\"older 
continuity of the kernels proven  in Section 9. We start with a corollary of Proposition 9.3 that will be needed in
 Appendix B.  For this,
 let $G\in C^0(\mathbb{T}^{3d})$
and consider the function
\qq
g(p,k)=\int G(\underline{k})\mu_{p, k}(d\underline{k})
\label{map}
\qqq
on $\mathbb{T}^{1+d}$, where $\mu_{p, k}(d\underline{k})$ is defined by (\ref{mu}):
\qq
\mu_{p,k} ( d
\underline{k})
=\Bigl(\sum^3_{1} s_{i}\om (k_{i}) + s_{4}\om (p-k)+ i\ep\Bigr)^{-1}\de(\sum^3_{1} k_{i}
-p-k)dk_1dk_2dk_3.
\label{mu2}
\qqq
Denote
\qq
g= I(G)
\label{CI}
\qqq
Then we have

\vskip 4mm

\no{\bf Corollary A.1.}  {\it

\vskip 2mm

\no (a) If $G$ is smooth, $I(G)$ is in $C^\al(\mathbb{T}^{1+d})$.

\vskip 2mm

\no (b)  Let $G(\underline{k})=H(k_1, k_2)h(k_3)$ with
$H$ smooth and $h\in C^0(\mathbb{T}^{d})$. Then $I(G)$ 
is in $C^\al(\mathbb{T}^{1+d})$ and
\qq
\|I(G)\|_\al\leq C(H)\|h\|_\infty.
\label{B1}
\qqq

\vskip 2mm

\no (c)  $I$ is a bounded map from
$C^0(\mathbb{T}^{3d})$
to $C^0(\mathbb{T}^{1+d})$. 

\vskip 2mm

\no Furthermore, if $G$
depends smoothly on some parameters so does $g$.
}

\vspace*{4mm}

\no{\bf Proof}. 
By comparing (\ref{mu2}) and (\ref{1}), we see that, if we replace $k_3$ in (\ref{mu2}) by $k_3+p$, we may identify $k_3$ here and $k'$ in (\ref{1}). But then, the function $G$ in (a) only changes $\rho_{\bf s}$
in (\ref{1}), and the statement (a) can be proven just as the one on $M(p,k)$ in Proposition 9.3. In (b), the function $H$ again affects only $\rho_{\bf s}$, and $h$ is the function on which the operator $K_p$ in Proposition 9.3 acts. Hence, (b) follows from the claims made on $K_p$. Finally (c) follows because we can bound the integral $I(G)$ by the sum norm of $G$ and the resulting integral is continuous on $\mathbb{T}^{1+d}$, for the same reason that $M(p,k)$ is continuous. \hfill$
\makebox[0mm]{\raisebox{0.5mm}[0mm][0mm]{\hspace*{5.6mm}$\sqcap$}}$
$
\sqcup$ 

\vspace*{4mm}

 We need one more regularity result for the analysis of
the $\theta$ in (\ref{thetadef}) that will be made in the Appendix B. For this, define the function
\qq
I(p)  = \int \de \bigl(\om (k_{1}) + \om (k_{2}) - \om (k_{3}) - \om (k_{4}) 
\bigr) \de \left(p - \sum^4_{1} k_{i}  \right)  \chi(\un{k}) d
\un{k}.
\label{(22b)}
\qqq
Then

\vspace*{4mm}

\no {\bf Lemma A.2.}  {\it Let $\chi$ in  (\ref{(22b)}) be smooth and
let
  $I'(p)$ be the discrete derivative on ${\pi  \over
2N}$ $\mathbb{Z}_{2N}$. Then, for some $\al>0$,
\qq
\|I'\|_{\al}\leq C
\label{(22a)}
\qqq
uniformly in $N$.}
 
\vspace*{4mm}


\no For the proof we need a further Lemma:

\vskip 0.4cm

\no{\bf Lemma A.3. } \ Let $\chi (\un{x})$ be smooth on $\mathbb{T}^3$ and
$F(\un{x}) = \sin (x_{1} + x_{2} + x_{3}) - \sin x_{1} - \sin x_{2} -
\sin x_{3}$. Then $g(p) \equiv \int \de (F(\un{x}) -p) \chi
(\un{x}) d\un{x}$ is smooth if $p\neq 0$ and 
$$
|g(p)| \leq C (\log p)^2,
$$  
$$
|g'(p)| \leq C \mid{\log p \over p}\mid.
$$

\vspace*{4mm}

\no {\bf Proof of Lemma A.2.}
In (\ref{(22b)}) write $k_{i} = {\pi\over 2} + k'_{i},
i=1,2$ and $k_{3} = -{\pi \over 2} + k'_{3}$ which
imply  $k_{4}= p-k'_{1}-k'_{2}-k'_{3}
- {\pi \over 2}$ and the argument of the $\de$-function equals
$$
\sum^3_{\al=2} \sin (k^\al_{1} + k^\al_{2} + k^\al_{3}) - \sin k^\al_{1} -
\sin k^\al_{2} - \sin k^\al_{3} + P,
$$
where
\qq
P = \sin (\textrm{k}_{1} + \textrm{k}_{2} + \textrm{k}_{3} - \textrm{p}) -
\sin \textrm{k}_{1} - \sin \textrm{k}_{2} - \sin \textrm{k}_{3}.
\label{(27)}
\qqq
Thus,
\qq
I (\textrm{p}) = \int d \un{\textrm{k}} \  J \Bigl(P
(\un{\textrm{k}}, \textrm{p}), \textrm{k}\Bigr),
\label{(28)}
\qqq
with
\qq
J (\la, \textrm{k}) &=& \int d \un{x} d \un{y} \  \de
\Bigl(F(\un{x}) + F (\un{y}) + \la\Bigr) \chi (\un{\textrm{k}},
\un{x}, \un{y})
\non
\\
&=& \int dt \int d \un{x} \ g(t, \un{\textrm{k}}, \un{x}) \de \Bigl(F(\un{x})
+ t + \la \Bigr),
\non
\qqq
where
$$
g  (t,  \un{\textrm{k}}, \un{x}) = \int \de \Bigl(F(\un{y})-t\Bigr) \chi  
 (\un{\textrm{k}}, \un{x}, \un{y}) d \un{y},
$$
 which by Lemma A.3. is smooth in all variables, for $t\neq 0$, and is bounded by 
  $|g| \leq C (\log t)^2$.

Similarily the $\un{x}$-integral yields
$$
J (\la,  \textrm{k}) = \int dtds \ h(t,s, \un{\textrm{k}}) \de (t+s+\la),
$$
with $h$ smooth if $t,s \neq 0$ and 
$$
|\p_{t}h| \leq C \left|{\log |t| \over t} (\log |s|)^2\right|,
$$
and similarily for $\p_{s}h$. Then one gets
\qq
|\p_{\la} J| \leq C(\log |\la|)^4.
\label{(29)}
\qqq
Consider first the $N\to\infty$ limit of (\ref{(28)}). Then by (\ref{(29)})
$|I'(p)| \leq C \int d \textrm{k} |\log P|^4$.

Since $P$ is an analytic function, $|\log P|^4$ is integrable. It is easy to
do the argument for the Riemann sum. In the same vein, one can extract a
little H\"older continuity for $I'(p)$. We leave the details for the reader. 
\hfill$
\makebox[0mm]{\raisebox{0.5mm}[0mm][0mm]{\hspace*{5.6mm}$\sqcap$}}$
$
\sqcup$ 

\vskip 2mm

\no {\bf Proof of Lemma A.3.} a) We have
$$
\p_{\al} F =  \cos (x_{1}+x_{2}+x_{3}) - \cos x_{\al},
$$
so $\nabla F = 0  \Longleftrightarrow \cos x_{1} = \cos x_{2} = \cos x_{3} =
\cos (x_{1} + x_{2} + x_{3})$ and thus $x_{1} = x_{2} = -x_{3}$ and
permutations thereof. At these points $F=0$. Hence $F(\un{x}) \neq 0 \Rightarrow
\nabla F (\un{x}) \neq 0$. Thus, given $\un{x} \in F^{-1} (p)$
there is a neighbourhood $U$ of $x$ and a smooth diffeomorphism: $\phi_{p}:
B_{\ep} (0) \to U$ such that $(F \circ \phi_{p})  (\un{y}) = p
+ y_{1}$ and $\phi_{p}$ is smooth in $p$. Given $\psi \in C^{\infty}_{0} (U)$,
$$
g_{\psi} (p) \equiv \int \de (F-p) \psi d\un{x} = \int dy_{2} ... dy_{d}
(\psi \circ \phi^{-1}_{p}) (0, y_{2},...,y_{d}) \det D\phi_{p}
$$
is smooth. By a partition of unity, we conclude that $g$ is smooth for $p \neq 0$.

b) The Hessian of $F$ is
$$
\p_{\al} \p_{\be} F = \de_{\al\be} \sin x_{\al} - \sin (x_{1}+x_{2}+x_{3}).
$$
At the critical points $x_{1} = x_{2} = -x_{3}$ it equals
$$
H =  - \sin x_{1} \left(\begin{array}{ccccccc}
0 & 1 & 1
\\
1 & 0 & 1
\\
1 & 1 & 2
\end{array}\right).
$$
H has eigenvalues 0, $\sin x_{1}$, $-3 \sin x_{1}$. By a partition of unity
argument it suffices to study two cases: 

1$^\circ$ $\chi$ has support in a
small ball around a critical point with $x_{1} \neq 0$, $x_{1} \neq \pi$. 

2$^\circ$ supp $\chi$
is a small ball around the origin, around $(\pi,\pi,-\pi)$ or permutations thereof.

\vskip 0.4cm

\no {Case $1^\circ$} \ By scaling and rotation, there exists a local coordinate $z
= (u,v,w)$ such that
\qq
F = uv + f(u,v,w)
\label{(24)}
\qqq
with $f$ analytic, $\CO (z^3)$ and $f(0,0,w) \equiv 0 \equiv D_{u,v}
f(0,0,w)$. Indeed, set $x_{1} = x + u + w, x_{2} = x+v+w$ and $x_{3} =x -w$,
then
\qq
F &=& \sin x (\cos (u+v+w) - \cos (u+w) - \cos (v+w) + \cos w) + \non\\
&& \cos x (\sin (u+v+w) - \sin (u+v) - \sin (v+w) + \sin w),  \non
\qqq
which, upon scaling, is of the form (\ref{(24)}). Writing $(u,v) \equiv y$, $f (u,v,w)
= (y,A(z)y)$ with $A=\CO(z)$. Since $uv$ is nondegenerate, there exist an
analytic diffeomorphism $g$ close to identity such that $F \circ g = uv$.

Hence
$$
g(p) = \int \de (uv-p) \tilde \chi (u,v,w) du dv dw,
$$
with $\tilde \chi \in C^\infty_{0} (B_{\ep} (0))$. Let $\phi = \int \tilde \chi dw$,
then, 
\qq
g(p) &=& - \int^\infty_{0} du \log u \p_{u} (\phi (u,p/u) + \phi (-u,-p/u))
\non\\
&=& -\log |p| \int^\infty_{0} du \p_{u} (\psi (pu,1/u) - \int^\infty_{0} du
\log u \p_{u} (\psi (pu, 1/u)),
\non
\qqq
with $\psi (u,v) = \phi (u,v) + \phi (-u,-v)$. Thus $g(p) = a(p) \log |p| + b
(p)$ with $a$ and $b$ smooth. The claim follows.

\vskip 0.4cm

\no {Case $2^\circ$} \ We may suppose $\un{x} = 0$, the other points being
similar. We have:
$$
 F = s_{1} c_{2} c_{3} + c_{1} s_{2} c_{3} + c_{1} c_{2} s_{3} - s_{1} s_{2}
s_{3} - s_{1} - s_{2} - s_{3} 
$$
where $s_{i} = \sin x_{i}$, $c_{i} = \cos x_{i}$. The diffeomorphism near
 the origin $y_{i} = \sin x_{i}$ leads to
$$
F = (y_{1} + y_{2}) (y_{2} + y_{3}) (y_{1} + y_{3}) +\CO (\un{y}^5),
$$
and letting $z_{1} = y_{1} + y_{2}$, $z_{2} = y_{2} + y_{3}$, $z_{3} = y_{1} +
y_{3}$, we get:
$$
F = z_{1}z_{2}z_{3} + f(\un{z}),
$$
with $f$ analytic in $B_{\ep}(0)$, $f = \CO (\un{z}^5)$, and symmetric
under permutations of coordinates. We want to bound
\qq
h(p) = \int \de (F (\un{z})-p) \psi (\un{z}) d\un{z},
\label{(24a)}
\qqq
for $\psi \in C^\infty_{0} (B_{\ep} (0))$ and $\ep$ small enough. By symmetry
we may insert into (\ref{(24a)}) 
$$6 \chi(|z_{1}| \leq |z_{2}| \leq
|z_{3}|).$$
 Expand
$$
f(\un{z}) = f_{0} (z_{2},z_{3}) + f_{1} (z_{2},z_{3}) z_{1} + f_{2}
(\un{z}) z^2_{1}
$$
Since $f_{1} = \CO (z_{2},z_{3})^4$, by a diffeomorphism $\phi$ close to
identity we have $(z_{2}z_{3} + f_{1}) \circ \phi = z_{2}z_{3}$ i.e. may
assume $f_{1} = 0$ and bound $\chi$ from above  by $\chi(|z_{1}|
\leq 2 |z_{2}| \leq 3 |z_{3}|)$:
\qq
&& \Bigl|h(p) \Bigr| \leq \int \de \Bigl(z_{1}z_{2}z_{3} + f_{0} (z_{2},z_{3}) +
f_2 (\un{z}) z^2_1 - p \Bigr) \tilde \psi (\un{z})\cdot \chi\Bigl(|z_{1}| \leq 2 |z_{2}| \leq 3 |z_{3}| \Bigr)
d\un{z},
\label{(25)}
\qqq
with $\tilde \psi \in C^\infty_{0} (B_{\ep} (0))$. On the support of
$\chi$ and $\tilde \psi$ $|f_{2}| \leq C |z_{3}|^3 \leq C\ep^2 |z_{3}|$
and thus $|z_{2}z_{3} + f_{2} z_{1}| \geq (1-\CO (\ep^2) | z_{2}z_{3}|$, since
$|z_{1}| \leq 2 |z_{2}|$. So, given $z_{2}, z_{3}$ and $p$, the argument of
the $\de$ function in (\ref{(25)}) either is nonzero for all $z_{1} \in B_{\ep} (0)$ or
vanishes at $z_{1} (z_{2}, z_{3}, p)$ satisfying
$$
{1\over 2} \left| {p-f_{0}\over z_{2}z_{3}}\right| \leq |z_{1}| \leq 2  \left| 
{p-f_{0} \over z_{2}z_{3}}\right|.
$$
Moreover $\Bigl|z_{2}z_{3} + \p_{1} (f_{2}z^2_{1})\Bigr| \geq \Bigl(1-\CO
(\ep^2)\Bigr) |z_{2}z_{3}|$. Thus
\qq
&& \Bigl|h(p) \Bigr| \leq C\int |z_{2}z_{3}|^{-1} \chi\Bigl(|z_{2}| \leq 
 2|z_{3}|\Bigr)  \chi\left(\left|{p-f_{0}\over z_{2}z_{3}}\right| < 
 C \ep\right)
  \tilde \psi (z_{1},z_{2},z_{3}) dz_{2}dz_{3}.
\label{(26)}
\qqq
Since $f_{0} (z_{2},z_{3}) = f_{0} (0,z_{3}) + z_{2} f_{3} (z_{2},z_{3})$
where $|f_{3}| \leq C z^4_{3} \leq C \ep^3 |z_{3}|$, the second factor $\chi$ in
(\ref{(26)}) is bounded by
$$
\chi\left(\left| {p-f_{0} (0, z_{3}) \over z_{2}z_{3}}  \right| < C 
 \ep\right). 
 $$
Now, $f_{0} (0,z_{3})$ is analytic, so 
$$
f_{0} (0,z_{3}) = a z^n_{3} \Bigl(1+\CO(z_{3})\Bigr),
$$
for some $a\neq 0$, $n<\infty$ (actually $n=5$). 

Insert this into (\ref{(26)}), write $\chi= \chi_{p} (z_{2}) + \Bigl(\chi-
\chi_{p} (z_{2})\Bigr)$, with
$$
\chi_{p} (z_{2}) = \chi\Bigl(|z_{2}| > C |p|^{{n-1\over n}}\Bigr).
$$
Then $h = h_{1} + h_{2}$ with
\qq
\Bigl|h_{1}(p)\Bigr| & \leq& \int {dz_{2}\over |z_{2}|}  \chi (z_{2})
\int {dz_{3}\over |z_{3}|} \chi\Bigl(|z_{3}| > \ha |z_{2}|\Bigr) \chi_p
\Bigl(\un{z}\in B_{\ep} (0) \Bigr)\non
\\
&\leq& C (\log p)^2. \non
\qqq
For $|z_{2}| \leq C |p|^{{n-1\over n}}$, we note that, since, in the support of $ \chi\left(\left| {p-f_{0} (0, z_{3}) \over z_{2}z_{3}}  \right| < C 
 \ep\right)$
$$
|p-az^n_{3}| \leq C \ep |z_{2}z_{3}|,
$$
we may write $z_{3} = (p/a)^{1/n} + x$, with
$$
\Bigl|{p\over a}\Bigr|^{{n-1\over n}} |x| \leq C \ep |z_{2}| \  |p|^{1/n},
$$
where $a$ is absorbed into $C$ in the RHS,
or
$$
|x| \leq C\ep |p|^{-{n-2\over n}} |z_{2}|.
$$
So, doing the $z_{3}$ integral, we get:
\qq
\Bigl|h_{2}(p)\Bigr| &\leq& C\ep \int {dz_{2}\over |z_{2}|} \chi
\Bigl(|z_{2}| < C \left|p\right|^{{n-1\over n}}\Bigr) |p|^{-1/n} |z_{2}| \
|p|^{-{n-2\over n}}
\non
\\
&\leq & C \ep.\non
\qqq
We conclude that $|g(p)| \leq C (\log p)^2$ i.e. the claim holds for $g$. The claim for $g'$ is
similar: $\p_{p}$ brings an extra $|z_{2}z_{3}|^{-1}$ and 
$$
\int {dz_{2}dz_{3}\over (z_{2}z_{3})^2} \chi\Bigl(|z_{2}z_{3}| > \ep
(p)\Bigr) \leq {C \over \ep} \left|{\log p \over p}\right|. 
$$
\hfill$
\makebox[0mm]{\raisebox{0.5mm}[0mm][0mm]{\hspace*{5.6mm}$\sqcap$}}$
$
\sqcup$ 

\vskip 0.4cm

\no {\bf Remark} \ $I(p)$ is \un{not} smooth for $p\neq 0$. By some algebra one
can show
$$
\nabla_{\textrm{k}} P=0, \ \ 
 P=0 \Longleftrightarrow \textrm{k}_{1} = \textrm{k}_{2} = \textrm{k}_{3}, \ \ 
\textrm{p} = 2 \textrm{k}_{1},
$$
and permutations. Thus zeros of $P$ and $\nabla_{\textrm{k}} P$ occur for
nonzero $\textrm{p}$ too.

\nsection{ Nonlinear  estimates}

\vskip 0.2cm

Here we prove the estimates on the nonlinear terms that were made in Section 10.
They are based mostly on an analysis of convolutions of functions in $E$ or $S$,
whose results were stated in Lemma 9.1. We start with the proof of this Lemma.

\vskip 0.2cm

\no {\bf Proof of Lemma 9.1}
Recall first that $j=j_+\sigma_++j_+\sigma_+$
and $T$ similarily. Let $T\ast j=j'$. Then
\qq
j'_+=T_+\ast j_++T_-\ast j_-\non\\
j'_-=T_+\ast j_-+T_-\ast j_+.\non
\qqq
Consider eg. $T_+\ast j_+$ and drop the $+$.

Recall that 
$j$  is defined through the 4-tuple $ {\bf j}$ in (\ref{jdef}). We need to define the 4-tuple $ {\bf j'}$.
We set
\qq
{j}'_0&=&\hf(T\ast j)(0),
\label{mult0}
\qqq
where the factor $\ha$ follows from our conventions  (\ref{jdef})  and 
(\ref{deltafunction}). Let, for $p\neq 0$,
\qq
{j}'_1&=&T\ast j_1
\label{mult1}\\
{ j}'_2&=&T\ast j_2
\label{mult2}\\
{_{j'_3}\over^{(Nd)^{3/2}}}&=&T\ast {_{j_3}\over^{(Nd)^{3/2}}}+T\ast  
{_{j_1}\over^{Nd}}-{_{1}\over^{Nd}}
(T\ast j_1).
\label{mult3}
\qqq

Let us estimate these in turn, using two simple observations:
\qq
\| \int f (p-p') g (p') dp' \|_{\alpha} \leq \|f\|_{\alpha} \|g\|_{L^1}
\label{c1}
\qqq
where $\|-\|_{L^1}$ is the $L^1$ norm, and
\qq
\| \int f (p-p') d (p')^{-1} d p' \|_{\alpha} \leq \|f\|_{\alpha}  
\label{c2}
\qqq
which holds because the Hilbert transform of a H\"older continuous function is H\"older
continuous (for $\alpha < 1$, see \cite{Ti}, Theorem 106), and  this is true  also when $p^{-1}$ is replaced
by $d(p)^{-1}$, and when we have  discrete sums as here.

Then, using
 $
\|T\|_{L^1}\leq C \|T\|_E,
$ (see (\ref{Br6})),
 and (\ref{c1}),
we get:
\qq
\|j'_1\|_\al\leq C\|T\|_E\| j_1\|_\al
\label{j'1}\\
\|j'_2\|_\al\leq C\|T\|_E\| j_2\|_\al.
\label{j'2}
\qqq
 For $j'_3$, we use  the identity
\qq
d(p) = d(p') + d(p-p') + d(p') d (p-p'),
\label{c5}
\qqq
to write:
 \qq
&&\Bigl( T\ast  
{_{j_1}\over^{Nd}}-{_{1}\over^{Nd}}
(T\ast j_1)\Bigr)(p)
= {_{1}\over^{Nd}} \Bigl( Nd (T\ast  
{_{j_1}\over^{Nd}})-
(T\ast j_1)\Bigr)(p)
\non\\
 &=&{_1\over^{Nd}}\Bigl(\int j_1(p-p')
 (d(p-p')^{-1}d(p')T(p')dp' + \int j_1(p-p')
d(p')T(p')dp'\Bigr)
\label{j3}
\qqq
Hence, using (\ref{tbound}), (\ref{j3}) is bounded by
\qq
C\|T\|_E\| j\|_S(N|p|)^{-1}\int({_1\over^{N|p'|}}+
N^{-1+\al/2}+
{_{1}\over^{(N|p'|)^{3/2}}}
)(1+
|p-p'|)^{-1})dp'
\label{j31a}
\qqq
The integral is bounded by
$$
C\log| Np|({_1\over^{N|p|}}+N^{-1+\al/2}).
$$
Combining with (\ref{j31a})
we arrive at the bound
\qq
|(\ref{j3})|\leq C\|T\|_E\| j\|_S{_{1}\over^{|Nd(p)|^{3/2}}}.
\label{j31b}
\qqq
Similar calculations give the bound
\qq
|T\ast {_{1}\over^{|Nd|^{3/2}}}|\leq \|T\|_E{_{1}\over^{|Nd|^{3/2}}}.
\non
\qqq
Combining with (\ref{j31b}) we get
\qq
\|j'_3(p)\|_\infty \leq C\|T\|_E\| j\|_S.
\label{j'3}
\qqq
Finally (\ref{mult0}) is bounded by
\qq
\hf\int|T(p)j (p)|dp\leq C\|T\|_E\| j\|_S
\label{j'4}
\qqq
using (\ref{Br0}), e.g.
$$
\int(N^2|p|^3)^{-1}dp\leq C,
$$
Estimates (\ref{j'1}) , (\ref{j'2}),  (\ref{j'3}) and  (\ref{j'4}) give  the claim
 (\ref{l1}).
 
 \vskip 2mm
 
 Let us next consider  part (c). We set
\qq
{j}'_0&=&\hf(j\ast k)(0)
\label{mult0a}
\qqq
and, for $p\neq 0$,
\qq
{j}'_1&=&Nd\Bigl(
{_{j_1}\over^{Nd}}\ast {_{k_1}\over^{ Nd}}+H_1\Bigr)+ {_{1}\over^{N}} (j_0k_1+k_0j_1)
\label{mult1a}\\
{ j}'_2&=&j\ast k_2+k\ast j_2-N^{-1+\al/2}j_2\ast k_2
\label{mult2a}\\
 {_{j'_3}\over^{(Nd)^{3/2}}}
&=&
 {_{j_3}\over^{(Nd)^{3/2}}}\ast  {_{k_3}\over^{(Nd)^{3/2}}} +H_2+{_{1}\over^{N}} (j_0k_3+k_0j_3),
\label{mult3a}
\qqq
where we write ${_{j_1}\over^{Nd}}\ast {_{k_3}\over^{(Nd)^{3/2}}}+
{_{k_1}\over^{Nd}}\ast {_{j_3}\over^{(Nd)^{3/2}}}=H_1+H_2$, and the $H_i$'s are defined 
after (\ref{Br410})  below.

To proceed, we need the following bounds: using  (\ref{c1}), (\ref{c2}), we get that,
\qq
h_{1} (p) \equiv  \int f (p-p') g (p') d(p')^{-1} dp',
\label{c3}
\qqq
satisfies
\qq
\|h_{1}\|_{\alpha} \leq C \|f\|_{\alpha} \|g\|_{\alpha}.
\label{c4}
\qqq
This holds because we can write in (\ref{c3}) $g(p') = g(p') -g(0) + g(0)$;
since $g$ is in $C^\alpha$,
$|g(p') - g(0)| |d(p')^{-1}| \leq \|g\|_{\alpha} |p'|^{-1+\alpha}$, i.e.
$(g(p') - g(0)) d(p')^{-1}$ is integrable; so (\ref{c4}) follows from
 (\ref{c2}) applied to the $g(0)$ term above and (\ref{c1}) to the $g(p') -g(0)$ term.

Using (\ref{c5}), we then get that
\qq
h_{2} (p) \equiv d(p) \int f(p-p') d (p-p')^{-1} g (p') d (p')^{-1} dp'
\label{c6}
\qqq
also satisfies
\qq
\|h_{2}\|_{\alpha} \leq C \|f\|_{\alpha} \|g\|_{\alpha}.
\label{c7}
\qqq
Now, consider (\ref{mult1a}). The bound (\ref{c7}) applied to the first term implies that its $C^\al$
norm is $\CO(N^{-1}) \|j_1\|_{\alpha} \|k_1\|_{\alpha}$.

Now, use  (\ref{c5}) to  write 
\qq
Nd({_{j_1}\over^{Nd}}\ast {_{k_3}\over^{(Nd)^{3/2}}})=
j_1\ast {_{k_3}\over^{(Nd)^{3/2}}}+
{_{j_1}\over^{Nd}}\ast {_{k_3}\over^{(Nd)^{1/2}}}
+{_{j_1}\over^{N}}\ast {_{k_3}\over^{(Nd)^{1/2}}}.
\label{Br410}
\qqq
Let  the sum of the first and third term in (\ref{Br410}),  plus the corresponding terms in 
$Nd({_{k_1}\over^{Nd}}\ast {_{j_3}\over^{(Nd)^{3/2}}})$ define  $Nd H_1$
and let $H_2= (Nd)^{-1}({_{j_1}\over^{Nd}}\ast {_{k_3}\over^{(Nd)^{1/2}}})$ plus the corresponding terms in 
${_{k_1}\over^{Nd}}\ast {_{j_3}\over^{(Nd)^{3/2}}}$.
Since $\| {_{k_3}\over^{(Nd)^{3/2}}}  \|_{L^1}\leq C N^{-1} \|k_3\|_\infty$, we get
from (\ref{c1}) that  
$$
\|j_1\ast {_{k_3}\over^{(Nd)^{3/2}}}\|_\al \leq 
\CO(N^{-1})\|j_1\|_\al \|k_3\|_\infty,
$$
which controls the contribution of the first term in (\ref{Br410}) to $Nd H_1$.
 This bound holds also  for the last term in (\ref{Br410}), of course (which is
even $\CO(N^{-3/2})\|j_1\|_\al \|k_3\|_\infty$). The terms ${_{1}\over^{N}} (j_0k_1+k_0j_1)
$ are trivially bounded, so we get:
\qq
\|Êj'_1Ê\|_\al \leq CN^{-1} \|Êj\|_S \|ÊkÊ\|_S.
\label{Br42}
\qqq

Using (\ref{c1}) and the fact that  $\| {_{j_1}\over^{(Nd)}}  \|_{L^1}\leq N^{-1} \log N \|Êj_1\|_\infty$
for all terms in   (\ref{mult2a}), we get
\qq  
\|Êj'_2\|_\al\leq CN^{-1+\al/2}.
 \label{Br43}
\qqq
  
 Consider now (\ref{mult3a}). Going back to the discrete sum (see (\ref{5.0})), it is easy to see that the first term is bounded  by $CN^{-1} |Np|^{-3/2} \|Êj_3\|_\infty \|Êk_3\|_\infty $, i.e. that the contribution of this term to  $|Êj'_3Ê(p)|$  is less than
$CN^{-1} \|Êj\|_S \|ÊkÊ\|_S$. Consider now $(Nd)^{-1}({_{j_1}\over^{Nd}}\ast {_{k_3}\over^{(Nd)^{1/2}}})$, contributing to $H_2$.
Going back to the discrete sum, it easy to bound it by $CN^{-1} |Np|^{-3/2}\log |Np|$.
Finally, the bound on ${_{1}\over^{N}} (j_0k_3+k_0j_3)$ is trivial and we get:
\qq
|Êj'_3Ê(p)|  \leq CN^{-1} \log (N|p|) \|Êj\|_S \|ÊkÊ\|_S.
\label{Br44}
\qqq
Finally, using repeatedly (\ref{Br0}), we get
\qq
|Êj'_0|  \leq CN^{-1}  \|Êj\|_S \|ÊkÊ\|_S.
\label{Br45}
\qqq   
 Combining (\ref{Br42}), (\ref{Br43}), (\ref{Br44}), (\ref{Br45}), proves part c) of the Lemma, since the 
 bound on $\|j'\|_S$ follows from the previous ones.
   
 Let us finally turn to (\ref{l2}).   Note that, using  (\ref{c5}),
 $$
 d(T\ast A)=dT\ast A+T\ast dA+dT\ast dA
 $$
 Since $dT\in S$ if $T\in E$ the claim, for $p\neq 0$, follows 
 from (a) and (c). For $p=0$, we simply apply (\ref{Br0}) to all the terms in $(T\ast A)(0)$. \hfill$
\makebox[0mm]{\raisebox{0.5mm}[0mm][0mm]{\hspace*{5.6mm}$\sqcap$}}$
$
\sqcup$ 
 
\vspace*{4mm}
Using this Lemma, it is rather easy to give the proof of the main estimate of section 9.

\vspace*{4mm}
\no {\bf Proof of Proposition 9.4.} (a) Apply (a) of Corollary A.1 to (\ref{udef})
and (\ref{vdef}) to get
\qq
m(p, k)=\int g(p, k, \underline{p})\prod^3_{i=1} F_{{i}} (2p_{i})
\de(2p-\sum_1^3 2p_i)d\underline{p}
\label{m}
\qqq
where $g$ is smooth in $\underline p$ and $C^\al$ in $p , k$. $m$ may now be defined
as an element of $E$ as in Lemma 9.1.b, the smooth function not affecting the bounds.

\no (b) If $m=u$ or if $m=v$ and $i\neq 3$ (in which case $i=1,2$, and, by symmetry, we can choose $i=1$) we have
\qq
m(p, k)=\int  \prod^3_{i=2} F_{{i}} (2p_{i})
(\int G\ f_1(p_1, k_1-p_1)\mu_{p, k}(d\underline{k}))
\de(2p-\sum_1^3 2p_i)d\underline{p}
\non
\qqq
with $G$ smooth, and $f_1$ in $\CS$. Using the representation 
(\ref{6(4b)}) for $f_1$ we write $G\ f_1(p_1, k_1-p_1)$ as a sum of terms, with singularities in $p_1$;
by Corollary A.1.b, we get that integrating each of those terms with $\mu_{p, k}(d\underline{k})$ gives rise to a function that is $C^\al$ in $p, k$. Thus,
\qq
\int G\ f_1(p_1, k_1-p_1)\mu_{p, k}(d\underline{k})=f(p_1,k;p,p_2,p_3)
\label{B2}
\qqq
where $f$ is in $\CS$ in the variables $p_1,k$ depending smootly on
$p_2,p_3$ and $C^\al$ in $p$.
The convolutions with the
$F_j$'s can then be estimated as in  Lemma 9.1.

If $m=v$ and $i= 3$ we need to study  
\qq
\int  \prod^2_{i=1} F_{{i}} (2p_{i})
 f(p_3, p-k-p_3)(\int 
G\mu_{p, k}(d\underline{k}))\de(2p-\sum_1^3 2p_i)d\underline{p},
\non
\qqq
where $f \in \CS$.
 Shifting $p_3$ by $p$, this becomes
\qq
\int  g(p, k,\underline{p})\prod^2_{i=1} F_{{i}} (2p_{i})
\de(\sum_1^3 2p_i) f_3(p+p_3, -k-p_3)
d\underline{p}
\non
\qqq
here $g$, defined in (\ref{map}) is, by Corollary A.1.a, $C^\al$ in $p, k$ and smooth in $\underline{p}$.
Performing the $p_i$ integrals for $i=1,2$, we get
\qq
\int  F(p, k, p_3) f(p+\hf p_3, -k-\hf p_3)
d{p_3}
\label{aaa}
\qqq
where $F$ is  $C^\al$ in the first two arguments 
and belongs to $E$ as a function of the third. We may proceed now as in
Lemma 9.1.(a) to define and estimate the quadruple $\bf g$ in $\CS$ 
corresponding to (\ref{aaa}). E.g. the  component of index $1$, see (\ref{mult1}),
is given by
\qq
g_1(p, k)=
\int  F(p, k, p_3) f_1(p+\hf p_3, -k-\hf p_3) dp_3
\label{B3}
\qqq
where $f_1$ is $C^{\al}$ in both arguments. Again, since $F$
is integrable in the third argument and  $C^{\al}$ in the others
the integral is  $C^{\al}$ in $p, k$. The other components
of $\bf g$ can be bounded as in Lemma 9.1.a.

(c) We have to specify again the quadruple ${\bf m}\in\CS$ 
corresponding to $m$. We define $m_i=0$ and
$m_3(p,k)$ to be the integral  (\ref{udef}) or   (\ref{vdef})
corresponding to $m$.  We use  Corollary A.1.c
to do the $k_i$ integrals and estimate the $p_i$
integrals by brute force. Since only the singularity
at $p_i=0$ (or, by periodicity at  $\pi$)
matters, we need the easy bounds (remember that the variable $p$ is
discrete!)
\qq
&&|Nd(p)|^{-1}\ast |Nd(p)|^{-1}\leq  CN^{-\hf}\log N|Nd(p)|^{-3/2}
\non\\
&&N^{-1+\al/2}\ast |Nd(p)|^{-1}\leq C\log NN^{-2+\al/2}\leq CN^{-\hf+\al/2}\log N|Nd(p)|^{-3/2}
\non\\
&&N^{-1+\al/2}\ast N^{-1+\al/2}\leq CN^{-\hf+\al}|Nd(p)|^{-3/2}
\non
\qqq
together with $|Nd(p)|^{-3/2}\leq|Nd(p)|^{-1}$
(since $N|p|\geq\pi$), which allows to use the first inequality here in order to bound 
 the convolutions with the last term in (\ref{6(4b)}),  to conclude
\qq
\|m(p,k)\|_\infty \leq C\ N^{-\hf+\al}\|f_k\|_E\prod_{l\neq k}\|f_l\|_\CS
\non
\qqq
which is the claim (\ref{e3}) since we defined ${\bf m}=(0, 0,0,m)$.
\hfill$
\makebox[0mm]{\raisebox{0.5mm}[0mm][0mm]{\hspace*{5.6mm}$\sqcap$}}$
$
\sqcup$ 
 
 \vspace*{4mm}
 
 Finally, we prove the estimates on the function
 $\th $ defined in (\ref{thetadef}).

\vspace*{4mm}
\no {\bf Proof of Proposition 9.7}

By the $(3\leftrightarrow 4)$ symmetry in
(\ref{5(21)}), we get (leaving out the factor $ {9\over 4}(2\pi)^{3d}  \la^2$),
\qq
\th (p)&=& i \sum_{{\bf s}} \int \prod^3_{1} W_{s_{i}} (p_{i}, k_{i}-p_{i})
s_{3}s_{4} \om (k_{3}) \om (k_{4})^{-1}
\non\\
&&\left[\rho \Bigl({_p\over^2}, {_p\over^2} - k_{3}\Bigr) - \rho
\Bigl({_p\over^2}, {_p\over^2} - k_{4}\Bigr) \right] \nu'_{{\bf s}pk}
(d\underline{p} \ d\underline{k})
\label{5(22)}
\qqq
and
\qq
\nu'_{{\bf s}pk} = \Bigl(\sum_{i} s_{i} \om (k_{i}) + i\ep\Bigr)^{-1} \de
(2p_{4})
\de (p-2\sum_{i} p_{i}) \de (p-\sum_{i} k_{i}) d{\underline p} d{\underline k}
\non
\qqq
Then the [ ] in (\ref{5(22)}) equals 
\qq
2 \left({1\over \om (k_{3})} - {1\over \om (k_{4})} \right) + (e^{ip}-1) \ r (p,k)
\label{5(23)}
\qqq
with $r$ smooth. The integral in (\ref{5(22)}) has singularities when
\qq
\sum s_{i} \om (k_{i}) = 0.
\label{5(241)}
\qqq
Recall that (\ref{5(241)}) forces 
\qq
\sum s_{i} = 0
\label{5(251)}
\qqq
Consider the {\bf s} such that (\ref{5(251)}) holds in (\ref{5(22)}), and, 
replace [ ] in (\ref{5(22)}) by the first term of (\ref{5(23)}). We define 
\qq
\th_{1} (p) &\equiv& 2i\sum_{{\sum s_i=0}} \int \prod_1^3W_{s_{i}}
 \left({s_{3}s_{4}\over \om (k_{4})} - 
  {s_{3} \om (k_{3}) s_{4} \over \om (k_{4})^2 }\right)
\cdot \nu'_{{\bf s} pk} (d \underline{p} \ d \underline{k}). 
\label{5(26)}
\qqq
By symmetry, we may replace $s_{3}$ by ${1\over 3} \sum^3_{i=1} s_{i}$ and, by
 (\ref{5(251)}) also by $-{1\over 3} s_{4}$. Again, by symmetry, $s_{3} \om
(k_{3})$ may be replaced by  ${1\over 3} \sum^3_{i=1} s_{i} \om (k_{i})$.
So, the parenthesis equals  ${-s_{4}\over3 \om (k_{4})^2} \sum^4_{i=1} s_{i} \om (k_{i})$, and the sum cancels the factor $\Bigl(\sum_{i} s_{i} \om (k_{i}) + i\ep\Bigr)^{-1} $ in $\nu'$.
Hence, (\ref{5(26)}) equals
\qq
\th_{1} (p) &=& -{2i \over 3} \sum_{{\sum s_i=0}}  \int \prod_1^3 
W_{s_{i}} {s_{4}\over \om (k_{4})^2} \de (2p_{4}) \de (p-2 \sum_{i} p_{i}) \de
(p-\sum_{i}  k_{i})  d \underline{p} \ d \underline{k}
\label{5(27)}
\qqq
We decompose $\theta$ as
\qq
\th (p) = \th_{1} (p) + \th_{2} (p) + p \th_{3} (p)
\label{5(28)}
\qqq
where $\th_{2}$ has the terms of (\ref{5(22)}) with $\sum s_{i} \neq 0$ and the first
 term of  (\ref{5(23)}) inserted, while
$\theta_3$ corresponds to the insertion  of the second term of  (\ref{5(23)}).

\vspace*{4mm}

Consider first $\th_{1}$ given by (\ref{5(27)}). Remember that 
$$
W_{s} (p,k-p) = Q (p,k-p) + i s \om (k)^{-1}  J (p,k-p).$$
The terms with an odd number of $ Q$ factors vanish by the ${\bf s} \to
- {\bf s}$
symmetry. Consider then the term linear in $J$. We insert $Q=Q_{0}+r$ and start with 
the term with no $r$. After shifting the $k_i$ variables by $2p_i$ (and using $2p_4=0$),
we obtain  a sum of terms of the form
\qq
&&
\int T\ast A^{\ast n_1}(2p_{1})  T\ast A^{\ast n_2}(2p_{2}) J (p_{3}, k_{3}+p_{3}) \prod^2_{i=1}
\Bigl(\om (k_{i} + 2p_{i}) + \om (k_{i})\Bigr)^{-2-n_{i}}\cdot
\non
\\
&&
\om (k_{4})^{-2} \om (k_{3}+2p_3)^{-1} \de (2p_{4}) \de \left(p-2 \sum p_{i}\right) \de \left(\sum k_{i}\right) d
\underline{k} \ d \underline{p}.
\label{7(15)}
\qqq
Now, use the fact that
$\om (k_{i} + 2p_{i}) - \om (k_{i} ) = (e^{2ip}-1) \CO (1)$, which implies
\qq
\Bigl(\om (k_{i} + 2p_{i}) + \om (k_{i})\Bigr)^{-2-n_{i}}=\Bigl(2\om (k_{i} )\Bigr)^{-2-n_{i}}
+ n_i (e^{2ip_i}-1) \CO (1).
\label{Br16}
\qqq
For the second term on the RHS of (\ref{Br16}), 
let us choose $i=1$, which we can do by symmetry, and insert it in
 (\ref{7(15)}), to obtain, after integrating over $k_1$, $k_2$, $k_4$:
\qq
&&
I_{1}(p) =n_1 \int (e^{2ip_{1}}-1) T\ast A^{\ast n_1} (2p_{1}) T\ast A^{\ast n_2}(2p_{2}) J
(p_{3},k_{3}+p_{3}) 
\cdot
\non
\\
&&
f (\underline{p}, k_{3}) \de (2p_{4}) \de \left(p-2\sum p_{i}\right) d
\underline{p} dk_{3},
\label{7(16)}
\qqq
with $f$ smooth. For the first  term on the RHS of (\ref{Br16}), we obtain:
\qq
&&
\tilde I_{1}(p) =\int T\ast A^{\ast n_1} (2p_{1}) T\ast A^{\ast n_2}(2p_{2}) J
(p_{3},k_{3}+p_{3})\cdot
\non
\\
&&
 \tilde f (\underline{p}, k_{3}) \de (2p_{4})\de \left(p-2\sum p_{i}\right) d
\underline{p} dk_{3},
\label{Br17}
\qqq
with $\tilde f$ smooth and even in $k_3$.
Doing the $k_3$-integral, we get, for  (\ref{7(16)}),
\qq
\int J
(p_{3},k_{3}+p_{3}) f (\underline{p}, k_{3})dk_3= g(2p_3, 2\underline{p})
\label{716a}
\qqq
where we used the $\pi$-periodicity of the result. $g$ is in $S$ as a function of $p_3$,
depending smoothly on $\underline{p}$, with $\|gÊ\|_S\leq C \|ÊJ\|_S$. For  (\ref{Br17}), we get, since $J$ is odd in $k_3$ and
$\tilde f$ even, that the integral vanishes if $ J
(p_{3},k_{3}+p_{3})$ is replaced by  $J
(p_{3},k_{3}) $, and thus, since $ \tilde f $ is smooth, the integral can be written as:
\qq
\int J
(p_{3},k_{3}+p_{3}) \tilde f (\underline{p}, k_{3})dk_3=\int J
(p_{3},k_{3}) \tilde f (\underline{p}, k_{3}-p_{3})dk_3= (e^{2ip_{3}}-1)
\tilde g(2p_3, 2\underline{p}),
\label{Br18}
\qqq
using again the $\pi$-periodicity of the result. $\tilde g$ is in $S$ as a function of $p_3$,
depending smoothly on $\underline{p}$, with $\| \tilde gÊ\|_S\leq C \|ÊJ\|_S$.
We write, using the constraints $\de(2p_4)$, $\de \left(p-2\sum p_{i}\right)$, 
\qq
e^{2ip_{3}}-1= (e^{ip}-1+1) (e^{-2ip_{1}}-1+1) (e^{-2ip_{2}}-1+1) -1.
\label{Br19}
\qqq
Expanding the product we see that the integral (\ref{Br17})
equals the sum of  terms of the form $I_1$ and of the form:
\qq
 I_2=d(p) \int T\ast A^{\ast n_1} (2p_{1}) T\ast A^{\ast n_2} (2p_2) \tilde g(2p_3, 2\underline{p})
\de (2p_{4})  \de \left(p-2\sum p_{i}\right) \  d\underline{p} \
dk_{3}.
\label{7(19)}
\qqq
 The integral in  (\ref{7(19)}) is a convolution of two functions in $E$ with one in $S$, hence, by Lemma 9.1, it is in $S$. The prefactor $d(p)$ cancels the $d^{-1}$, so that this contribution satisfies
 \qq
\| d^{-1}I_2 \|_S &\leq (C\de_{n_{1}+n_{2},0}\|t\|_S+(C\|A\|_E)^{n_{1}+n_{2}})
\|J\|_S, \ \ \ 
\label{Br20}
\qqq
and also $I_2(0) =0$.

Going back to $I_1$, see (\ref{7(16)}), (\ref{716a}), we obtain:
\qq
I_{1} (p)=n_1 \int f(p_{1}) F(p_{2}) g
(p_{3},\underline{p}) \de(2p_4)\de \left(p-\sum p_{i}\right) d
\underline{p} 
\label{B4}
\qqq
with $f\in S$ and $F\in E$.  So, we have a convolution of two elements of $S$ and one of $E$, i.e. the convolution of two elements of $S$.
Going back to the definition (\ref{6(4b)}),  we see that we can write
\qq
I_1=I'_1+I''_1 + I'''_1
\label{B5}
\qqq
corresponding to the $j'_i$, $i=1,2,3$ terms in the convolution of two elements of $S$
that are bounded in part c) of Lemma 9.1. From that Lemma, we get:
\qq
 |I'_{1} (p)| &\leq &{1\over N^2|d(p)| } (C\de_{n_{1}+n_{2},0}\|t\|_S+(C\|A\|_E)^{n_{1}+n_{2}})
\|J\|_S, \ \ \ 
\label{I'a}
\qqq
If we identify $d(p)^{-1}I'_1$ with an element of $S$
of the form $(0,  0, 0, \star)$,   we get from  
(\ref{I'a}) and  $(N|d(p)|)^{-1/2}\leq C$,

\qq
\| d^{-1}I'_1 \|_S &\leq (C\de_{n_{1}+n_{2},0}\|t\|_S+(C\|A\|_E)^{n_{1}+n_{2}})
\|J\|_S. \ \ \ 
\label{Br31}
\qqq
Next, we get, also from Lemma 9.1.c, together with the definition (\ref{6(4b)}):
\qq
\| I''_{1} \|_\al &\leq &N^{-2+\al} 
(C\de_{n_{1}+n_{2},0}\|t\|_S+(C\|A\|_E)^{n_{1}+n_{2}})
\|J\|_S,
\label{I2a}
\qqq
Now,
identify  $d(p)^{-1}I''_1$ with an element of $S$
of the form $(0, \star, 0, 0)$, we get, writing $d(p)^{-1}I''_1= (Nd(p))^{-1}NI''_1$,
that 
\qq
\| d^{-1}I''_1 \|_S &\leq N^{-1+\al} (C\de_{n_{1}+n_{2},0}\|t\|_S+(C\|A\|_E)^{n_{1}+n_{2}})
\|J\|_S. \ \ \ 
\label{I2b}
\qqq
For $I'''_1$, we use (\ref{br52}), and identify $d(p)^{-1}I'''_1$ with an element of $S$
of the form $(0, 0, 0, \star)$. Since $(N|d(p)|)^{-1}\log (N |p|)\leq C$, we get
\qq
\| d^{-1}I'''_1 \|_S &\leq (C\de_{n_{1}+n_{2},0}\|t\|_S+(C\|A\|_E)^{n_{1}+n_{2}})
\|J\|_S. \ \ \ 
\label{I2d}
\qqq
Combining these estimates, we get:
\qq
\| d^{-1}I_1 \|_S &\leq (C\de_{n_{1}+n_{2},0}\|t\|_S+(C\|A\|_E)^{n_{1}+n_{2}})
\|J\|_S. \ \ \ 
\label{I2c}
\qqq
We also get, by (\ref{Br50}):
\qq
|I_1(0)|\leq C N^{-1}.
\label{br40}
\qqq

\vspace*{4mm}

The terms with one $J$ and one $r$ are sums of terms of the form:
\qq
&&
\int T\ast A^{\ast n_1} (2p_{1}) r  (p_{2}, k_{2}-p_{2}) J (p_{3}, k_{3}-p_{3})
\Bigl( \om ( 2p_{1}-k_{1})+\om (k_{1}) \Bigr)^{-2-n_{1}}
\non
\\
&&
\om (k_{3})^{-1} \om (k_{4})^{-2}\de(2p_4) \de \left(p-2 \sum p_{i}\right) \de \left(p- \sum k_{i}\right) d
\underline{p} \ d \underline{k}
\non
\qqq
Doing the $k_{1}$ and $k_{4}$ integrals, and shifting $k_2$, $k_3$, this equals
\qq
I_{3}= \int T\ast A^{\ast n_1} (2p_{1}) r (p_{2}, k_{2}) J (p_{3},k_{3}) f
(\underline{p}, k_{2}, k_{3}, p) \de (2p_{4}) \de \left(p-\sum p_{i}\right) d 
\underline{p}\ dk_{2} \  dk_{3}
\label{7(17)}
\qqq
with $f$ smooth.  We have $T\ast A^{\ast n_1} \in E$, $r,J\in \CS$. Proceeding as with $I_1$, we get
\qq
\|d^{-1} I_{3} \|_S &\leq & (C\|A\|_E)^{n_{1}} \|r\|_S \ \| J \|_S.
\label{I'b}
\qqq
We also have, by (\ref{Br50}):
\qq
|I_3(0)|\leq C N^{-1}.
\label{br41}
\qqq

\vspace*{4mm}

In a similar way, we may analyse $\th_{2}$ i.e. (\ref{5(22)}) with $\sum s_{i}
\neq 0$ and the first term in (\ref{5(23)}) inserted. We note that the terms that are odd
in $Q$ vanish since, for those terms,  because of the ${\bf s} \to {-\bf s}$ symmetry, 
the measure is proportional to $\de \Bigl(\sum s_{i} \om
(k_{i})\Bigr)$, which vanishes for $\sum s_i \neq 0$.

Starting again with the term linear in $J$ and with $r=0$, it is given by
\qq
\int  T\ast A^{\ast n_1}(2p_{1}) T\ast A^{\ast n_2} (2p_{2}) J (p_{3},k_{3}) h
(p_{3}+k_{3},k_{3})\de (2p_{4}) \de \left(p-2\sum p_{i}\right) \  d\underline{p} \ dk_{3}
\label{7(18)}
\qqq
with $h(p,k) = h(-p,-k)$ smooth. By oddness of $J$ in $k$, we may replace $h$ by
$h(k_{3}+p_{3}, k_{3}) - h (k_{3}-p_{3},k_{3})$ i.e., near $p_{3}=0$, $h$ is
$\CO (p_{3})$. Similarily, using $J(p+\pi, k+\pi) = J (p,k)$ and the
$2\pi$-periodicity of $h$, 
\qq
&& \int J (p_{3},k_{3}) h (p_{3}+k_{3}) dk_{3} = {1\over 2} \int J
(p_{3}-\pi,k_{3}) \left[h(k_{3}+p_{3}-\pi,  k_{3}+\pi) \right.
\left. - h (k_{3} - (p_{3}-\pi), k_{3}+\pi)\right]
\non
\qqq
i.e. (\ref{7(18)}) may be written as
\qq
 \int T\ast A^{\ast n_1} (2p_{1}) T\ast A^{\ast n_2} (2p_2) (e^{2ip_{3}}-1) J (p_{3},k_{3})
 \tilde h(p_{3},k_{3}) \de (2p_{4}) \de \left(p-2\sum p_{i}\right) \  d\underline{p} \
dk_{3}.
\label{Br4}
\qqq
Writing $e^{2ip_{3}}-1$ as in (\ref{Br19}), and 
 expanding the product, we see that the integral (\ref{Br4})
equals the sum of  terms of the form $I_1$ and of the form $I_2$, i.e.
\qq
\tilde  I_2=d(p) \int T\ast A^{\ast n_1} (2p_{1}) T\ast A^{\ast n_2} (2p_2) J (p_{3},k_{3})
 \tilde h(p_{3},k_{3}) \de (2p_{4}) \de \left(p-2\sum p_{i}\right) \  d\underline{p} \
dk_{3}
\label{7(19b)}
\qqq
with $\tilde h$ smooth. The integral in  (\ref{7(19b)}) is in $S$ and the prefactor $d(p)$ cancels the $d^{-1}$, so that $\tilde I_2$ has the same bound as in (\ref{Br20}).
Finally, the term with one $J$ and one $r$ is again of
the form (\ref{7(17)}). 

The remaining terms in $\th_{1}$ and $\th_{2}$ are of type $J^3$ and $Jr^2$.
These are bounded by brute force by
\qq
{\log N \over N^2} \|J\| (\| r \|^2 + \| J \|^2),
\label{brute}
\qqq
and considered as elements of $\CS$ of the form $(0, 0, 0, \star)$. Since $|d|^{-1}\leq  CN$ we
obtain by combining eqs. (\ref{I2c}), (\ref{I'b}), (\ref{7(19b)}) and
(\ref{brute})
\qq
\| d^{-1} (\th_1+\th_2) \|_{S} &\leq& C (\|t\|_S+\|A\|_E+\|r\|_S+\|J\|^2_S)\|J\|_S.
\label{theta121}
\qqq

We still need to estimate $p\th_{3} (p)$ in (\ref{5(28)}), i.e. the contribution to 
(\ref{5(22)}) of the second term in 
(\ref{5(23)}), which we can write as:
\qq
\om^{-1} (k_{3} - p) - \om^{-1} (k_{3}) - (3 \leftrightarrow 4) = p f (k_{3},
k_{4})+ \CO(p^2),
\label{Br32}
\qqq
with $f$ odd.
Consider the terms where $W_{s_{i}} = Q_{0},   \ \ \forall i$.

We get a sum of terms of the form
\qq
&&
\int \prod^3_{i=1} F_{n_{i}} (2p_{i}) \om^{-2-n_{i}} (p_{i}, k_{i} - p_{i})
s_{3}s_{4} \om (k_{3}) \om (k_{4})^{-1}.
\non\\
&&
 \left[\om (k_{3} -p)^{-1} - \om (k_{3})^{-1} + \om (k_{4} -p)^{-1} - \om
(k_{4})^{-1} \right] \nu' (d\un{p} \  d \un{k}).
\label{(20)}
\qqq
 Write:
\qq
\om^{-2-n_{i}}  (p_{i}, k_{i} - p_{i}) = \om (k_{i})^{-2-n_{i}} + (e^{2ip_{i}} -1)n_i
\CO (1).
\non
\qqq
Therefore (\ref{(20)}) gives rise to two contributions: the one coming from 
$\om (k_{i})^{-2-n_{i}} $;  after inserting (\ref{Br32}) in the  [-] in (\ref{(20)}), and writing
$p=-i d(p) + \CO(p^2)$, this contribution can be written as, :
\qq
\ha d(p)(F_{n_{1}} \ast F_{n_{2}} \ast F_{n_{3}}) (p)\bigl(I(p) - I (-p)\bigr) + \CO(p^2)(F_{n_{1}} * F_{n_{2}} * F_{n_{3}}) (p),
\label{(21)}
\qqq
with $F_{n}=T\ast A^{\ast n}$,
\qq
I(p)  = \int \de \bigl(\om (k_{1}) + \om (k_{2}) - \om (k_{3}) - \om (k_{4}) 
\bigr) \de \left(p - \sum^4_{1} k_{i}  \right) \phi (\un{k}) d
\un{k},
\label{(22)}
\qqq
where $\phi$ is smooth, and $I$ is odd (since
$ f (k_{3},
k_{4})$ above, and hence $\phi$, is odd). By Lemma 9.1,
$(F_{n_{1}} \ast F_{n_{2}} \ast F_{n_{3}}) (p)$ is in $E$, so $d(p)(F_{n_{1}} \ast F_{n_{2}} \ast F_{n_{3}}) (p)$ is in $S$.
By Lemma A.2., the first term in the RHS of 
 (\ref{(21)}) multiplied by $d^{-1}$ is in  $S$ and so is the last one, since $\CO(p^2)(F_{n_{1}} \ast F_{n_{2}} \ast F_{n_{3}})(p)$ is $N^{-1}$ times a $C^\al$ function.
 The second contribution, coming from $(e^{2ip_{i}} -1)n_i
\CO (1)$, is of the form:
\qq
d(p)n_i \int \prod^3_{i=1} F_{n_{i}} (2p_{i}) (e^{2ip_{1}}-1) \psi (\un{p},
\un{k},p) \nu' (d \un{p} \ d \un{k}),
\label{(23)}
\qqq
where $\psi$ is smooth. The integral in (\ref{(23)}) is in $S$ with norm bounded by $(C \|A\|_E)^{\sum n_i}$.
The other terms in $\th_3$ are simpler to bound, and we get:
\qq
\| d^{-1} \th_3 \|_{S} &\leq& C (\|t\|_S+\|A\|_E).
\label{theta3}
\qqq
Of course, $\th_3(0)=0$.
(\ref{theta121}) and 
 (\ref{theta3}) together with (\ref{7(19)}), (\ref{br40}), (\ref{br41}), yield the claims.
\hfill$
\makebox[0mm]{\raisebox{0.5mm}[0mm][0mm]{\hspace*{5.6mm}$\sqcap$}}$
$
\sqcup$

\vspace*{8mm}

\no {\bf {Acknowledgments.}}
We thank Joel Lebowitz, Rapha\"el Lefevere, Jani Lukkarinen, Alain Schenkel  and Herbert Spohn
for useful discussions.  A.K.  thanks the Academy of Finland
for funding.

\end{document}